\documentstyle[psfig,epsfig,12pt,phy]{book}

\makeatletter
\def\nothing#1{}
\newdimen\earraycolsep
\def\eqnarray{\let\@currentlabel\theequation
\global\@eqnswtrue\m@th
\global\@eqcnt\z@\tabskip\@centering\let \\\@eqncr
$$\halign to\displaywidth\bgroup\@eqnsel\hskip\@centering
$\displaystyle\tabskip\z@{##}$&\global\@eqcnt\@ne
\hskip 2\earraycolsep \hfil$\displaystyle{##}$\hfil
&\global\@eqcnt\tw@ \hskip 2\earraycolsep $\displaystyle\tabskip\z@{##}$\hfil
\tabskip\@centering&\llap{##}\tabskip\z@\cr}
\renewcommand{\theequation}{\arabic{equation}}
\renewcommand{\thetable}{\arabic{table}}
\renewcommand{\thefigure}{\arabic{figure}}
\def\title{\chapter}
\renewcommand\chapter{\ifodd\c@page\clearpage\else\cleardoublepage\fi
		    \global\@topnum\z@
		    \@afterindenttrue
		    \secdef\@chapter\@schapter}
\def\@makechapterhead#1{%
  \vspace*{120\p@}%
  {\parindent \z@ \raggedright \reset@font
    \bfseries #1\par
    \nobreak
    \vskip 36\p@
  }}
\def\author#1
{{\pretolerance=10000 \raggedright \advance 
\leftskip by 1in \noindent #1 \vskip 1pc}}
\def\affiliation#1{{\advance\leftskip by 1in \noindent #1 \vskip -1pc}}

\def\tablenote#1
{\setbox0=\hbox{$^{\hbox{\scriptsize #1}}$}\noindent\hangindent=\wd0 \box0}
\renewcommand\section{\@startsection{section}{1}{\z@}{2pc \@plus 1ex minus
    .2ex}{1pc \@plus .2ex}{\reset@font\normalsize\bfseries}}
\renewcommand\subsection{\@startsection{subsection}{2}{\z@}{1pc \@plus 1ex
    minus.2ex}{1pc \@plus .2ex}{\reset@font\normalsize\bfseries}}
\renewcommand\subsubsection{\@startsection{subsubsection}{3}{\parindent}
	{1pc \@plus 1ex minus.2ex}{-0.5em}{\reset@font\normalsize\bfseries}}

\def\AmS{{\protect\the\textfont2 A\kern-.1667em\lower.5ex\hbox{M}\kern-.125emS}}

\def\p@LaTeX{{\family{times}\series{m}\shape{n}
\selectfont L\kern-.36em\raise.3ex
\hbox{\scriptsize A}\kern-.15em T\kern-.1667em\lower.7ex\hbox{E}\kern-.125emX}}

\newlength{\colwidth}

\def\@oddhead{\hfil}
\def\@evenhead{\hfil}
\def\@oddfoot{{\bfseries\hfil\thepage}}
\def\@evenfoot{{\bfseries\thepage\hfil}}
\def\fnum@figure{\footnotesize\raggedright{\bfseries \figurename~\thefigure.}}
\def\fnum@table{\normalsize\raggedright{\bfseries \tablename~\thetable.}}
\long\def\@makecaption#1#2{\vskip 10\p@ {#1 #2\par}}
\long\def\@makefntext#1{\setbox0=\hbox{$\m@th^{\@thefnmark}$}\noindent
\hangindent=\wd0 \box0 #1}
\def\centerfig#1#2#3#4{\vspace*{#2}\relax\centerline
{\hbox to#1{\special{#4:#3.#4 x=#1, y=#2}\hfil}}}
\newbox\@atbox
\long\def\atable#1#2#3{\begin{table}[tbp]\centering\footnotesize
\setbox\@atbox\hbox{#2}
\parbox{\wd\@atbox}{\caption{#1}}\par\smallskip #2
\par\smallskip\parbox{\wd\@atbox}{\raggedright #3}
\end{table}}
\def\@bibitem{\noindent \hangindent=2pc \hangafter=1}
\def\thebibliography{%
\section*{REFERENCES}%
\bgroup\footnotesize
\def\newblock{\hskip .11em plus.33em minus.07em}%
\let\bibitem\@bibitem}
\def\endthebibliography{\par\egroup}
\def\@nbibitem#1{\noindent \hangindent=2pc \hangafter=1
\refstepcounter{enumi}\hbox to 2pc{\arabic{enumi}.\hfil}%
\immediate\write\@auxout{\string\bibcite{#1}{\arabic{enumi}}}}
\def\numbibliography{%
\section*{REFERENCES}%
\bgroup\footnotesize
\setcounter{enumi}{0}%
\def\newblock{\hskip .11em plus.33em minus.07em}%
\let\bibitem\@nbibitem}
\def\endnumbibliography{\par\egroup}
\makeatother

\catcode`@=11
\newbox\Lbox
\newbox\Rbox
\newdimen\Twofigdimen
\Twofigdimen=\textwidth
\advance\Twofigdimen by -0.25truein
\divide\Twofigdimen by 2
\def\Lcaption[#1]#2{%
   \stepcounter{figure}%
   \def\@currentlabel{\thefigure}%
   \addcontentsline{lof}{figure}{\numberline{\thefigure}{\ignorespaces #1}}
   \setbox\Lbox=\vtop{\hsize=3in
      \raggedright\tolerance=10000
      \noindent Figure~\thefigure: #2}%
}
\def\Rcaption[#1]#2{%
   \stepcounter{figure}%
   \def\@currentlabel{\thefigure}%
   \addcontentsline{lof}{figure}{\numberline{\thefigure}{\ignorespaces #1}}
   \setbox\Rbox=\vtop{\hsize=3in
      \raggedright\tolerance=10000
      \noindent Figure~\thefigure: #2}%
   \bigskip
   \centerline{\copy\Lbox\hfill\copy\Rbox}%
}
\def\Twofigs#1#2{%
   \hbox to \textwidth{\hbox to \Twofigdimen{\hss{#1}\hss}\hfil
      \hbox to \Twofigdimen{\hss{#2}\hss}}}
\catcode`@=12
\def\mup{$\mu^+$}
\def\mum{$\mu^-$}
\def\({ \left( }
\def\){ \right) }
\def\b{\begin{equation}}
\def\e{\end{equation}}
\def\={\ =\ }

\def\+{\ +\ }
\def\-{\ -\ }

\def\Ls{\cal L \rm}

\def\mumu{$\mu^+\mu^-$}
\def\ee{$e^+e^-$}
\def\pp{$pp$}
\def\ppbar{$p\bar{p}$}

\def\GeV{{\rm GeV}}
\def\TeV{{\rm TeV}}

\def\fb{{\rm fb}}
\def\simge{                       
 \mathrel{\rlap{\raise 0.511ex
	\hbox{$>$}}{\lower 0.511ex \hbox{$\sim$}}}}
\def\simle{
    \mathrel{\rlap{\raise 0.511ex
	\hbox{$<$}}{\lower 0.511ex \hbox{$\sim$}}}}
\def\Fig#1{Fig.~\ref{fg.#1}}

\def\sdc{silicon drift detector}

\def\mumu{$\mu ^+\mu ^-$\ }

\def\E0c{\frac{E_0}{c}}
\def\1s2{\frac{1}{\sqrt{2}}}
\def\cross{\!\times\!}

\begin{document}
\title{MUON-MUON AND OTHER HIGH ENERGY COLLIDERS}
\author{R. B. Palmer, J. C. Gallardo
} 
\affiliation{Center for Accelerator Physics\\
Brookhaven National Laboratory \\  
Upton, NY 11973-5000, USA}
\tableofcontents
\listoffigures
\listoftables
\clearpage

\section{COMPARISON OF COLLIDER TYPES}
\vskip -1pc
\subsection{Introduction}

Before we discuss the muon collider in detail, it is useful to look
at the other types of colliders for comparison.
In this chapter we consider the high energy physics advantages, disadvantages 
and luminosity requirements of hadron (\pp, \ppbar), of lepton (\ee, \mumu) 
and photon-photon colliders. Technical problems in obtaining  increased energy 
in each type of machine are presented. Their relative size, and probable 
relative costs  are discussed. 
\subsection{Physics Considerations}
\vskip -1pc
\subsubsection{General.}
Hadron-hadron colliders (\pp \, or \ppbar) generate interactions between  the
many constituents of the  hadrons (gluons, quarks and antiquarks); the initial 
states are not defined and most interactions occur at relatively  low energy,
generating a very large background of uninteresting  events. The rate of the
highest energy events is higher for antiproton-proton machines, because  
the antiproton contains valence antiquarks that can annihilate on the quarks 
in the proton. But this is a small effect for  colliders above a few TeV, when 
the interactions are dominated by interactions between quarks and antiquarks 
in their seas, and between the gluons. In either case the individual 
parton-parton interaction energies (the energies used for physics) are a relatively 
small fraction of the total center of mass energy. This is a disadvantage when compared with lepton machines. An advantage, however, is 
that all final states are accessible. Many, if not most, initial  discoveries 
in Elementary Particle Physics have been made with these machines. 

In contrast, lepton-antilepton collider generate interactions between the fundamental 
point-like  constituents in their beams, the reactions generated are  
relatively simple to understand, the full machine energies are available for 
``physics", and there is negligible background of low  energy events. If the center of 
mass energy is set equal to the  mass of a suitable state of interest, then 
there can be a large  cross section in the {\bf s}-channel, in which a single 
state is  generated by the interaction. In this case, the mass and quantum  
numbers of the state are constrained by the initial beams. If the  energy 
spread of the beams is sufficiently narrow, then precision  determination of 
masses and widths are possible. 

A gamma-gamma collider, like the lepton-antilepton machines, would also have 
all the machine energy  available for physics, and would have well defined 
initial states, but these states would be different from those with the lepton 
machines, and thus be complementary to them. 

For most purposes (technical considerations aside) \ee and \mumu  colliders
would be equivalent. But in  the particular case of {\bf s}-channel Higgs boson
production,  the cross section, being proportional to the mass squared, is 
more than 40,000 times greater for muons than electrons.  When technical
considerations are included, the situation is  more complicated. Muon beams are
harder to polarize and muon  colliders will have much higher backgrounds from
decay  products of the muons. On the other hand muon collider  interactions
will require less radiative correction and will have  less energy spread from
beamstrahlung.

Each type of collider has its own advantages and disadvantages 
for High Energy Physics: they would be complementary. 
\subsubsection{Required Luminosity for Lepton Colliders.} 

   In lepton machines the full center of mass of the leptons is 
available for the final state of interest and a ``physics 
energy" $E_{\rm phy}$ can be defined that is equal to the total center of mass 
energy. 
 \b
E_{\rm phy}\ =\ E_{c\ of\ m}
 \e

\begin{figure}[hbt!] 
\centerline{\epsfig{file=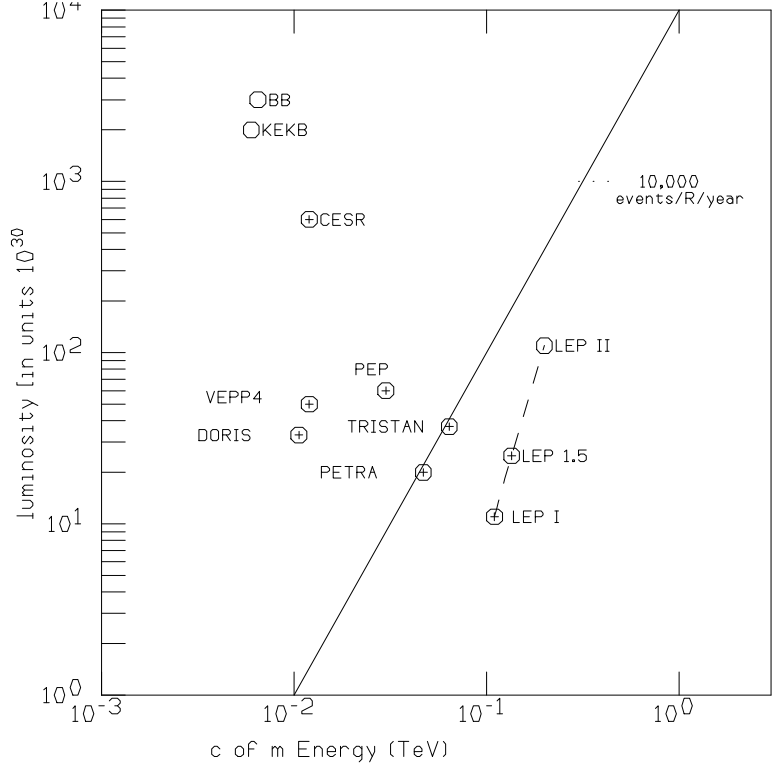,height=4.0in,width=4.0in}}
 \caption{Luminosity of lepton colliders as a 
function of Energy
 \label{lumfig} }
\end{figure}

   Since fundamental cross sections fall as the square of the 
center of mass energies involved, so, for a given rate of events, 
the luminosity of a collider must rise as the square of its 
energy. A reasonable target luminosity is one that would give 
10,000 events per unit of R per year (the cross section for lepton pair 
production is one R, the total cross section is about 20 R, and somewhat 
energy dependent as new channels open up): 
\b
\Ls_{req.}\ \approx \  10^{34}\ (cm^{-2} s^{-1})\  \left( {E_{phy} \over 1\ 
(TeV)} \right)^2 \label{reqlum} 
\e
   Fig.~\ref{lumfig} shows this required luminosity, together with crosses at 
the approximate achieved luminosities of some lepton colliders. Target 
luminosities of possible future colliders are also given as circles. 
\subsubsection{The Effective Physics Energies of Hadron Colliders.}
Hadrons, being composite, have their energy divided between 
their various constituents. A typical collision of 
constituents will thus have significantly less energy than that 
of the initial hadrons. Studies done in Snowmass 82 and 96 
suggest that, for a range of studies, and given the required luminosity (as 
defined in Eq.~\ref{reqlum}), then the hadron machine's effective 
``physics" energy is between about  1/3 and 1/10 of its total. We will take a value of 1/7: 
  $$
E_{\rm phy}(\Ls=\Ls_{req.})\ \approx\ {E_{c\ of\ m} \over 7}
  $$
The same studies have also concluded that a factor of 10 in 
luminosity is worth about a factor of 2 in effective physics energy, this 
being approximately equivalent to: 
  $$
E_{\rm phy}(\Ls)\= E_{\rm phy}(\Ls=\Ls_{req.})\ \({ \Ls \over \Ls_{req} }\)^{0.3}
  $$
From which, with Eq.~\ref{reqlum}, one obtains:
 \b
E_{\rm phy} \approx  
  \( {E_{c\ of\ m}\over 7 (TeV)} \)^{0.6}
               \( {\Ls \over 10^{34}(cm^{-2}s^{-1})} \)^{0.2}  (TeV)
\label{Eeffeq}
 \e
\begin{table}[thb!]  
\centering \protect
\caption{Effective Physics Energy of Some Hadron Machines}
\begin{tabular}{lccc}
\hline 
Machine & C of M Energy & Luminosity & Physics Energy \\
         &  TeV         &$cm^{-2}s^{-1}$& TeV   \\
\hline
ISR & .056 & $10^{32}$ & 0.02 \\
TeVatron & 1.8 & $7\times 10^{31}$ & 0.16 \\
LHC   & 14 & $10^{34}$ & 1.5 \\
VLHC   & 60 & $10^{34}$ & 3.6 \\
\hline
\end{tabular}
\label{physE}
\end{table}
Tb.~\ref{physE} gives some examples of this approximate ``physics" energy.
It must be emphasized that this effective physics energy is not a well 
defined quantity. It should depend on the physics being studied. The initial 
discovery of a new quark, like the top, can be made with a significantly lower 
``physics" energy than that given here. And the capabilities of different types 
of machines have intrinsic differences. The above analysis is useful only in 
making very broad comparisons between machine types.
\subsection{Hadron-Hadron Machines}
\vskip -1pc
 \subsubsection{Luminosity.}
An antiproton-proton collider requires only one ring, compared  with the two
needed for a proton-proton machine (the antiproton has the opposite charge to 
the proton and can thus rotate in the same magnet ring in the opposite 
direction - protons going in opposite directions require two rings with 
bending fields of the opposite sign), but the  luminosity of an antiproton-
proton collider is limited by the  constraints in antiproton production. A 
luminosity of at least $10^{32}\ {\rm cm}^{- 2}{\rm s}^{-1}$ is expected at 
the antiproton-proton Tevatron; and a luminosity of $10^{33}\ {\rm cm}^{- 
2}{\rm s}^{-1}$ may be achievable, but  LHC, a proton-proton machine, is 
planned to have a luminosity  of $10^{34}\ {\rm cm}^{-2}{\rm s}^{-1}$. Since 
the required luminosity rises with energy, proton-proton machines seem to be 
favored for future hadron colliders. 

   The LHC and other future proton-proton machines might even\cite{lumlim}  
 be upgradable to $10^{35}\ {\rm cm}^{-2}{\rm s}^{-1}$, but radiation damage 
to a detector would then be a severe problem. The 60 TeV Really 
Large Hadron Colliders (RLHC: high and low field versions) discussed at 
Snowmass are  being designed as proton-proton machines with luminosities of  
$10^{34}\ {\rm cm}^{-2}{\rm s}^{-1}$ and it seems reasonable to assume that 
this is the highest practical value. 
\subsubsection{Size and Cost.}
The size of hadron-hadron machines is limited by the field of  the magnets used
in their arcs. A cost minimum is obtained when a  balance is achieved between
costs that are linear in length, and  those that rise with magnetic field. The
optimum field will  depend on the technologies used both for the the linear 
components (tunnel, access, distribution, survey, position  monitors,
mountings, magnet ends, etc) and those of the magnets  themselves, including
the type of superconductor used. 

The first hadron collider, the 60 GeV ISR at CERN, used  conventional iron pole
magnets at a field less than 2~T. The  only current hadron collider, the 2 TeV
Tevatron, at FNAL, uses NbTi  superconducting magnets at approximately
$4\,{}^\circ K$ giving a bending field of about 4.5 T. The  14 TeV Large 
Hadron Collider (LHC), under construction at CERN,  plans to use the same 
material at $1.8\,{}^\circ K$ yielding bending fields of about $8.5\,{\rm T}.$ 

 \begin{figure}[hbt!] 
\centerline{\epsfig{file=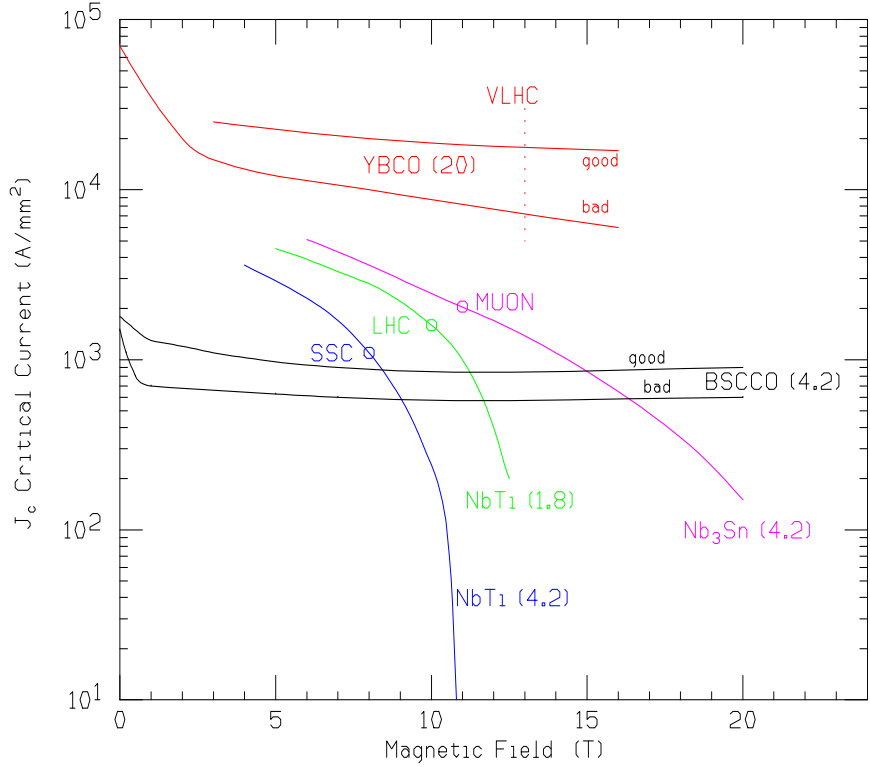,height=4.5in,width=4.0in,angle=90}}
\caption{Critical current densities of superconductors as a function of
magnetic field. \label{super} }
\end{figure} 
Future colliders may use new materials allowing even higher  magnetic fields.
Fig.~\ref{super} shows the critical current densities of various superconductors
as a function of magnetic field. The numbers in parenthesis refer to the
temperatures in ${}^{\circ}$ K. {\it good} and {\it bad} refer to the best and
worst performance according to the orientation of the tape with
respect to the direction of the magnetic field.
Model magnets have been made with ${\rm Nb_3Sn},$ and  studies are underway on the
use of high T$_c$ superconductor.  {Bi$_2$Sr$_2$Ca$_1$Cu$_2$O$_8$} (BSCCO)
material is currently available in useful  lengths as powder-in-Ag tube
processed tape. It has a higher  critical temperature and field than
conventional superconductors,  but, even at $4\,{}^\circ K,$ its current
density is less than ${\rm Nb_3Sn}$  at all fields below 15~T. It is thus unsuitable
for most accelerator magnets. In contrast YBa$_2$Cu$_3$O$_7$ (YBCO) 
material has a current density above that for Nb$_3$Sn ($4\,{}^\circ K$), at
all fields and temperatures below $20\,{}^\circ K.$  But this material must be
deposited on specially treated  metallic substrates and is not yet available in
lengths greater  than 1~m. It is reasonable to assume, however, that it will be 
available in useful lengths in the not too distant future.  

\begin{figure}[htb!] 
\centerline{\epsfig{file=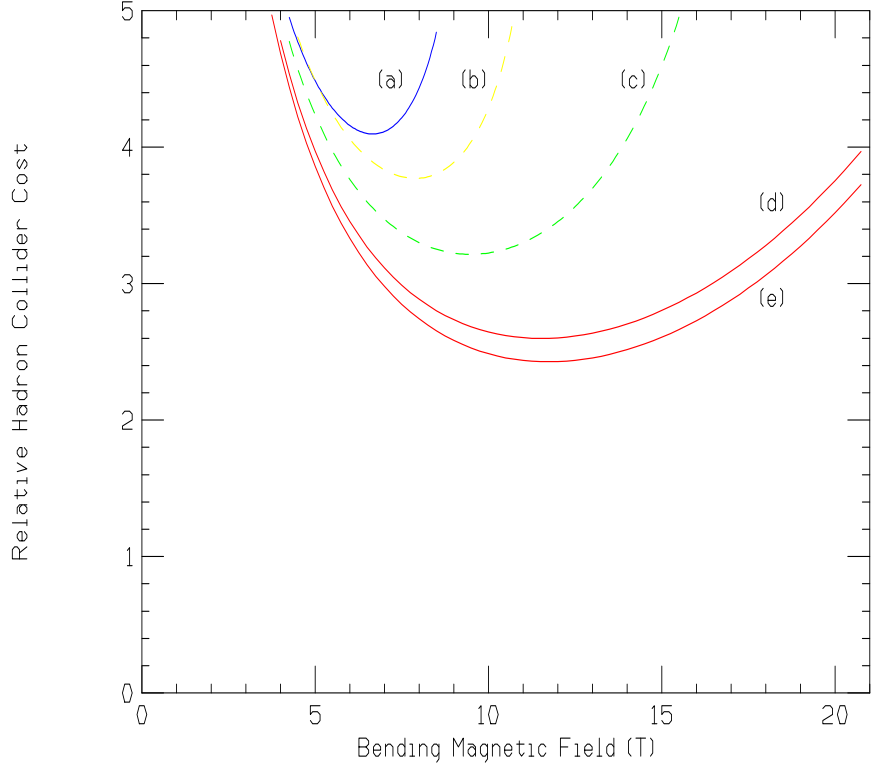,height=4.5in,width=4.0in,angle=90}}
\caption[Relative costs of a collider as a function of its bending magnetic 
field, for different superconductors and operating temperatures. ]{Relative costs of a collider as a function of its bending magnetic 
field, for different superconductors and operating temperatures.
Costs are given for NbTi at
(a) $4\,{}^\circ K$, and (b) $1.8\,{}^\circ K$,  (c) Nb$_3$\,Sn at
$4.3\,{}^\circ K$, and (d) and (e) YBCO High T$_c$ at  $20\,{}^\circ K.$ 
NbTi  and Nb$_3$\,Sn  costs per unit weight were taken to be the  same; YBCO
was taken to be either equal to NbTi (in (d)), or 4 times NbTi (in (e)). \label{mag} }
\end{figure} 
A parametric study was undertaken to learn what the use of  such
materials might do for the cost of colliders. 2-in-1 cosine  theta
superconducting magnet cross sections (in which the two magnet coils are 
circular in cross section, have a cosine theta current distributions and are 
both enclosed in a single iron yoke) were calculated using  fixed criteria for 
margin, packing fraction, quench protection,  support and field return. 
Material costs were taken to be linear  in the weights of superconductor, 
copper stabilizer, aluminum  collars, iron yoke and stainless steel support 
tube. The  cryogenic costs were taken to be inversely proportional to the 
operating temperature, and linear in the outer surface area of  the cold mass. The values of the cost dependencies were scaled from LHC  estimates. 

Results are shown in Fig.~\ref{mag}. Costs were calculated  assuming NbTi at
(a) $4\,{}^\circ K$, and (b) $1.8\,{}^\circ K,$  Nb$_3$\,Sn at (c) 
$4.3\,{}^\circ K$ and YBCO High T$_c$ at $20\,{}^\circ K$ (d) and (e).  
NbTi  and Nb$_3$\,Sn  costs per unit weight were taken to be the  same; YBCO
was taken to be either equal to NbTi (in (d)), or 4 times NbTi (in (e)).
It is seen that the optimum field moves from about 6~T for  NbTi at
$4\,{}^\circ K$ to about 12~T for YBCO at $20\,{}^\circ K$;  while the total
cost falls by almost a factor of 2.

One may note that the optimized cost per  unit length remains approximately 
constant. This might have been  expected: at the cost minimum, the cost of 
linear and field  dependent terms are matched, and the total remains about 
twice  that of the linear terms.  

The above study assumes this particular  type of magnet and
may not be indicative of the optimization for  radically different designs. A
group at FNAL\cite{pipe} is  considering an iron dominated, alternating
gradient, continuous,  single turn collider magnet design (Low field RLHC). Its
field would  be only 2~T and circumference very large (350 km for 60 TeV), but 
with its simplicity and with tunneling innovations, it is hoped to  make its
cost lower than the smaller high field designs. There  are however greater
problems in achieving high luminosity with  such a machine than with the higher
field designs. 
 \subsection{Circular \ee Machines}
\vskip -1pc
\subsubsection{Luminosity.}
 The luminosities of most circular electron-positron  colliders has
been between $10^{31}$ and $10^{32}\,{\rm cm}^{-2}{\rm s}^{-1}$ (see Fig.\ref{lumfig}),  CESR is fast
approaching $10^{33}\,{\rm cm}^{-2}{\rm s}^{-1}$ and machines are  now being constructed with even high
values. Thus, at least in  principle, luminosity does not seem to be a
limitation (although it may be noted that the 0.2~TeV electron-positron
collider LEP has a  luminosity below the requirement of Eq.\ref{reqlum}).

  \subsubsection{Size and Cost.}
At energies below 100 MeV, using a reasonable bending  field, the size
and cost of a circular electron machine is  approximately proportional to its
energy. But at higher energies,  if the bending field $B$ is maintained, the energy lost
$\Delta  V_{\rm turn}$ to synchrotron radiation rises rapidly 
 \b
 \Delta  V_{\rm turn}\ \propto \ {E^4 \over R\ m^4}\ \propto \ {E^3\ B \over m^4}
\label{syncheq}
 \e
and soon becomes excessive ($R$ is the radius of the ring). A cost minimum is then obtained when the cost of
the ring is balanced by the cost of the  rf needed to replace the synchrotron
energy loss. If the ring  cost is proportional to its circumference, and the rf
is  proportional to its voltage then the size and cost of an  optimized machine
rises as the square of its energy. This  relationship is well demonstrated by
the parameters of actual  
machines as shown later in  Fig.~\ref{length}. 

   The highest circular \ee collider is the LEP at CERN which has a 
circumference of 27 km, and will achieve a maximum center of mass 
energy of about 0.2 TeV. Using the predicted scaling, a 0.5 TeV 
circular collider would have to have a 170 km circumference, and 
would be very expensive. 

 \subsection{\ee Linear Colliders}
\vskip -1pc
 \subsubsection{Size and Cost.}
So, for energies much above that of LEP (0.2 TeV) it is probably
impractical to build 
a circular electron collider. The only  possibility then
is to build two electron 
linacs facing one  another. Interactions occur at the center, and the 
electrons,  after they have interacted, must be discarded. 
\subsubsection{Luminosity.}
 The luminosity $\Ls$ of a linear collider can be written: 
 \b
   \label{lumeq}
   \Ls \= {1 \over 4\pi E}\ \ {N \over \sigma_x}\ 
    {P_{\rm beam} \over \sigma_y}
   \ \ n_{\rm collisions}
 \e
where $\sigma_x$ and 
$\sigma_y$ are average beam spot sizes including any pinch 
effects, and we take $\sigma_x$ to be much 
greater than $\sigma_y$. $E$ is the 
beam energy, $P_{\rm beam}$ is the total beam power,
and, in this case, $n_{\rm collisions}=1$.
This can be expressed\cite{yokoyachen} as,
 \b
\label{constraint}
\Ls \ \approx \ {1 \over 4\pi E}\ \ {n_\gamma \over 2 r_o \alpha \ U(\Upsilon)}\ \
       \ \ {P_{\rm beam} \over \sigma_y}
 \e
where the quantum correction $U(\Upsilon)$ is given by 
 \b
   U(\Upsilon)\ \approx \ \sqrt{{1 \over {1+\Upsilon^{2/3}}}  }
 \e
with
 \b
   \Upsilon\ \approx \ {2 F_2 r_e^2 \over \alpha} 
   \ {N \ \gamma \over \sigma_z \ \sigma_x}
 \e
$F_2\approx 0.43$, $r_o$ is the classical 
electromagnetic radius, $\alpha$ is 
the fine-structure constant, and $\sigma_z$ is the rms bunch length. 
The quantum correction $\Upsilon$
is close to unity for all 
proposed machines with energy less than 2 TeV, and this term is often 
omitted\cite{peskin}. 
Even in a 5 TeV design\cite{me}, an $\Upsilon$ of 21
gives a suppression factor of only 3. 

$n_{\gamma}$ is the 
number of photons emitted by one electron as it passes through the 
other bunch. 
If $n_\gamma$ is significantly greater than one, then problems are incountered 
with 
backgrounds of electron pairs and mini-jets, or
with unacceptable beamstrahlung energy loss.
Thus $n_\gamma$ 
can be taken as a rough criterion of these effects and
constrained to a fixed value. We then find:
 $$
\Ls  \ \propto {1 \over  E}\ \  {P_{beam} \over \sigma_y\ U(\Upsilon)}
 $$
which may be compared to the required luminosity that increases 
as the square of energy, giving the requirement: 
 \b
\label{Ecubed}
{P_{\rm beam} \over \sigma_y\ U(\Upsilon)}\  \propto \ E^3.
 \e
It is this requirement that makes it hard to design very high 
energy linear colliders. High beam power demands high efficiencies and heavy wall power consumption. A small $\sigma_y$ requires tight tolerances, low beam emittances and strong final focus and a small value of $U(\Upsilon)$ is hard to obtain because of 
its weak dependence on $\Upsilon$ ($\propto \Upsilon^{-1/3}$).

\subsubsection{Conventional RF.}

\begin{figure}[hbt!] 
\centerline{\epsfig{file=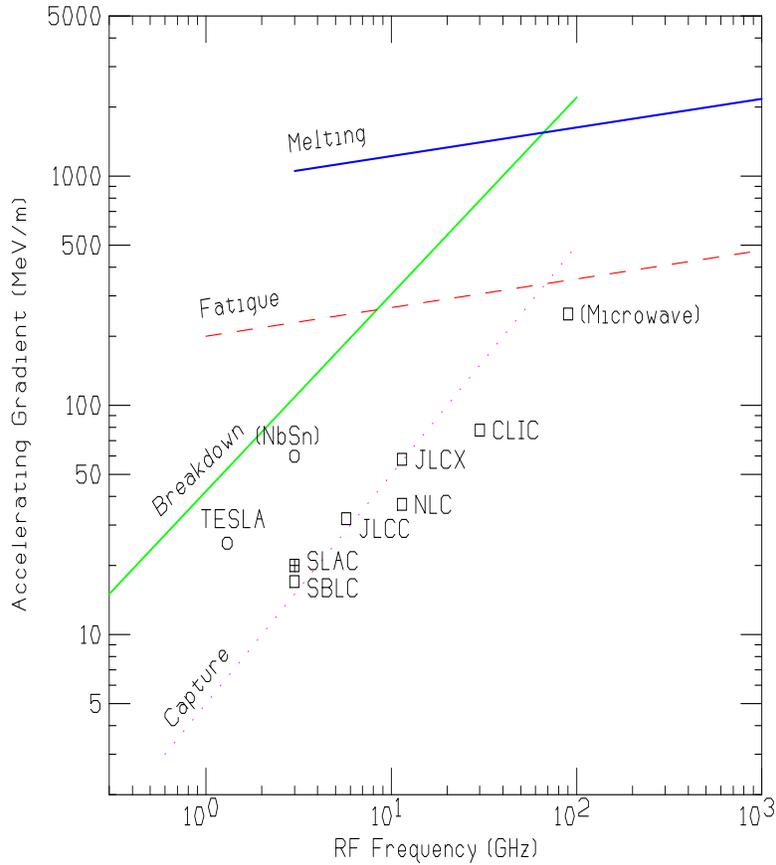,height=4.5in,width=4.0in}}
 \caption{Gradient values and limits in linear 
collider electron linacs, superconducting examples are indicated as circles
 \label{grad} }
 \end{figure}

 The gradients for structures have limits that are frequency dependent. 
Fig.~\ref{grad} shows the gradient limits from breakdown, fatigue and dark 
current capture, plotted against the operating rf frequency. Operating 
gradients and frequencies of several linear collider designs\cite{bluebook} 
are also indicated. 

One sees that for conventional structure designs (indicated as squares in 
Fig.~\ref{grad}), the proposed gradients fall well below the limits, except 
for the dark current capture threshold. Above this threshold, in the absence 
of focusing fields, dark current electrons emitted in one cavity can be 
captured and accelerated down the entire linac causing loading problems. We 
note, however, that the superconducting TESLA design is well above this limit, 
and a detailed study\cite{akasaka}  has shown that the quadrupole fields in a 
focusing structure effectively stop the build up of such a current. 

The real limit on accelerating gradients in these designs come from a trade 
off between the cost of rf power against the cost of length. The use of high 
frequencies reduces the stored energy in the cavities, reducing the rf costs 
and allowing higher accelerating gradients: the optimized gradients being 
roughly proportional to the frequency. One might thus conclude then that 
higher frequencies should be preferred. There are however counterbalancing 
considerations from the requirements of luminosity. 
 \begin{figure}[hbt!] 
\centerline{\epsfig{file=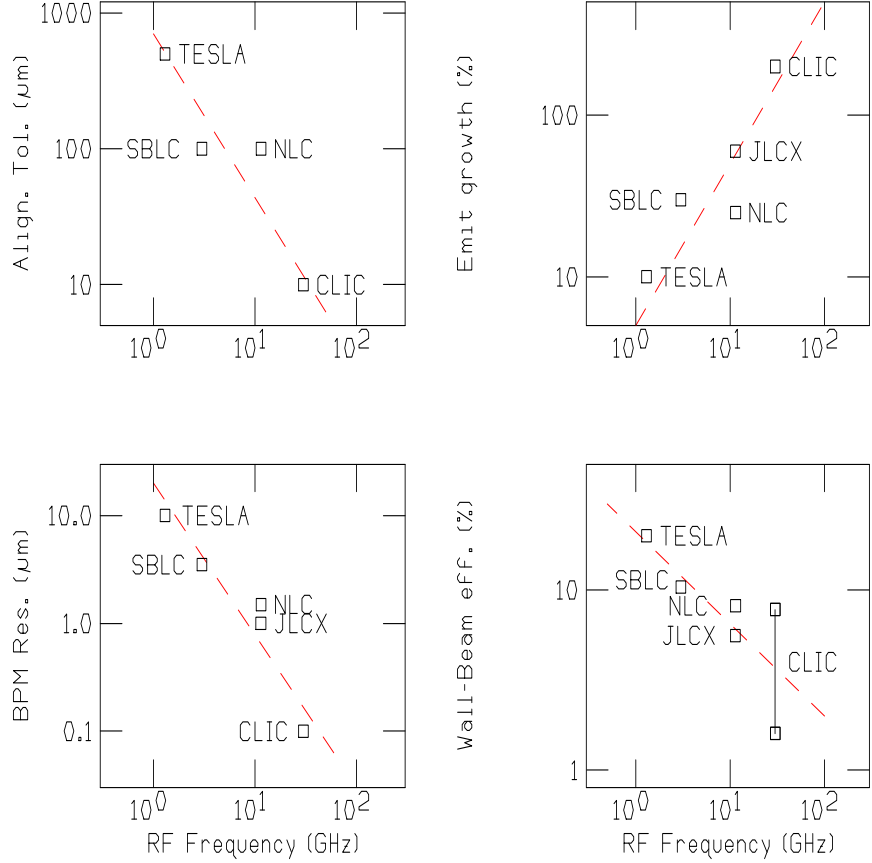,height=4.5in,width=4.5in}}
 \caption{Dependence of some sensitive parameters 
as a function of linear collider rf frequency.
 \label{freq} }
 \end{figure} 
    
Fig.~\ref{freq}, using parameters from the linear collider 
proposals \cite{bluebook}, plots some relevant parameters against 
the rf frequency. One sees that as the frequencies rise, 
 \begin{itemize}
 \item the required alignment tolerances are tighter;
 \item the resolution of beam position monitors must also be better; and
 \item despite these better alignments, the calculated  emittance growth 
during acceleration is worse; and
 \item the wall-power to beam-power efficiencies are also less.
 \end{itemize}
Thus while length and cost considerations may favor high 
frequencies, yet luminosity considerations demand lower 
frequencies. 
\subsubsection{Superconducting RF.}

         If, however, the rf costs can be reduced, for instance when 
superconducting  cavities are used, then there will be no trade off between rf 
power cost and length and higher gradients should be expected to lower the 
length and cost. 
The removal of the constraint applied by rf power considerations is evident 
for the TESLA gradient plotted in Fig.~\ref{grad}. Its value is well above 
the trend of conventional rf designs. Unfortunately the gradients achievable 
in currently operating niobium superconducting cavities is lower than that 
planned in the higher frequency conventional rf colliders. Theoretically the 
limit is about 40~MV/m, but practically 25~MV/m is as high as seems possible. 
Nb$_3$Sn and high Tc materials may allow higher field gradients in the future. 
A possible value for Nb$_3$Sn is also indicated on Fig.~\ref{grad}. 

In either case, the removal of the requirements for huge peak rf power allows 
the choice of longer wavelengths (the TESLA collaboration is proposing 23 cm 
at 1.3 GHz) and greatly relieves the emittance requirements and tolerances, 
with no loss of luminosity. 

At the current 25 MeV per meter gradients, the length and cost of a 
superconducting machine is probably higher than for the conventional rf designs. With
greater luminosity more certain, its proponents can argue that it is worth it the greater price.
If higher gradients become possible, using new superconductors, 
then the advantages of a superconducting 
solution could become overwhelming.   
\subsubsection{At Higher Energies.}
\begin{figure}[ht!] 
\centerline{\epsfig{file=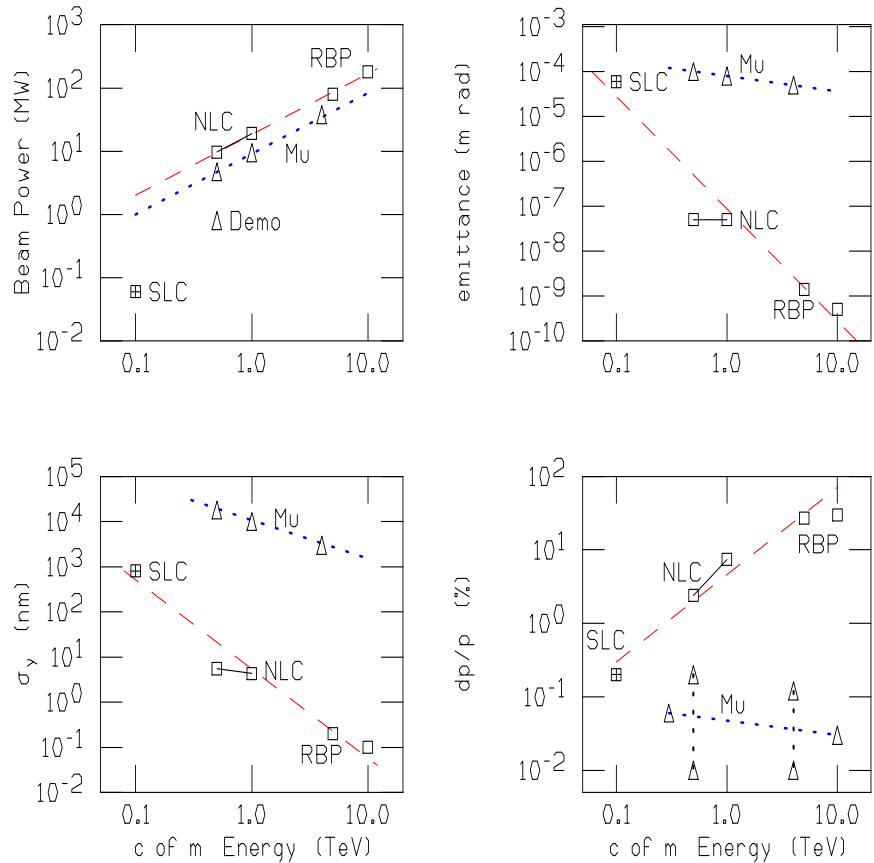,height=4.5in,width=4.5in}}
 \caption{Dependence of some sensitive 
parameters on linear collider energy, with comparison of same parameters for 
\mumu colliders. 
 \label{energy} }
 \end{figure}

    At higher energies (as expected from Eq.~\ref{Ecubed}),  
obtaining the required luminosity gets harder. Fig.\ref{energy} 
shows the dependency of some example machine 
parameters with energy.  SLC is taken as the example at 0.1 TeV, 
NLC parameters at 0.5 and 1 TeV, and 5 and 10 TeV examples are 
taken from a review paper by one of the authors\cite{me}. One sees 
that: 
 \begin{itemize}
 \item the assumed beam power rises approximately as $E$;
 \item the vertical spot sizes fall approximately as $E^{-2}$;
 \item the vertical normalized emittances fall even faster: $E^{-2.5}$; 
and
 \item the momentum spread due to beamstrahlung has been allowed to rise almost linearly with $E$. 
 \end{itemize}

   These trends are independent of the acceleration method, 
frequency, etc, and indicate that as the energy and required 
luminosity rise, so the required beam powers, efficiencies, 
emittances and tolerances will all get harder to achieve. The 
use of higher frequencies or exotic technologies that would allow 
the gradient to rise, will, in general, make the achievement of 
the required luminosity even more difficult. It may well prove
impractical to construct linear electron-positron colliders, with 
adequate luminosity, at energies above a few TeV. 
 \subsection{$\gamma-\gamma$ Colliders}

   A gamma-gamma collider\cite{telnov} would use opposing electron 
linacs, as in a linear electron collider,  but just prior to the 
collision point, laser beams would be Compton backscattered off the 
electrons to generate photon beams that would collide at the IP instead of the 
electrons. If suitable geometries are used, the mean 
photon-photon energy could be 80\% or more of that of the 
electrons, with a luminosity about 1/10th. 

   If the electron beams, after they have backscattered the 
photons, are deflected, then backgrounds from beamstrahlung can 
be eliminated. The constraint on ${N / \sigma_x}$ in Eq.\ref{lumeq} is thus removed and one might hope that higher 
luminosities would now be possible by raising $N$ and lowering 
$\sigma_x$. Unfortunately, to do this, one needs sources of 
larger number of electron bunches with smaller emittances, and one must 
find ways to accelerate and focus such  beams without excessive 
emittance growth. Conventional damping rings will have difficulty 
doing this\cite{mygamma}. Exotic electron sources would be needed, and 
methods using lasers to generate\cite{palmerchen} or cool\cite{telnovcool} 
 the electrons and positrons are under 
consideration.  

   Thus, although gamma-gamma collisions can and should be made 
available at any future electron-positron linear collider, to add 
physics capability, they may not give higher luminosity for a 
given beam power. 
\subsection{\mumu Colliders}
   There are two advantages of muons, as opposed to 
electrons, for a lepton collider. 
 \begin{itemize}
\item
   The synchrotron radiation, that forces high energy electron 
colliders to be linear, is (see Eq. \ref{syncheq}) inversely 
proportional to the fourth power of mass: It is negligible in muon 
colliders with energy less than 10 TeV. Thus a muon collider, up 
to such energy, can be circular. In practice this means in 
can be smaller. The linacs for a 0.5 TeV NLC would be 20 km 
long. The ring for a muon collider of the same energy would be 
only about 1.2 km circumference. 
\item
   The luminosity of a muon collider is given by the same formula 
(Eq.~\ref{lumeq}) as given above for an electron positron collider, but there 
are two significant changes: 1) The classical radius $r_o$ is now that for the 
muon and is 200 times smaller; and 2) the number of collisions a bunch can 
make $n_{collisions}$ is no longer 1, but is now related to the average 
bending field in the muon collider ring, with 
 $$
n_{collisions} \ \approx \ 150  \ B_{ave}
 $$
With an average field of 6 Tesla, $n_{collisions}\approx 900$.
Thus these two effects give muons an {\it in principle} luminosity 
advantage of more than $10^5$. 
 \end{itemize} 

As a result of these gains, the required beam power, spot sizes, emittances
and energy spread are far less in \mumu colliders than in \ee machines of the 
same energy. The comparison is made in Fig.~\ref{energy} above.

But there are problems with the use of muons:
 \begin{itemize}
 \item Muons can be best be obtained from 
the decay of pions, made by higher energy protons impinging on a target.
A high intensity proton source is thus required and  
very efficient 
capture and decay of these pions is essential.
 \item Because  the muons are made with very large emittance, 
they must be cooled and this must be done very rapidly because of their short lifetime. Conventional synchrotron, electron, or stochastic cooling is
too slow. Ionization cooling is the only clear 
possibility, but does not cool to very low emittances.
 \item Because of their short lifetime, conventional 
synchrotron acceleration would be too slow. Recirculating 
accelerators or pulsed synchrotrons must be used.
 \item
Because they decay while stored in the collider, muons radiate
the ring and 
detector with their decay products. Shielding is essential and backgrounds
will certainly be significant.
 \end{itemize}

These problems and their possible solutions will be discussed in more detail 
in the following chapters. Parameters will be given there of a 4 TeV center of 
mass collider, and of a 0.5 TeV demonstration machine.
\subsection{Comparison of Machines}
\vskip -1pc
\subsubsection{Length.}
In Fig.~\ref{length}, the effective physics energies (as defined by 
Eq.~\ref{Eeffeq}) of representative machines 
are plotted against their total tunnel lengths. We note:
 \begin{figure}[hbt!] 
\centerline{\epsfig{file=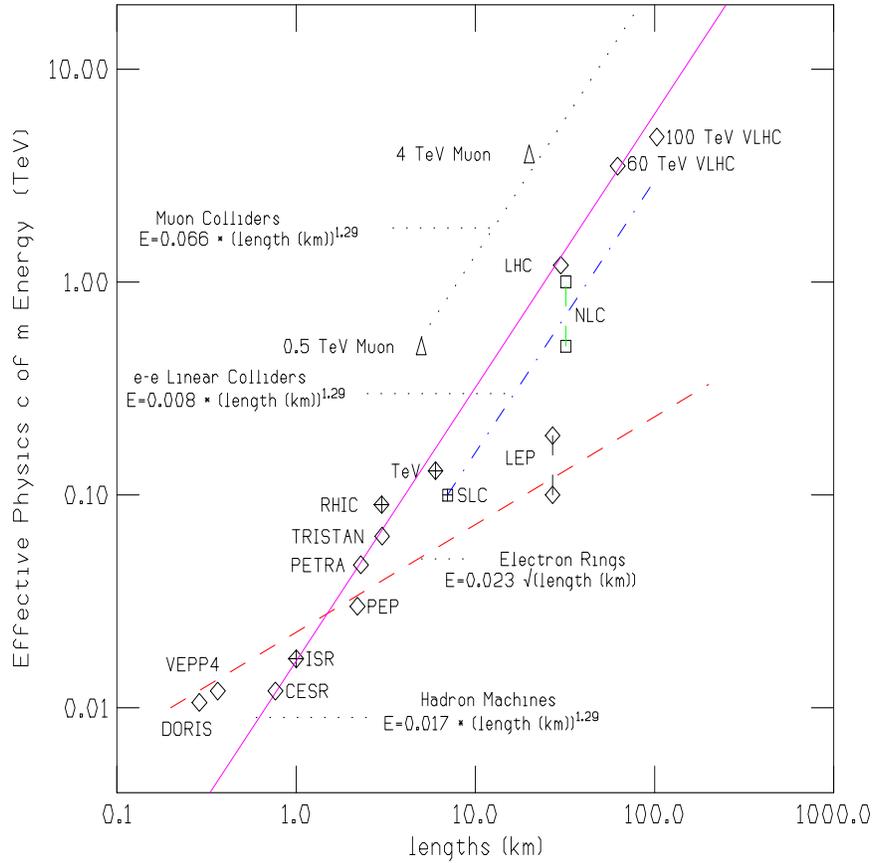,height=4.5in,width=4.5in}}
 \caption{Effective physics energies of 
colliders as a function of their total length.
 \label{length} }
\end{figure} 
 \begin{itemize}
 \item  Hadrons Colliders:
 It is seen that the energies of machines rise with their size, but 
that this rise is faster than linear ($E_{\rm eff}\propto L^{1.3}$). This
extra rise is a reflection of the steady increase in bending magnetic 
fields used as technologies and materials have become available. 

 \item Circular Electron-Positron Colliders:
The energies of these machines rise approximately as the square root of their 
size, as expected from the cost optimization discussed above.

 \item Linear Electron-Positron Colliders:
The SLC is the only existing machine of this type and only one example of a 
proposed machine (the NLC) is plotted. The line drawn has the same slope as 
for the hadron machines and implies a similar rise in accelerating 
gradient, as technologies advance. 

 \item Muon-Muon Colliders:
Only the 4 TeV collider, discussed above, and the 0.5 TeV {\it  demonstration
machine}  have been plotted. The line  drawn has the same slope as for the
hadron machines. 
\end{itemize}
            
\begin{figure}[hbt!] 
\centerline{\epsfig{file=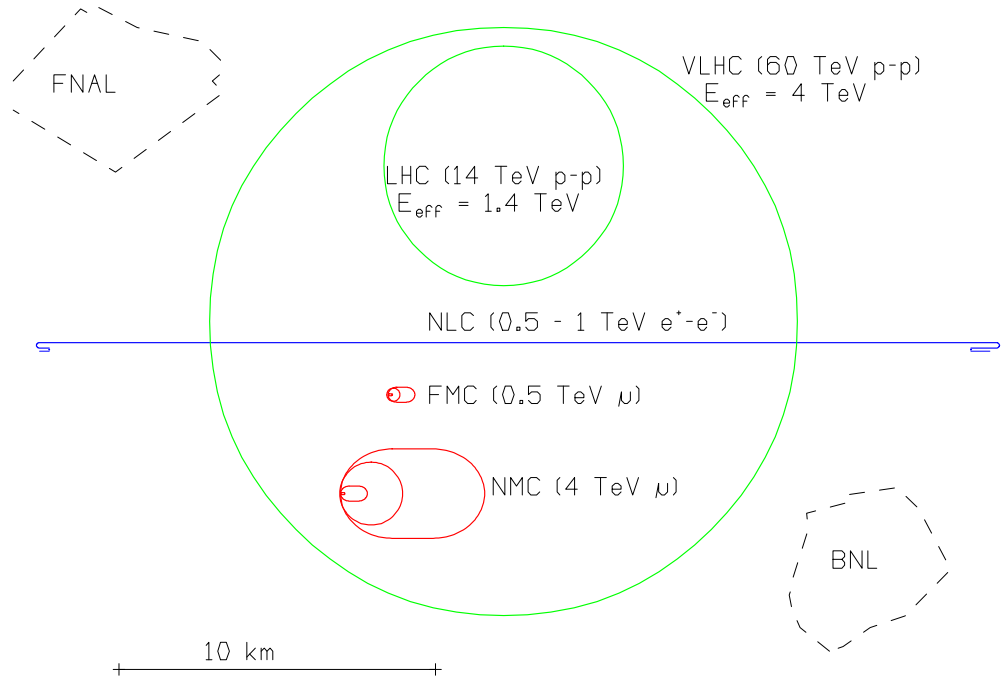,height=5.18in,width=3.52in,angle=90}}
 \caption{Approximate sizes of some possible future colliders.
 \label{examples} }
\end{figure}

   It is noted that the muon collider offers the greatest energy per unit 
length. This is also apparent in Fig.~\ref{examples}, in which the 
footprints of a number of proposed machines are given on the same scale. But 
does this mean it will give the greatest energy per unit of cost ? 
  \subsubsection{Cost.}
\begin{figure}[thb!] 
\centerline{\epsfig{file=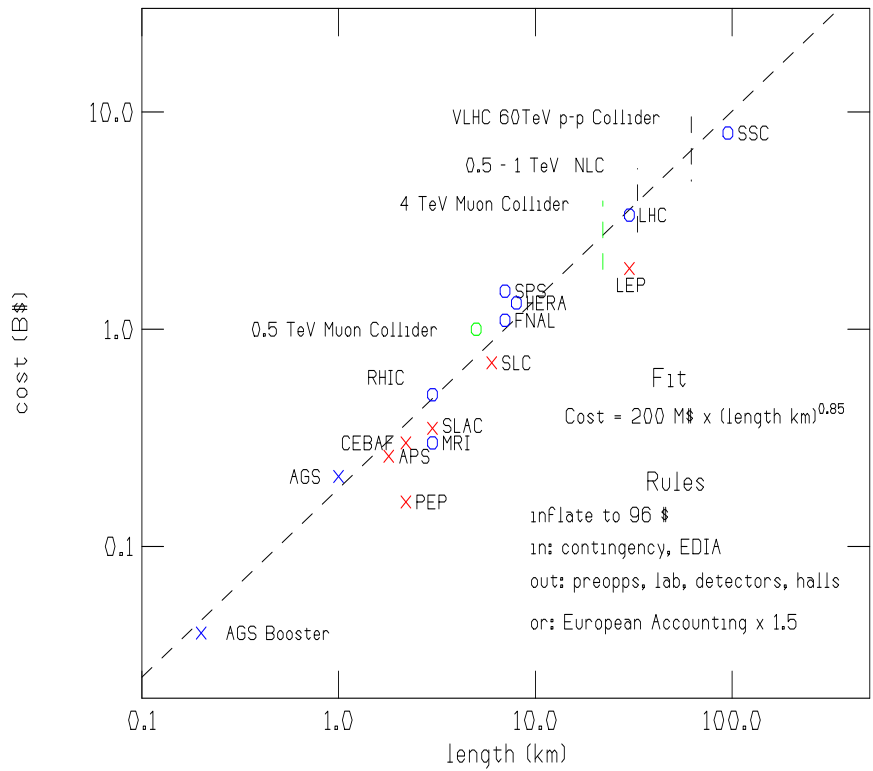,height=4.5in,width=4.0in,angle=90}}
 \caption{Costs of some machines as a function of their total lengths.
 \label{cost} }
\end{figure}

 Fig.~\ref{cost} plots the cost of a sample of machines 
against their size. Before examining this plot, be warned:
the numbers you will see will not be the ones you are familiar 
with. The published numbers for different projects use different 
accounting procedures and include different items in their costs. 
Not very exact corrections and escalation have been made to obtain 
estimates of the costs under fixed criteria: 1996~\$'s, US 
accounting, no detectors or halls. The resulting numbers, as 
plotted, must be considered to have errors of at least $\pm$ 20\%. 

The costs are seen to be surprisingly well represented by a 
straight line. Circular electron machines, as expected, lie 
significantly lower. The only plotted muon collider 
(the 0.5 TeV demonstration machine's very preliminary cost 
estimate)  lies above the line. But the clear indication 
is that length is, or at least has been, a good estimator of 
approximate cost. It is interesting to note that the fitted line 
indicates costs rising, not linearly, but as the 0.85\,th power of 
length. This can be taken as a measure of economies of scale. 

\subsection{Conclusions}
Our conclusions for this chapter, with the caveat that they are indeed only 
our opinions, are: 
 \begin{itemize}
 \item   The LHC is a well optimized and appropriate next step towards high 
{\it effective physics} energy.

\item   A Very Large Hadron Collider with energy greater than the 
SSC (e.g. 60 TeV c-of-m) and cost somewhat less than the SSC, may 
well be possible with the use of high T$_c$ superconductors that 
may become available. 

\item   A ``Next Linear Collider" is the only clean way to 
complement the LHC with a lepton machine, and the only way to do 
so soon. But it appears that even a 0.5 TeV collider will be more 
expensive than the LHC, and it will be technically 
challenging: obtaining the design luminosity may not be easy. 

\item Extrapolating conventional rf \ee linear colliders to 
energies above 1 or 2 TeV will be very difficult. Raising the rf 
frequency can reduce length and probably cost for a given energy, 
but obtaining  luminosity increasing as the square of energy, as 
required, may not be feasible. 

\item   Laser driven accelerators are becoming more realistic and 
can be expected to have a significantly lower cost per TeV. But 
the ratio of luminosity to wall power and the ability to preserve very small 
emittances, is likely to be significantly worse than for 
conventional rf driven machines. Colliders using such technologies are thus 
 unlikely to achieve very high luminosities and are probably unsuitable for 
higher (above 2 TeV) energy physics research. 

\item A higher gradient superconducting Linac collider using Nb$_3$Sn or
high T$_c$ materials, if it becomes technically possible, could be the only 
way to attain the required luminosities in a higher energy \ee collider.

\item   Gamma-gamma collisions can and should be obtained at any 
future electron-positron linear collider. They would add physics 
capability to such a machine, but, despite their freedom from the 
beamstrahlung constraint, may not achieve higher 
luminosity.

\item   A Muon Collider, being circular, could be far smaller 
than a conventional electron-positron collider of the same 
energy. Very preliminary estimates suggest that it would also be 
significantly cheaper. The ratio of luminosity to wall power for 
such machines, above 2 TeV, appears to be better than that for 
electron positron machines, and extrapolation to a center of mass 
energy of 4 TeV or above does not seem unreasonable. If research 
and development can show that it is practical, then a 
0.5 TeV muon collider could be a useful complement to \ee colliders, and, 
at higher energies (e.g. 4 TeV), could be a viable alternative. 
   \end{itemize} 
\section{PHYSICS CONSIDERATIONS}
\vskip -1pc
\subsection{Introduction}
The physics opportunities and possibilities of the muon collider have been 
well documented in the Feasibility Study\cite{book} and by additional papers\cite{ref100}.  
For most reactions the physics capabilities of $\mu^+\mu^-$ and
$e^+e^-$ colliders with the same energy and luminosity are similar,
so that the choice between them will depend mainly on the feasibility
and cost of the accelerators.
But for some reactions, the larger muon mass does provide some 
advantages: 

\begin{itemize}
\item The suppression of synchrotron radiation induced by the opposite bunch 
(beamstrahlung) allows, in principle, the use of beams with very low momentum 
spread 
\item QED radiation is reduced by a factor of $[\ln(\sqrt{s}/m_\mu) /
\ln(\sqrt{s}/m_e)]^2$, leading to smaller $\gamma\gamma$ backgrounds
and a smaller effective beam energy spread.
\item $s$-channel Higgs production is enhanced by a factor of $(m_\mu/m_e)^2
\approx 40000$.
\item The suppression of synchrotron radiation, allowing acceleration and 
storage of muons in a ring, combined with the suppression of beamstrahlung, 
may allow the construction of \mumu colliders at higher energy than \ee 
machines.
\end{itemize}

The disadvantages are:

\begin{itemize}
\item Less polarization appears practical in a \mumu collider than in an \ee 
machine, and some luminosity loss is likely.
\item The \mumu machine will have considerably worse background and probably 
require a shielding cone, extending down to the vertex, that takes up a larger 
solid angle than that needed in an \ee collider.
\end{itemize}

In the following sections we will give examples of physics for which there 
is a advantage in \mumu. These examples are taken from the discussion 
in section II of the \mumu Collider Feasibility Study \cite{book}. For a 
discussion of the other physics, SUSY particle identification in particular, 
the reader is refered to the physics sections of the Next Linear Collider 
Zeroth Order Design Report (ZDR)\cite{ZDR}. 
\subsubsection{Precision Threshold Studies.}
\begin{figure}[hbt!]
\begin{center}
\centerline{\epsfig{file=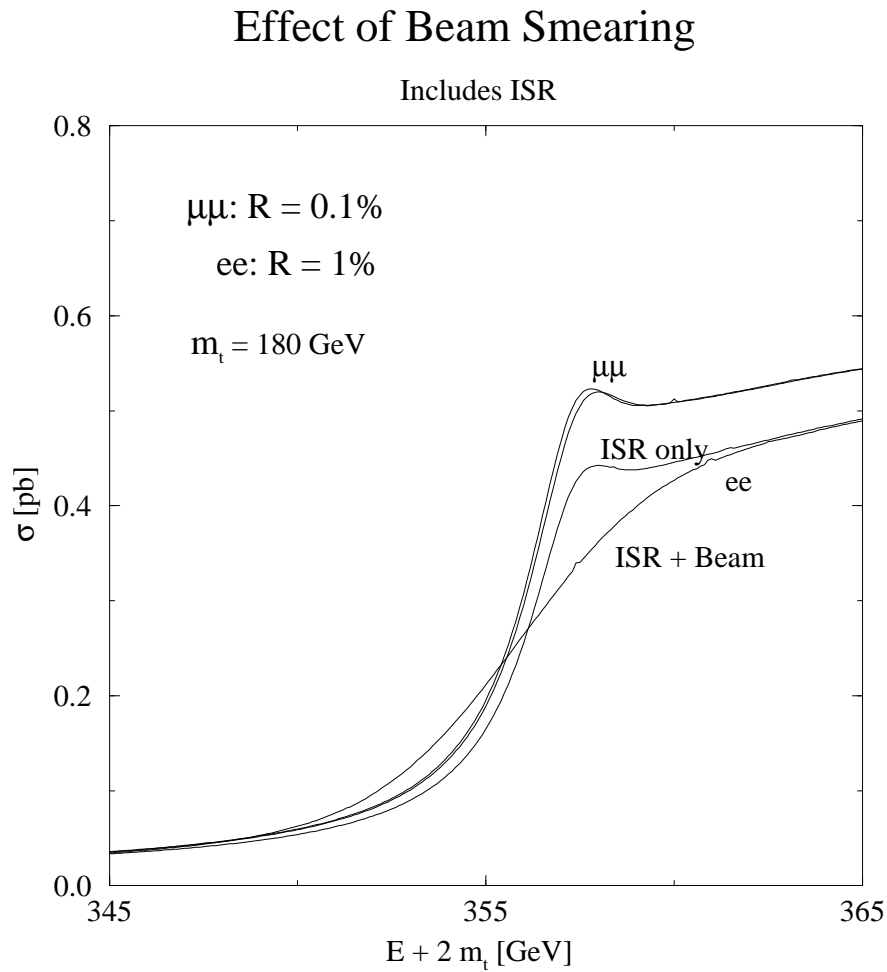,width=12cm}}
\begin{minipage}{12cm}       
\bigskip
\caption[The threshold curves are shown for
\mumu and \ee machines including ISR and with and without
beam smearing.]
{{\baselineskip=0pt
The threshold curves are shown for
\mumu and \ee machines including ISR and with and without
beam smearing. Beam smearing has only a small effect
at a muon collider, whereas at an electron collider the threshold region is
significantly smeared (An rms energy spread of $1\,\%$ is assumed for \ee and $0.1\, \%$ for \mumu ). }}
\label{thresh}
\end{minipage}
\end{center}
 \end{figure}

The high energy resolution and suppression of Initial State Radiation (ISR) in 
a \mumu collider makes it particularly well suited to threshold studies. As an 
example, Fig.~\ref{thresh} shows the threshold curves for top quark production 
for both \mumu and \ee machines, with and without beam smearing. (An rms energy 
spread of 1 \% is assumed for \ee and 0.1 \% for \mumu). The rms mass 
resolution $\Delta m_t$ obtained with $10\,{\rm fb}^{-1}$in a $\mu^+\mu^-$ Collider is estimated to be 
$\pm\, 0.3\,{\rm  GeV}.$ This can be compared with 4 GeV for the Tevatron, 2 GeV for the 
LHC, and 0.5 GeV for NLC. 
\subsubsection{Studies of Standard Model, or SUSY Model Light, Higgs {\it h}.}
The feature that has attracted most theoretical interest is the
possibility of s-channel studies of 
Higgs production.  This is possible with $\mu$'s, but not 
with e's, due to the strong coupling of muons to the Higgs channel that is 
proportional to the mass of the lepton. If the Higgs sector is more complex 
than just a simple standard model (SM) 
Higgs, it will be necessary to measure the widths and 
quantum numbers of any newly discovered particles to ascertain the nature of 
those particles and the
structure of the theory.  In addition to the increased coupling strength of 
the muons, the beamstrahlung is much 
reduced for muons  allowing much better definition of the 
beam energy.  
\begin{figure}[!htb]
\centerline{\epsfig{file=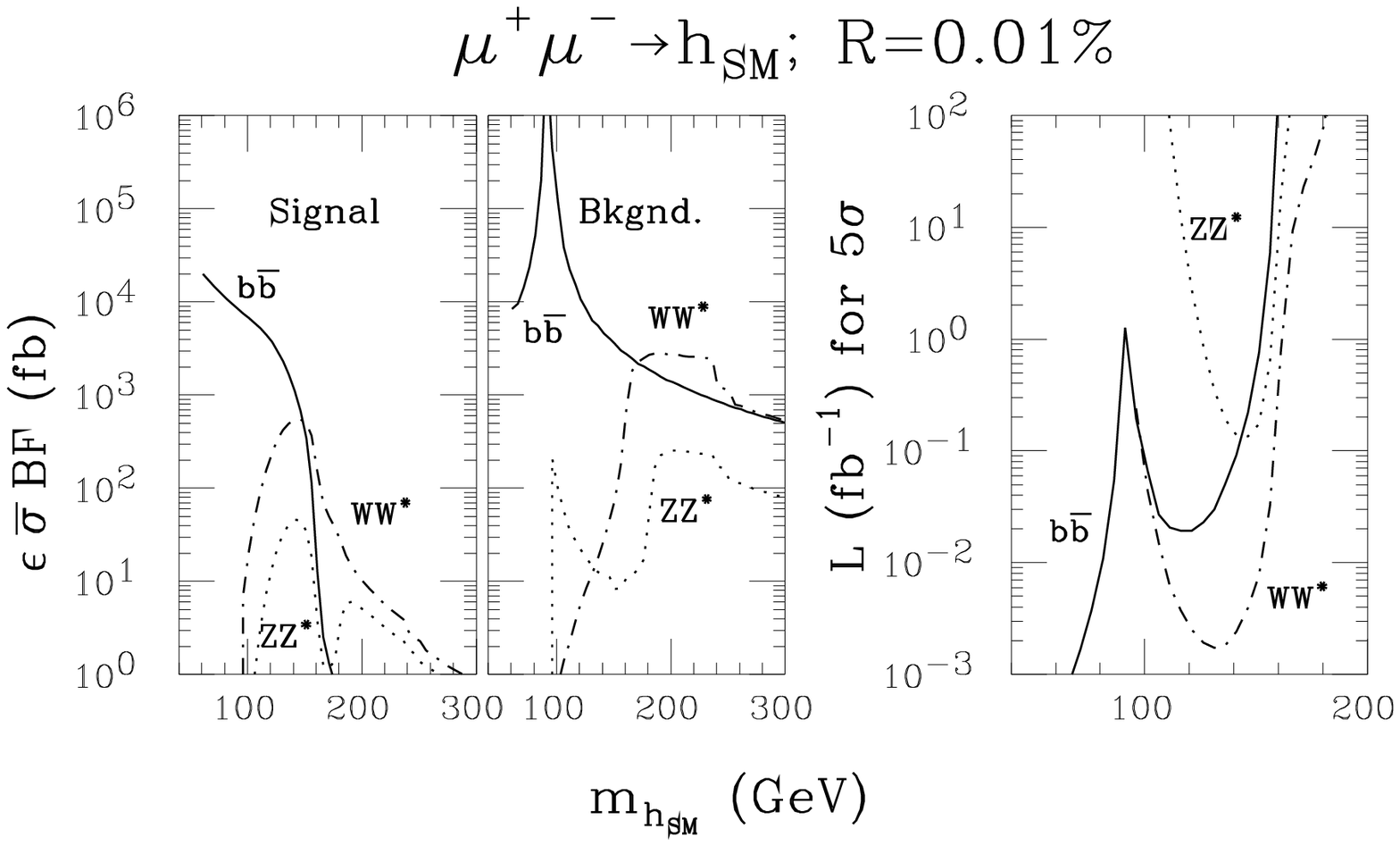,width=12.2cm}}
\vskip .3in
\begin{center}
\centerline{\epsfig{file=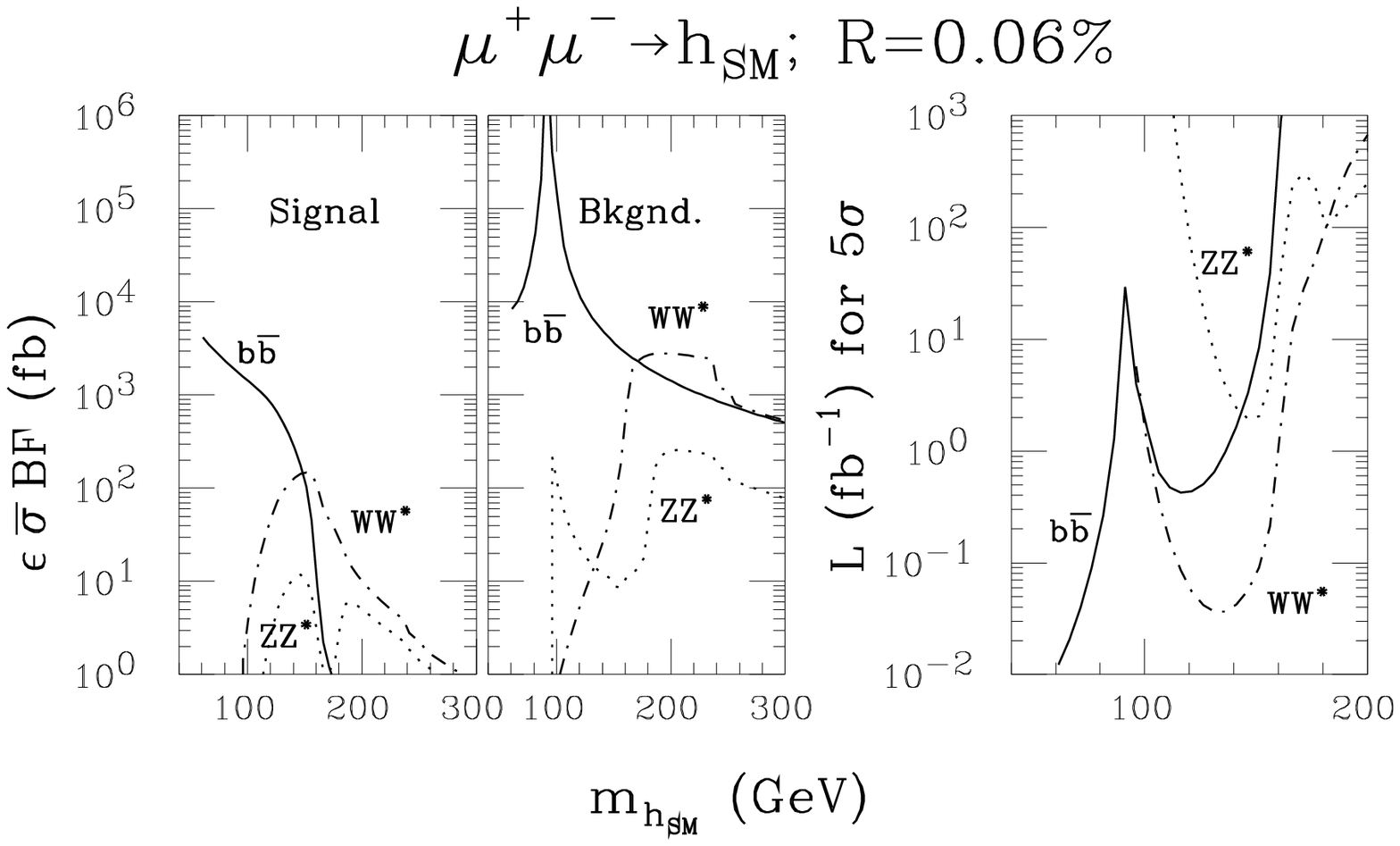,width=11cm}}
\begin{minipage}{11cm}       
\caption[ The (a) $\hsm$ signal and (b) background
cross sections ]
{ The (a) $\hsm$ signal and (b) background
cross sections, $\epsilon \overline\sigma \br(X)$,
for $X=b\anti b$, and useful (reconstructable, non-$4j$) 
$W\wstarp$ and $Z\zstarp$ final states
(including a channel-isolation efficiency of $\epsilon=0.5$)
versus $\mhsm$ for SM Higgs $s$-channel production.
Also shown: (c) the corresponding luminosity required for
a $S/\protect \sqrt B =5$ standard deviations signal in each
of the three channels. Results for $R=0.01\%$ and $R=0.06\%$ are  
given.\label{Higgs}}
\end{minipage}
\end{center}
\end{figure}

The cross sections for Higgs production with a \mumu collider are substantial. 
Fig.~\ref{Higgs} shows a) the Higgs signal, b) the background, and c) the 
luminosity required for a $5\,\sigma$ signal significance, for two different rms 
energy spreads of the muon beam: 0.01 \% and 0.06 \%. Signals are shown for 
three final states: $b\bar{b}$, WW$^{(*)}$  and ZZ$^{(*)}$ (reconstructable, 
non- 4 jet, with channel isolation efficiency $\epsilon$ = 0.5). It is seen 
that:

\begin{itemize}
\item For an rms energy resolution of 0.01 \%, a luminosity of only 0.1~ 
$fb^{-1}$ is required to yield a detectable signal for all $m_{h_{SM}}$ above 
the current LEP limit, except in the region of the Z peak, where 1~$fb^{-1}$ 
is required.
\item For an rms energy resolution of 0.06 \%, the luminosity required is 20-
30 times larger, indicating that the higher resolution is desirable even at 
significant loss of luminosity.
\end{itemize} 
\begin{figure}[tbh!]
\begin{center}
\centerline{\epsfig{file=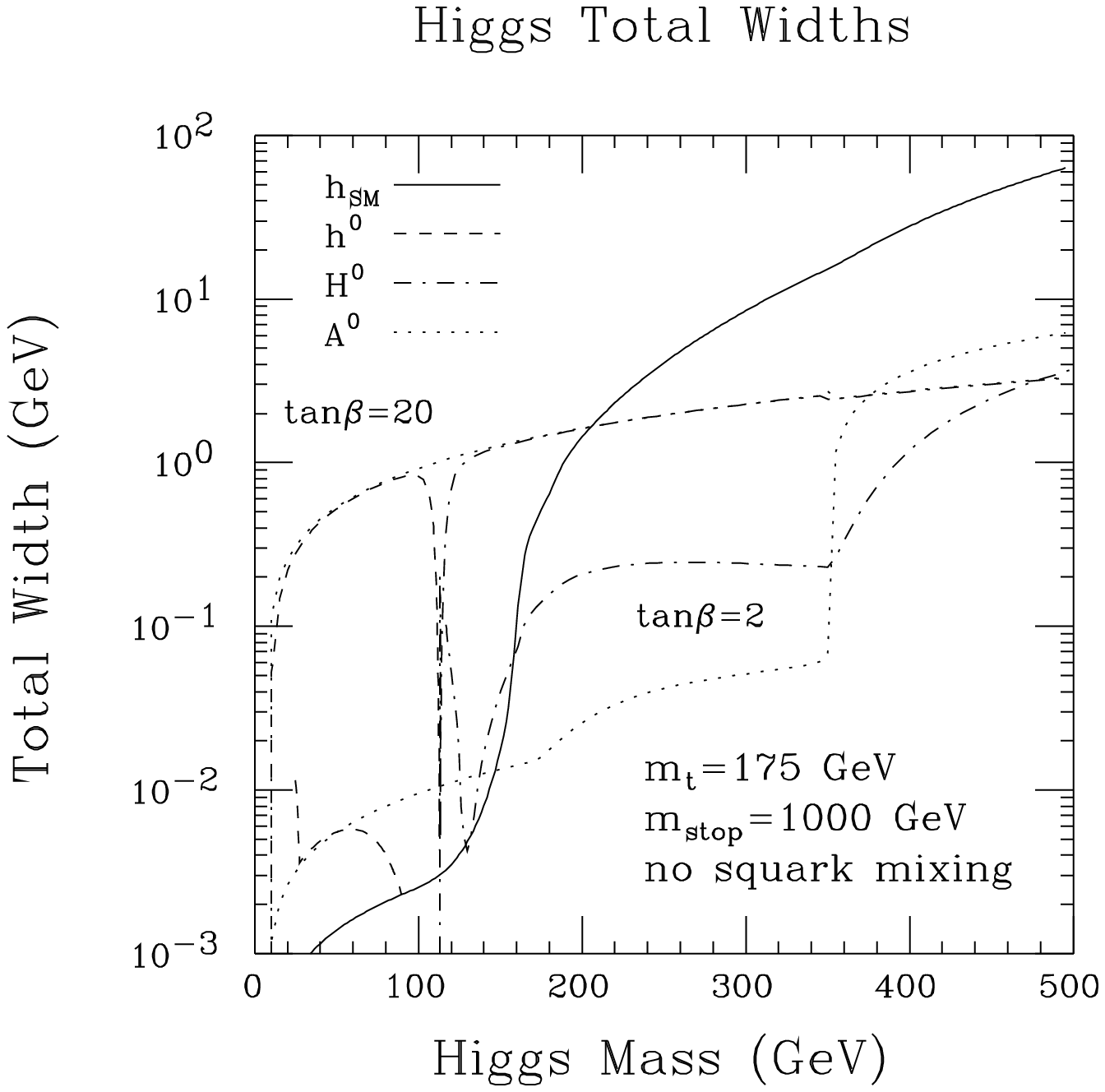,width=12cm}}
\begin{minipage}{12cm}       
\smallskip
\caption[Total width vs mass of the SM and MSSM Higgs bosons
for $\mt=175\gev$. ]
{{\baselineskip=0pt
Total width vs mass of the SM and MSSM Higgs bosons
for $\mt=175\gev$.
In the case of the MSSM, we have plotted results for
$\tan\beta =2$ and 20, taking $\mstop=1\tev$ and
including two-loop corrections.}}
\label{higgswidths}
\end{minipage}
\end{center}
\vspace*{-.2in}
\end{figure}

Fig.~\ref{higgswidths} shows the total widths of standard model and MSSM 
Higgs. In the case of MSSM masses are plotted for the stop quark mass $m_{stop}$ = 1 TeV, tan 
$\beta$ = 2 and 20. Two loop corrections have been included, but no squark 
mixing or SUSY decay channels.

The standard model Higgs with mass below $m_t$ is seen to be very narrow. For 
110 GeV it is $\approx$ 3 MeV. A Supersymmetric model Higgs would be wider, 
but might be only a little wider. It could be important to measure the width 
of a low mass Higgs to determine its character. It has been shown that a 
muon collider with an rms energy spread of 0.01 \% could measure the width of 
a 110 GeV standard model Higgs to $\pm 1\,{\rm MeV}$ with only 2 inverse femtobarns. 
Only if the Higgs mass is close to that of the Z does it become difficult to 
make such a determination without a large amount of data (200 inverse femtobarns). This could be a very important measurement (that could not be done in 
any other way) since it would destinguish clearly the nature of the boson 
seen. Together with branching ratio measurements (also possible with a muon 
collider), it could even predict the 
mass of the other SUSY Higgs bosons: H and A. 
\subsubsection{Studies of SUSY Model Heavy Higgs Particles: {\it H} and {\it A}.}
The H and A SUSY Higgs bosons are expected to be significantly heavier than
the lightest {\it h}, and might have quite similar masses. If tan $\beta$ is 
small $(< 3)$ then they can easily be identified at the LHC, but may not be 
identified there 
if tan $\beta$ is large. They could be searched for in an 
\ee machine in \ee$\rightarrow H,A$, (h,A, or h, A are depressed) but only up 
to about $m_{H,A} \approx \sqrt{s}/\sqrt{2}$ or even less. In a muon collider, 
on the other hand, they could, providing tan $\beta$ is large $(>3),$ be easily  observed in the s-channel, up to masses equal to $\sqrt{s}$. 
         \subsubsection{Studies of Non-SUSY Model Strong WW Interactions.}
 If SUSY does not exist and we are forced to a much higher mass scale to study 
the symmetry breaking process then a 4 TeV 
muon collider is a viable choice to 
study WW scattering as it becomes a strong reaction. 

Fig.~\ref{fg.hma1} shows the mass distribution for the $1\,\TeV$ Higgs signals and 
physics backgrounds from PYTHIA in a toy detector, which includes segmentation 
of $\Delta \eta = \Delta\phi = 0.05$ and the angular coverage, $20^\circ < 
\theta < 160^\circ$, assumed in the machine background calculations. Since the 
nominal luminosity is $1000\,\fb^{-1}$, there are $\simge 1000$ events per bin 
at the peak. The loss in signal from the $20^\circ$ cone is larger for this 
process than for $s$-channel processes but is still fairly small, as can be 
seen in Fig.~\ref{fg.hmath}. The dead cone has a larger effect on $\gamma\gamma \to 
WW$ and thus the accepted region has a better signal to background ratio. 
\begin{figure}
\Twofigs{\epsfysize=3in\epsfbox{4tev1.ai}}{\epsfysize=3in\epsfbox{thmax.ai}} 
\Lcaption[Signals and physics backgrounds for a $1\,\TeV$ Higgs boson.]
{Signals and physics backgrounds for a $1\,\TeV$ Higgs boson
at a $\mu\mu$ collider, including the effect of a $20^\circ$ dead
cone around the beamline.
\label{fg.hma1}}
\Rcaption[Signals and physics backgrounds for a $1\,\TeV$ Higgs boson vs.
$\theta_{\rm min}$.]{$WW \to WW$ signal and $\gamma\gamma \to
 WW$ background vs.{}
the minimum angle, $\theta_{\rm min}$, of the $W$.\label{fg.hmath}}
\end{figure}
%
%
It would be desirable to separate the $WW$ and $ZZ$ final
states in purely hadronic modes by reconstructing the masses. Whether
this is possible or not will depend on the details of the calorimeter
performance and the level of the machine backgrounds. If it is not,
then one can use the $\sim12\%$ of events in which one $Z \to ee$ or
$\mu\mu$ to determine the $Z$ rate.  Clearly there is a real challenge
to try to measure the hadronic modes.

The
 background from $\gamma\gamma$ and $\gamma Z$ processes is
smaller at a muon collider than at an electron collider but not
negligible. Since the $p_T$ of the photons is usually very small while
the $WW$ fusion process typically gives a $p_T$ of order $M_W$, these
backgrounds can be reduced by making a cut $p_{T,WW} > 50\,\GeV$.
This cut keeps most of the signal while
significantly reducing the physics background. 
\subsubsection{Summary.}

For many reactions, SUSY particle discovery for example, an \ee collider, with 
its higher polarization and lower background, would be preferable to a \mumu 
machine of the same energy and luminosity. There are however specific 
reactions, s-channel Higgs production for example, where the \mumu machine 
would have unique capabilities. Ideally both machines would be built and they 
would be complementary. Whether both machines could be built, at both moderate 
and multi TeV energies, and whether both could be afforded, remains to be 
determined. 

There are several hardware questions that must be carefully studied.  The 
first is the question of the luminosity available when the beam momentum 
spread is decreased.  In addition there will have to be good control of the 
injected beam energy as there is not time to make large adjustments in the 
collider ring. Precision determination of the energy and energy spread will 
be mandatory: presumably by the study of spin precession.  Finally, the 
question of luminosity vs. percent polarization needs additional study;  
unlike the electron collider, both beams can be polarized but as shown later 
in this report, but the luminosity decreases as the 
polarization increases. 

\section{MUON COLLIDER COMPONENTS}
\vskip -1pc
\subsection{Introduction}

The possibility of muon colliders was introduced by Skrinsky et 
al.\cite{ref2} and Neuffer\cite{ref3} and has been aggressively 
developed over the past two years in a series of meetings and 
workshops\cite{ref4,ref5,ref6,ref7}. 

\begin{figure}
\centerline{\epsfig{file=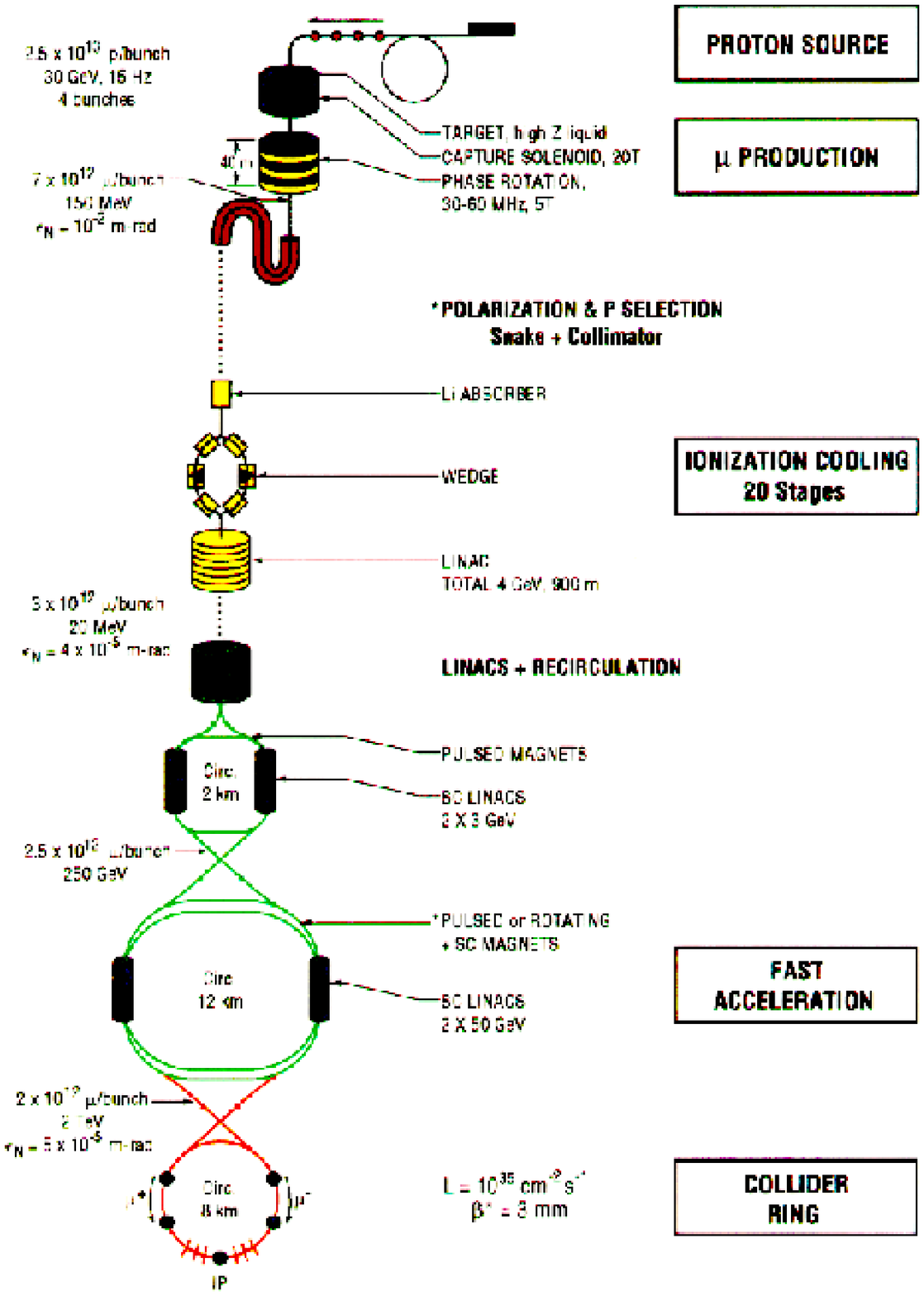,height=5.1in,width=4.0in}}
 \caption{Overview of a 4 TeV Muon Collider
 \label{overview}}
 \end{figure}

\begin{figure}[hbt!]
\centerline{\epsfig{file=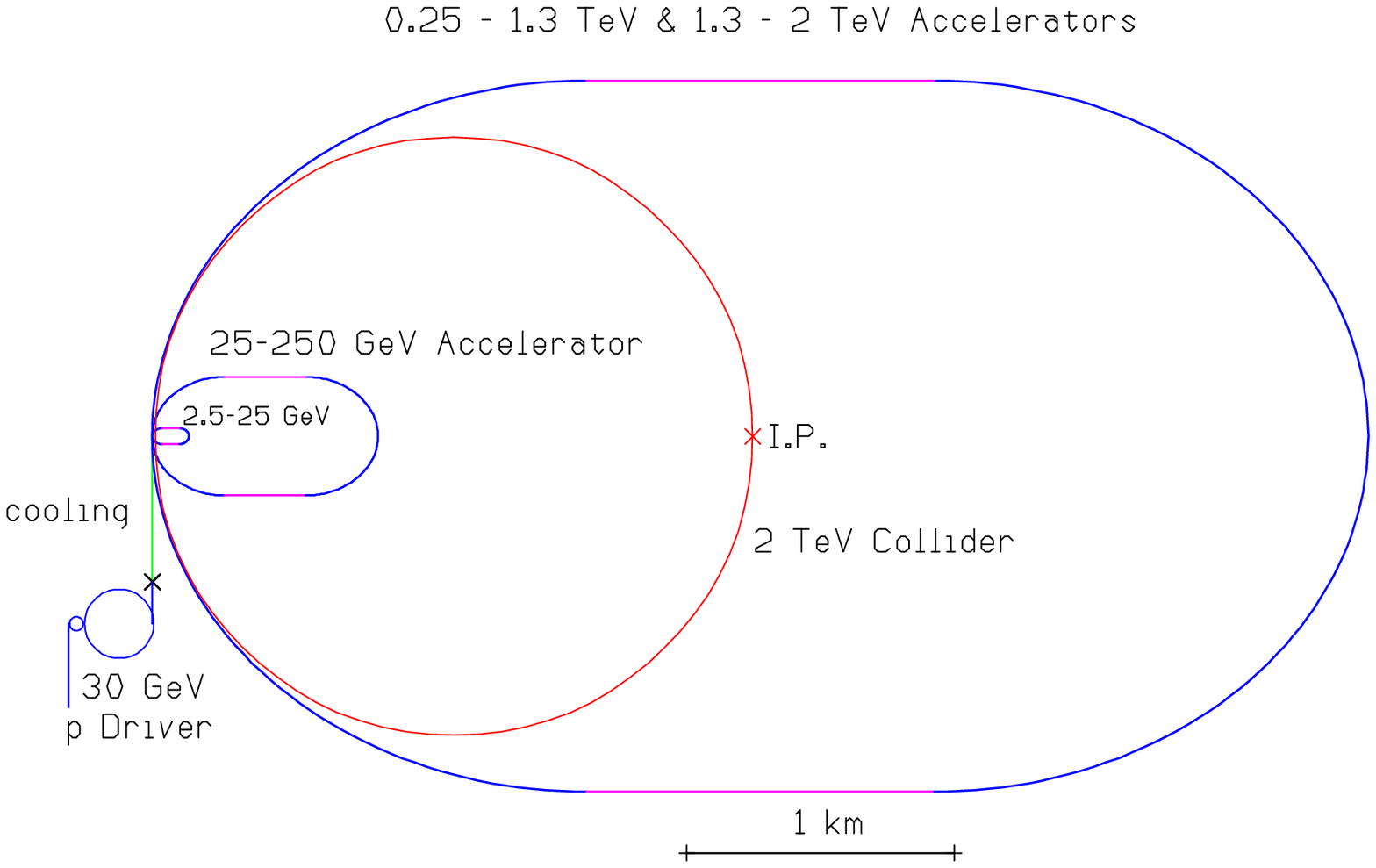,height=3.5in,width=5.5in}}
\caption{Layout of the collider and accelerator rings.
 \label{accpict}}
 \end{figure}

A collaboration, lead by BNL, FNAL and LBNL,  with contributions from 
18 institutions has been studying a 4 TeV, high luminosity  scenario 
and presented a Feasibility Study\cite{book} to the  1996 Snowmass 
Workshop. The basic parameters of this machine are shown 
schematically in Fig.~\ref{overview} and given in Tb.~\ref{sum}.
Fig.~\ref{accpict} shows a possible layout of such a machine. 

Tb.~\ref{sum} also gives the parameters of
a 0.5 TeV demonstration machine based on the AGS as an 
injector. It is assumed that a demonstration version based on upgrades 
of the FERMILAB, or CERN machines would also be possible. 
\begin{table}[thb!]  
\centering \protect
\caption{Parameters of Collider Rings
\label{sum}}
\begin{tabular}{llcc}
\hline
c-of-m Energy              & TeV      & 4     & .5       
\\
\hline
Beam energy                & TeV      &     2    &   .25 
   \\
Beam $\gamma$              &          &   19,000 &  
2,400   \\
Repetition rate            & Hz       &    15    &    
2.5    \\
Proton driver energy       & GeV      & 30 & 24 \\
Protons per pulse          &          &  $10^{14}$ & 
$10^{14}$\\
Muons per bunch            & $10^{12}$  &   2    &    4  
    \\
Bunches of each sign       &          &   2      &    1  
    \\
Beam power                 & MW       &   38     &   .7 
\\
Norm. {\it rms} emit.   $\epsilon_N$   &$\pi$ mm mrad  & 
 50  &  90    \\
Bending Field              &  T   &    9    &    9     
\\
Circumference              &  Km      &    8    &    1.3 
   \\
Ave. ring  field $B$    & T   & 6    &   5       \\
Effective turns          &       & 900    &   800   \\
$\beta^*$ at intersection   & mm     &   3   &   8     
\\
{\it rms} I.P. beam size        & $\mu m$&   2.8 &  17   
 \\
Chromaticity             &         & 2000-4000 & 40-80 
\\
$\beta_{\rm max}$ & km   & 200-400 & 10-20 \\
Luminosity &${\rm cm}^{-2}{\rm s}^{-1}$& 
$10^{35}$&$10^{33}$\\
\hline
\end{tabular}
\end{table}

The main components of the 4 TeV collider would be:
\begin{itemize}
\item  A proton source with KAON like parameters (30 GeV, $10^{14}$ 
protons per pulse, at 15 Hz). 

\item A liquid metal target surrounded by a 20~T hybrid solenoid to make and capture pions. 

\item A 5 T solenoidal channel within a sequence of rf cavities 
to allow the pions to decay into muons and, at the same 
time, decelerate the fast ones that come first, while 
accelerating the lower momentum ones that come later. Muons from 
pions in the 100-500 MeV range emerge in a 6 m long bunch at 150 
$\pm$ 30 MeV.

\item A solenoidal snake and collimator to select 
the momentum, and thus polarization, of the muons. 

\item A sequence of 20 ionization cooling stages, each consisting of: 
 a) energy loss material in a strong focusing environment for 
transverse cooling; b) linac reacceleration and c) lithium wedges in a 
dispersive environment for cooling in momentum space. 

\item A linac and/or recirculating linac pre-accelerator, followed by
a sequence of pulsed field synchrotron accelerators using 
superconducting linacs for rf.

\item An isochronous collider ring with locally corrected low beta 
($\beta$=3 mm) insertion.
\end{itemize}
\subsection{Proton Driver}
The specifications of the proton drivers are given in Tb~\ref{driver}. 
In the 4 TeV example, it is a high-intensity (4 bunch, $2.5\times  
10^{13}$ protons per pulse) 30 GeV proton synchrotron. The preferred 
cycling rate would be 15 Hz, but for a demonstration machine using the 
AGS\cite{ref8}, the repetition rate would be limited to 2.5 Hz and 
the energy to $24\,$GeV. For the lower energy machine, 2 final 
bunches are employed (one to make $\mu^-$'s and the other to make 
$\mu^+$'s). For the high energy collider, four are used  (two $\mu$ 
bunches of each sign). 

 \begin{table}[bth!]
\centering
 \caption{Proton Driver Specifications
\label{driver}} 
 \begin{tabular}{llcc} 
\hline 
           &          & 4 TeV    & .5 TeV Demo \\ 
\hline 
Proton energy  & GeV      &    30    &   24  \\
 Repetition rate            & Hz       &    15    &   2.5 \\
 Protons per bunch          & $10^{13}$&   2.5    &   5   \\
 Bunches                    &          &   4      &      2   \\
 Long. phase space/bunch    & eV s    &   5    &    10  \\
 Final {\it rms} bunch length     & ns     &      1     &    1    \\
 \hline 
\end{tabular} 
\end{table}

 \begin{figure}[t!] 
\centerline{\epsfig{file=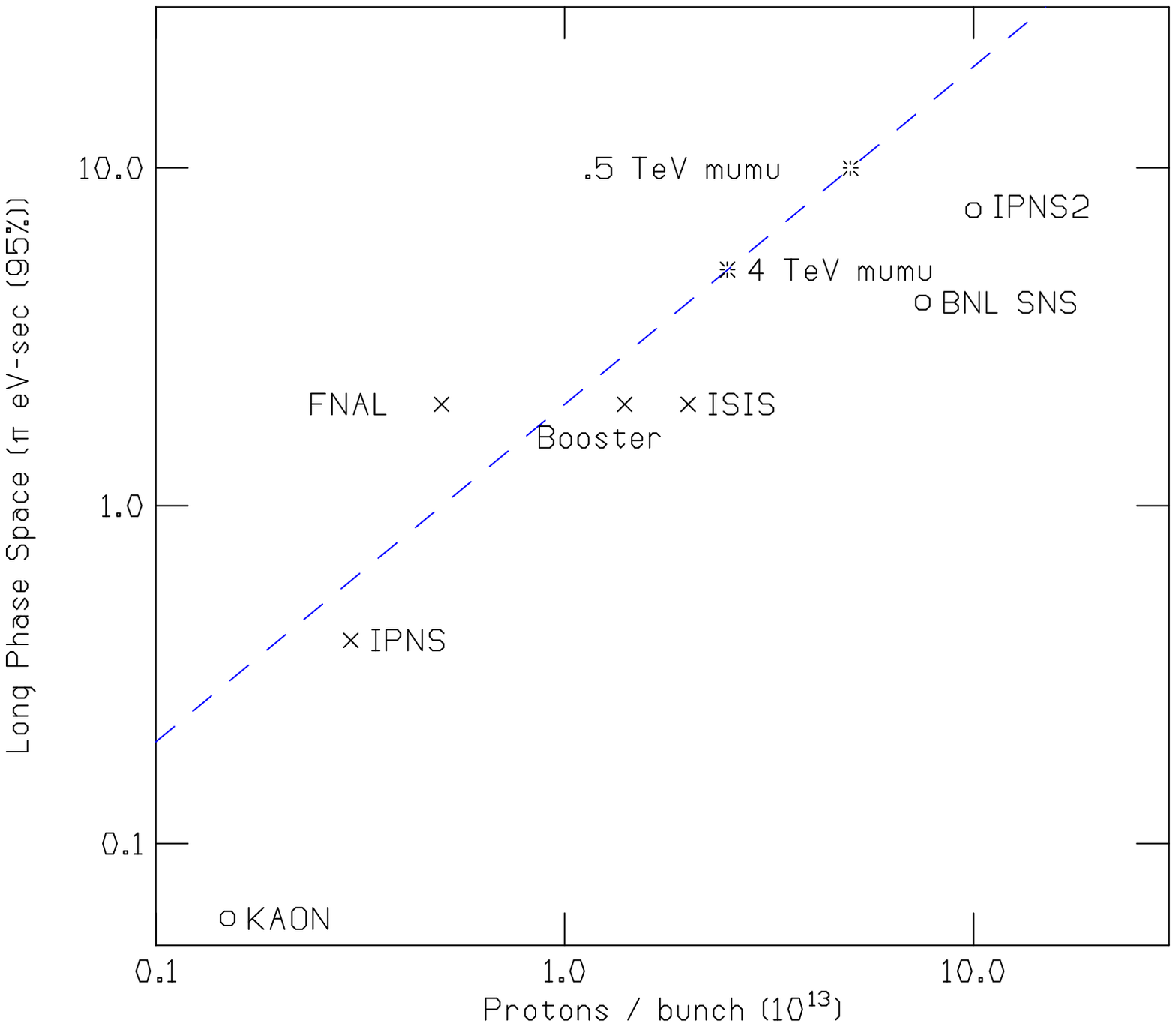,height=4.5in,width=4.5in}}
 \caption[Longitudinal Phase Space of Bunches vs. number of Protons ]
{Longitudinal Phase Space of Bunches vs. number of Protons; {\it x}'s indicate 
existing machines, circles proposed machines, and stars the values used here
 \label{phasespace}}
 \end{figure} 

In order to reduce the cost of the muon phase rotation section,  
minimize the final muon longitudinal phase space and maximize the achievable polarization,, it appears that 
the final proton bunch length should be of the order of 1 ns. Is this 
practical ? 

There appears to be a relationship between the number of 
protons in a bunch and the longitudinal phase space of that bunch that 
can be maintained stability in a circular machine. 
Fig.~\ref{phasespace} shows values obtained and those planned in a 
number of machines. The conservative assumption is that phase space 
densities will be similar to those already achieved: around 2 eV seconds 
per $10^{13}$ protons, as indicated by the line in Fig.~\ref{phasespace}. 
The required bunches of $2.5\ 10^{13}$ protons would 
thus be expected to have a phase space of $5\,{\rm eV s}$ (at 95\%) = 
$6\pi\,\sigma_t\sigma_E\ {\rm eV s}$ rms. A 1 ns rms bunch at 30 GeV 
with this phase space will have an rms momentum spread of $0.8\,\%,$  
($2\, \%$ at 95\%), and the space charge tune shift just  before 
extraction would be $\approx 0.5.$ Provided the rotation can be 
performed rapidly enough, this should not be a problem. For the 0.5 
TeV machine the bunch intensity, and thus area, would be double, leading 
to a final  spread of 1.6 \% rms (4 \% at 95 \%). 

An attractive technique\cite{ref11} for bunch compression would be 
to generate a large momentum  spread with modest rf at a final energy 
close to transition. Pulsed quads would then be employed to move the 
operating point away from transition, resulting in rapid compression. 

Earlier studies had suggested that the driver could be a 10 GeV 
machine with the same charge per fill, but a repetition rate of 
$30\,$Hz. This specification was almost identical to that 
studied\cite{ref9} at ANL for a spallation neutron source. Studies at 
FNAL\cite{ref10} have further established that such a specification is 
reasonable. But if 10 GeV protons are used, then approximately twice as 
many protons per bunch are required for the same pion production: $5\times 
10^{13}$ per bunch for the 4 TeV case, $1\times 10^{14}$ per bunch for the  
0.5 TeV case; the phase space of the bunches would be expected to be 
twice as big and the resulting \% momentum spread for the 1 ns bunch 6 
times as large: i.e. 12 \% (at 95\%) which may be hard to achieve.  
For the 0.5 TeV specification, this rises to 24 \%: {\it clearly 
unreasonable}. 
\subsection{Target and Pion Capture}
\vskip -1pc
\subsubsection{Pion Production.}

\begin{figure}[hbt!]
\centerline{\epsfig{file=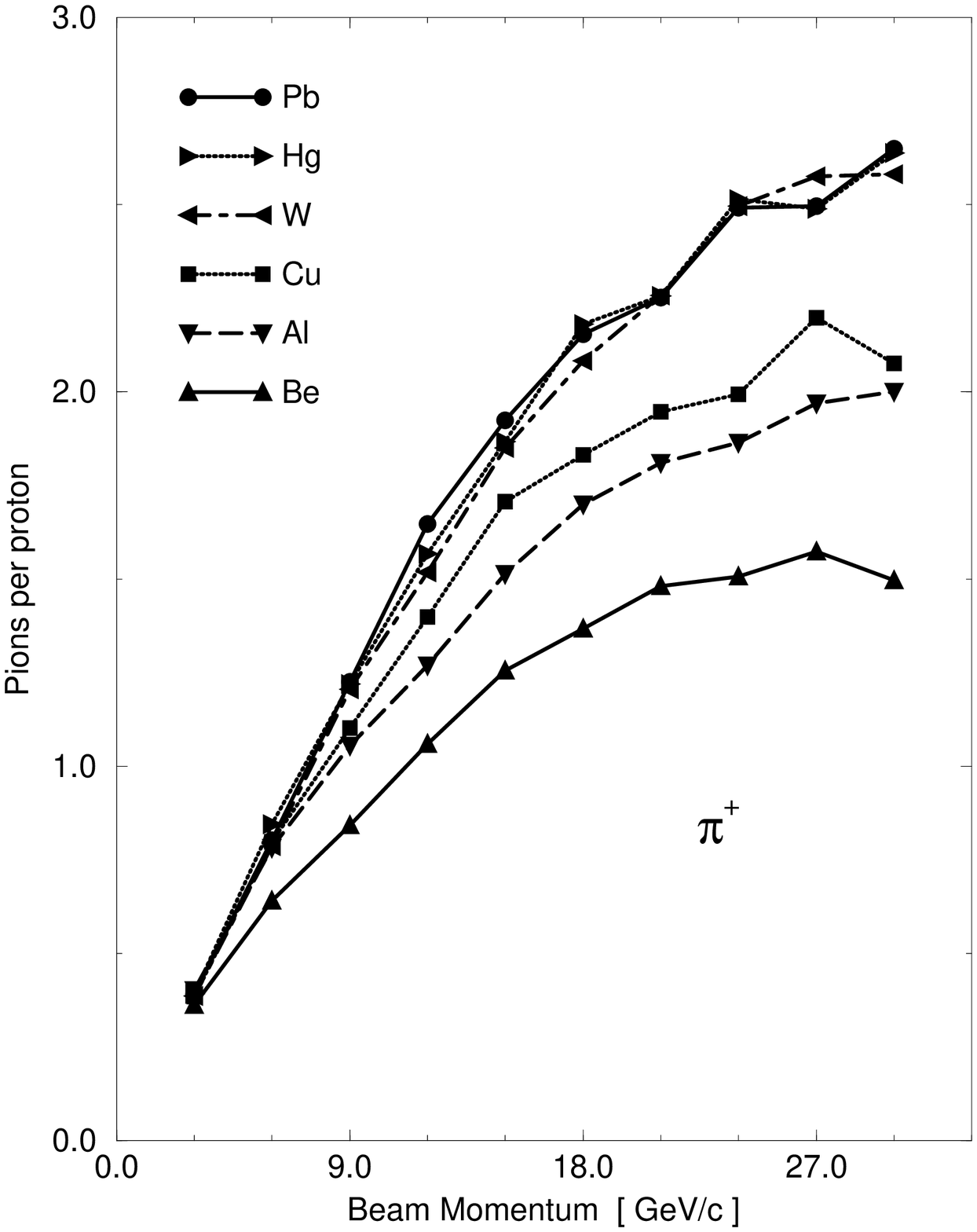,height=4.0in,width=3.5in}}
\caption{ARC forward $\pi^+$ production vs. proton energy and target material.
\label{pionproda}}
\end{figure}

Predictions of the nuclear Monte-Carlo program ARC\cite{ref12} suggest 
that $\pi$ production is maximized by the use of heavy target 
materials, and that the production is peaked at a relatively low pion 
energy ($\approx 100\,$MeV), substantially independent of the initial 
proton energy. Fig.\ref{pionproda} shows the forward $\pi^+$ 
production as a function of proton energy and target material; the 
$\pi^-$ distributions are similar. 

\begin{figure}[hbt!]
\centerline{\epsfig{file=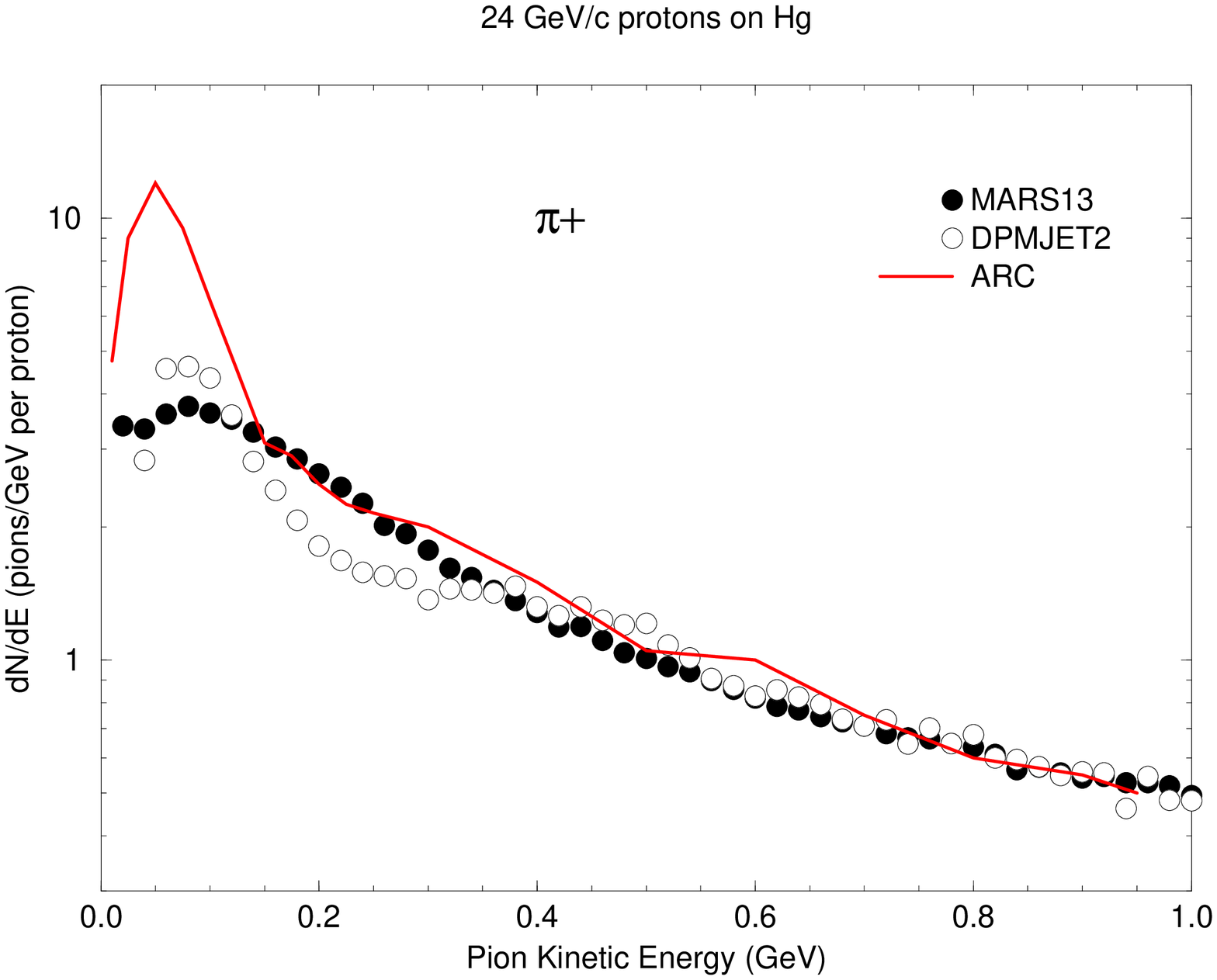,height=4.0in,width=3.5in}}
\caption{$\pi^{+}$ energy distribution for $24\,$GeV protons on Hg. 
 \label{pionprodp}}
 \end{figure}

Other programs\cite{ref13},\cite{ref14} do not predict such a large low energy 
peak,(see for instance Fig.~\ref{pionprodp})
and there is currently very little data to indicate which is right. An 
experiment (E910)\cite{ref15a}\cite{ref15b}, currently running at the AGS, should decide this question, 
and thus settle at which energy the capture should be optimized. 
\subsubsection{Target.}
For a low repetition rate the target could probably be made of Cu, 
approximately 24 cm long by 2 cm diameter. A study\cite{ref15} 
indicates that, with a 3 mm rms beam, the single pulse instantaneous 
temperature rise is acceptable, but, if cooling is only supplied from 
the outside, the equilibrium temperature, at our required repetition 
rate, would be excessive. Some method must be provided to give cooling 
within the target volume. For instance, the target could be made of a 
stack of relatively thin copper disks, with water cooling between 
them. A graphite target could be used, but with significant loss of 
pion production, or a liquid metal target. Liquid lead and gallium are 
under consideration. In order to avoid shock damage to a container, 
the liquid could be in the form of a jet. 

\begin{figure}[hbt!]
\centerline{\epsfig{file=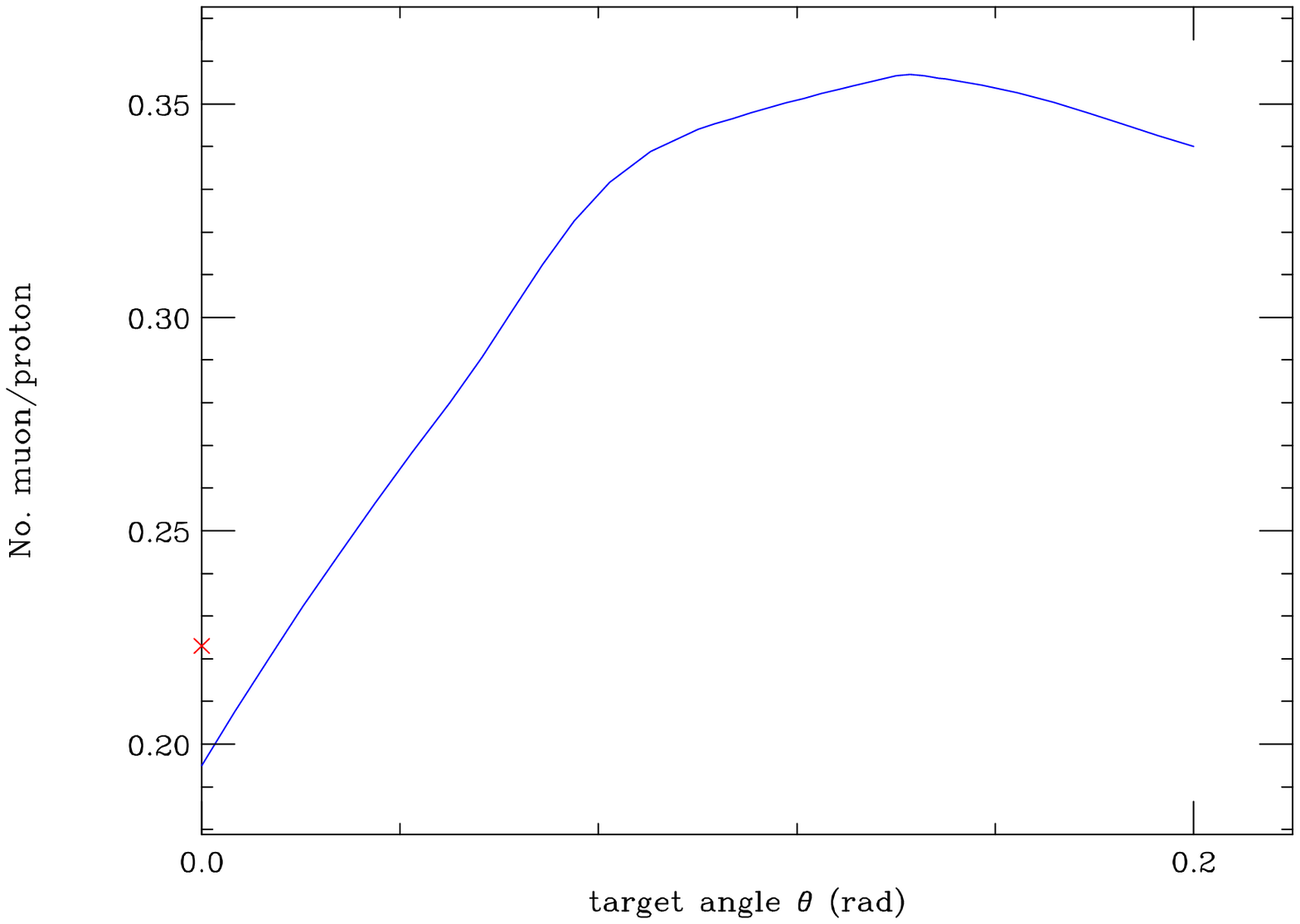,height=3.5in,width=3.5in}}
\caption[The muon to proton ratio as a 
function of the skew angle for a target whose length and transverse position 
has been reoptimized for the skew case ]
{The muon to proton ratio as a 
function of the skew angle for a target whose length and transverse position 
has been reoptimized for the skew case. The single {\it x} indicates the production 
ratio at zero angle with the original optimization. 
 \label{skew}}
 \end{figure}

   It appears that for maximum muon yield, the target (and incoming 
beam) should be at an angle to the axis of the solenoid and outgoing 
beam. The introduction of such an angle reduces the loss of pions when 
they reenter the target after being focused by the solenoid. A Monte 
Carlo simulation\cite{bob&juan} gave a muon production  increase of 60 
\% with at an angle 150 milliradians. The simulation assumed a copper 
target (interaction length 15 cm), ARC\cite{ref12} pion production 
spectra, a fixed pion absorption cross section, no secondary pion 
production, a 1 cm target radius, and the capture solenoid, decay 
channel, phase rotation and bunch defining cuts described below. 
Fig.~\ref{skew} shows the final muon to proton ratio as a function 
of the skew angle for a target whose length (45 cm) and transverse 
position (front end displaced - 1.5 cm from the axis) had been 
reoptimized for the skew case. The single X indicates the production 
ratio at zero angle with the original optimization (target length 30 
cm, on axis). One notes that the reoptimized target length is 
 3 interaction lengths long, and thus absorbs essentially 
all of the initial protons. 
\subsubsection{Capture.}
\begin{figure}[htb!]
\centerline{\epsfig{file=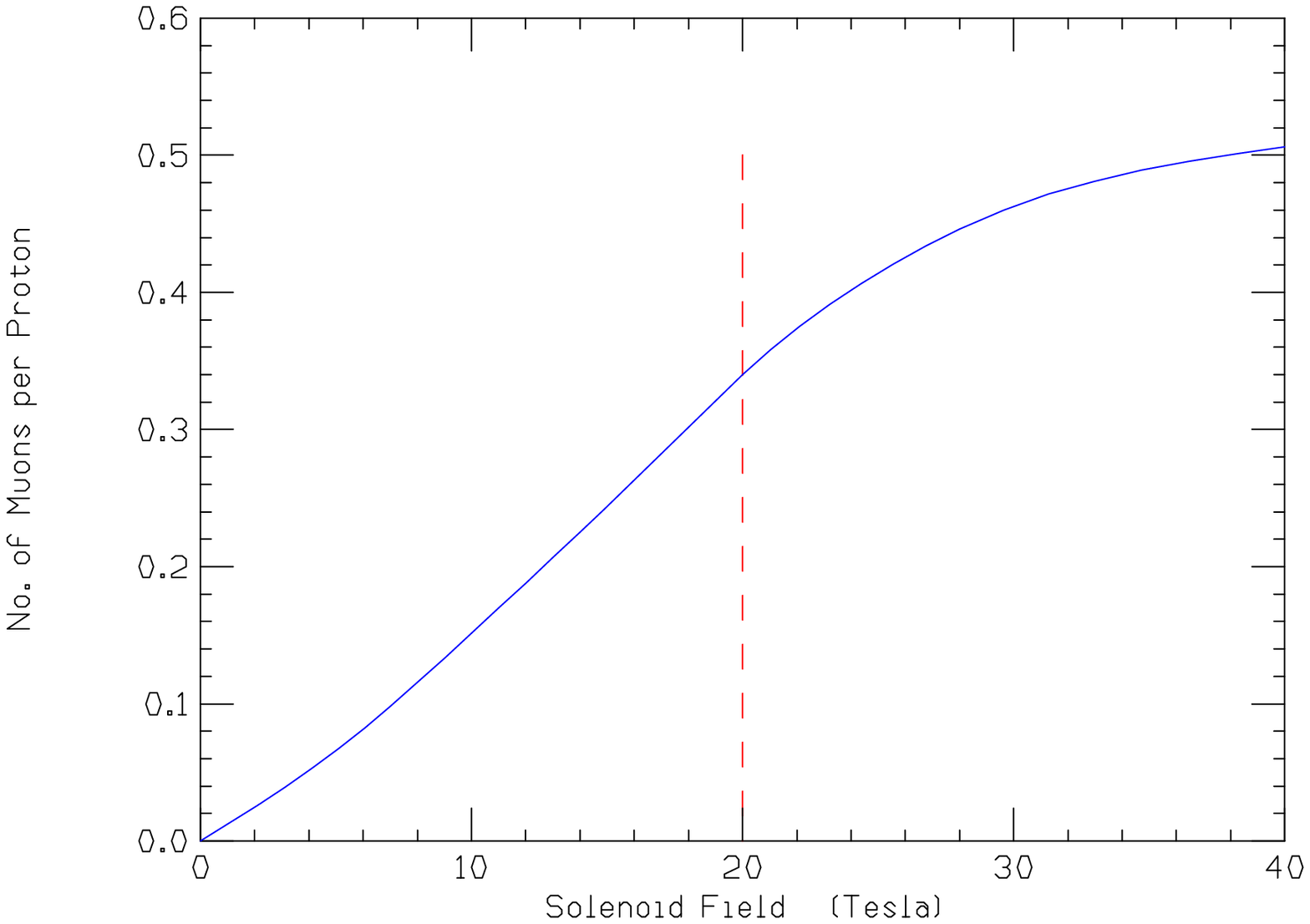,height=3.5in,width=3.5in}}
\caption{The muon to proton ratio as a 
function of capture solenoid field
 \label{muvsb}}
 \end{figure}

Several capture methods were studied\cite{myoldcapture}. Pulsed horns 
were effective at the capture of very high energy pions. Multiple lithium 
lenses were more effective at lower pion energies, but neither was as 
effective as a high field solenoid at the 100 MeV peak of the pion 
spectrum. Initially, a $15\,{\rm cm}$ diameter, $28\,T$ field was considered. 
Such a magnet could probably be built using superconducting outer 
coils and a Bitter, or other immersed sheet conductor inner coil, but 
such an immersed coil would probably have limited life\cite{ref16}. 
A 15 cm diameter, $20\,T$ solenoid could use a more conventional hollow 
conductor inner coil and was thus chosen despite the loss of 24 \% 
pion capture (see Fig.~\ref{muvsb}) 

\begin{figure}[hbt!]
\centerline{\epsfig{file=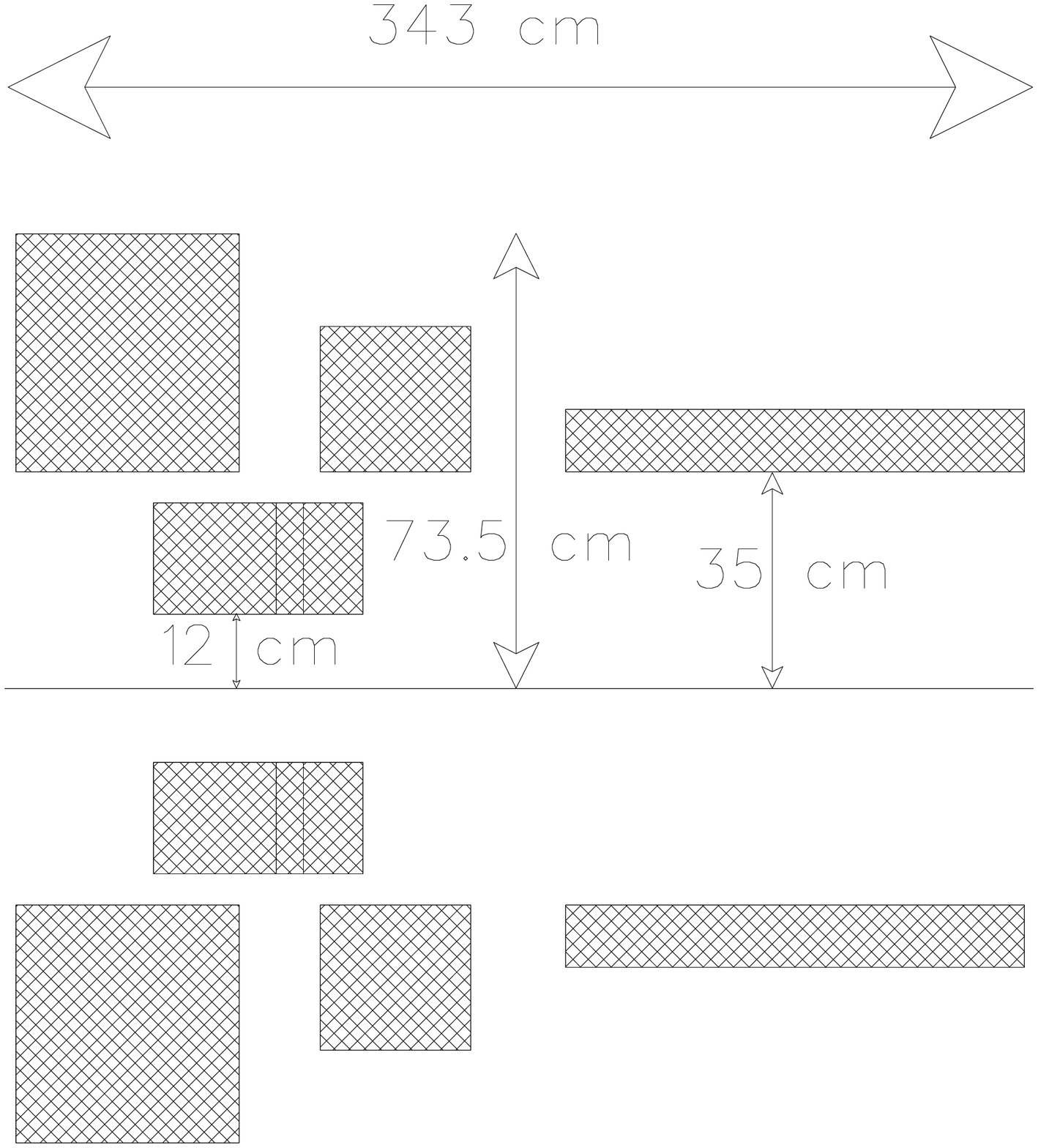,height=4.0in,width=4.0in}}
\caption{Schematic of a hybrid magnet solenoid system for $\pi$ capture and 
matching.
 \label{bitter}}
 \end{figure}

A preliminary design\cite{ref16} (see Fig.~\ref{bitter}) has an inner 
Bitter magnet with an inside diameter of 24 cm (space is allowed for a 
4 cm heavy metal shield inside the coil) and an outside diameter of 60 
cm; it provides half (10T) of the total field, and would consume 
approximately 8 MW. The superconducting magnet has a set of three 
coils, all with inside diameters of 70 cm and is designed to give 10 T 
at the target and provide the required tapered field 
to match into the periodic superconducting 
solenoidal decay channel ($5\,$T and radius $=15\,$cm). A similar 
design has been made at LBL\cite{mikegreens}.

A new design\cite{weggelnew} using a hollow conductor insert is now in 
progress. The resistive coil would give 6 T and consume 4 MW. The 
superconducting coils will supply 14 T.

Monte Carlo studies indicate a yield of 0.4--0.6 muons, of each sign, per 
initial proton, captured in the decay channel. Surprisingly, this conclusion 
seems relatively independent of whether the system is optimized for energies 
of 50 to 500 MeV (using ARC), or 200 to 2000 MeV (using MARS). 
\subsubsection{ Use of Both Signs.}
Protons on the target produce pions of both signs, and a solenoid will capture 
both, but the required subsequent phase rotation rf systems will have opposite 
effects on each. One solution is to break the proton bunch into two, aim them 
on the same target one after the other, and adjust the rf phases such as to 
act correctly on one sign of the first bunch and on the other sign of the 
second. This is the solution assumed in the parameters of this paper. 

A second possibility would be to separate the charges into two channels, 
delay the particles of one charge by introducing a chicane in one of the 
channels, and then recombine the two channels so that the particles of the 
two charges are in line, but separated longitudinally (i.e. box cared). Both 
charges can now be phase rotated by a single linac with appropriate phases of 
rf. 

A third solution is to separate the pions of each charge prior to the use
of rf, and feed the beams of each charge into different channels. 

In either of the latter two solutions, there is a problem in separating the 
beams. After the target, and prior to the use of any rf or cooling, the beams 
have very large emittances and energy spread. Conventional charge separation 
using a dipole is not practical. But if a solenoidal channel is bent, then the 
particles trapped within that channel will drift\cite{ref15},\cite{drift}, in a direction 
perpendicular to the bend (this effect is discussed in more detail in the section on Options below). With our parameters this drift is dominated by a 
term (curvature drift) that is linear with the forward momentum of the 
particles, and has a  direction that depends on the sign of the charges. If 
sufficient bend is employed\cite{ref15}, the two charges could be 
separated by a septum and captured into two separate channels. When these 
separate channels are bent back to the same forward direction, the momentum 
dispersion is separately removed in each new channel. 

Although this idea is very attractive, it has some problems:

\begin{itemize}
\item If the initial beam has a radius r=$0.15\,$m, and if the momentum range
to be accepted is $F={p_{{\rm max}}\over p_{{\rm min}}}=3,$ then the required
height of the solenoid just prior to separation is 2(1+{\it F})r=$1.2\,$m. Use of a
lesser height will result in particle loss. Typically, the reduction in yield
for a curved solenoid compared to a straight solenoid is about $25\,\%$ (due to
the loss of very low and very high momentum pions), but this must be weighed
against the fact that both charge signs are captured for each proton on target.
\item The system of bend, separation, and return bend will require significant
length and must occur prior to the start of phase rotation (see below).
Unfortunately, it appears that the cost of the phase rotation rf is 
strongly dependent on keeping this distance as short as possible. 
 \end{itemize}
Clearly, compromises will be involved, and more study of this concept is 
required.                          
\subsection{Phase Rotation Linac}
The pions, and the muons into which they decay, have an energy spread from
about 0 - 500 MeV, with an rms/mean of $\approx 100 \%$, and peak at
about 100 MeV. It would be difficult to handle such a wide spread in any
subsequent system. A linac is thus introduced along the decay channel, with
frequencies and phases chosen to deaccelerate the fast particles and
accelerate the slow ones; i.e. to phase rotate the muon bunch. Tb.~\ref{rot}
gives an example of parameters of such a linac. It is seen that the lowest
frequency is 30 MHz, a low but not impossible frequency for a conventional
structure.

\begin{table}[htb!]
\centering \protect
\caption{Parameters of Phase Rotation Linacs
\label{rot}}
\begin{tabular}{cccc}
\hline
          &        &           &           \\
Linac     & Length & Frequency & Gradient  \\
	  &  m     &   MHz     &  MeV/m    \\
\hline
1         &  3    &   60     &   5      \\
2         &  29   &   30     &   4       \\
3         &  5    &   60      &   4      \\
4         &  5    &   37     &   4       \\
\hline
\end{tabular}
\end{table}

\begin{figure}[htb!]
\centerline{\epsfig{file=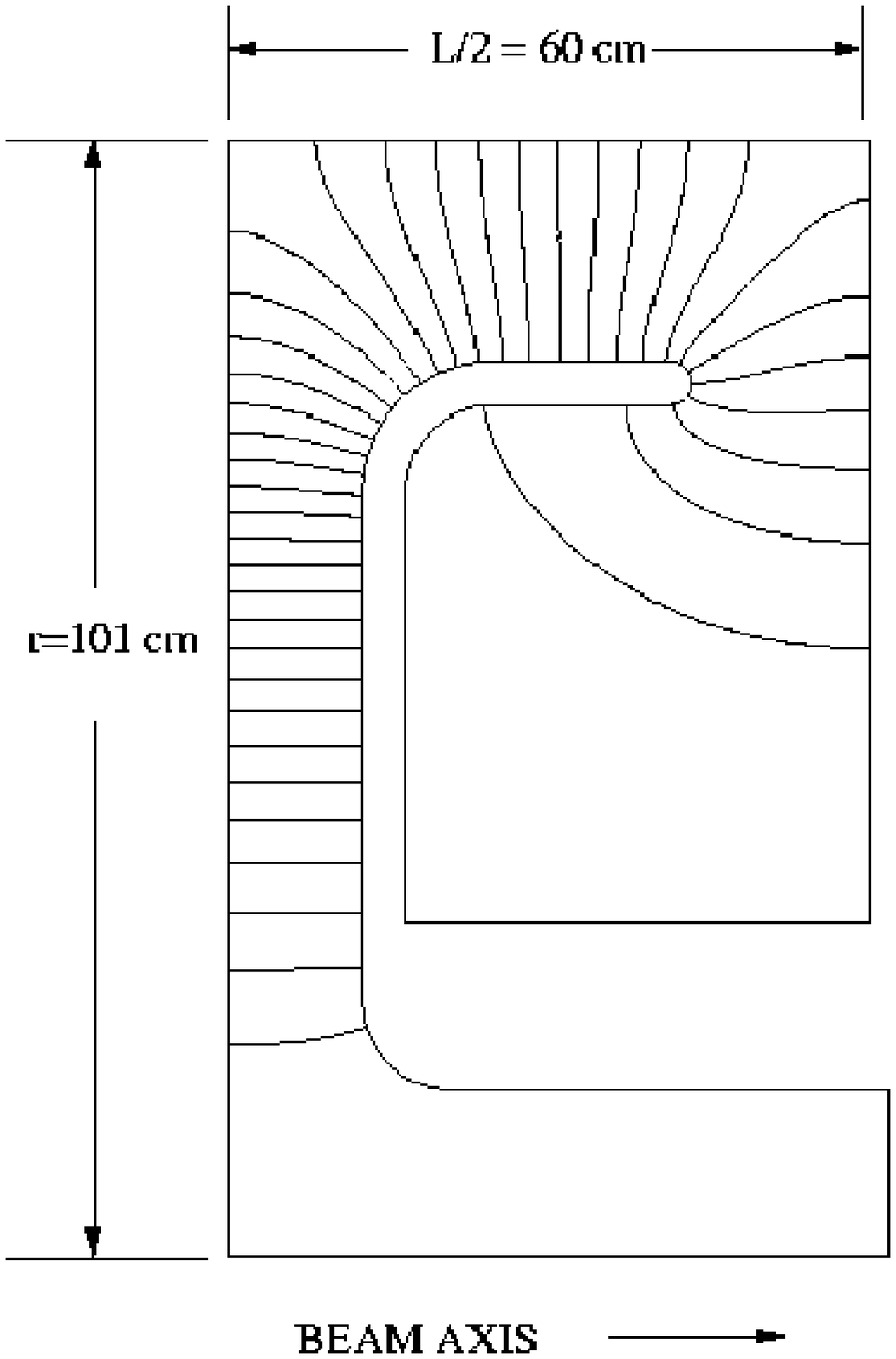,height=4.25in,width=4.25in}}
\vskip 1cm
\caption{30 MHz cavity for use in phase rotation and early stages of cooling.
\label{30MHz}}
\end{figure}

A design of a reentrant 30 MHz cavity is shown in Fig.\ref{30MHz}. Its
parameters are given in Tb.\ref{30MHzt}.
\begin{table}[htb!]
\centering
\caption{Parameters of 30 MHz rf Cavity
\label{30MHzt}}
\begin{tabular}{llc}
\hline
Cavity Radius    & cm & 101 \\
Cavity Length  &   cm & 120  \\
Beam Pipe Radius  &  cm  &  15  \\
Accelerating Gap  &  cm  &  24  \\
Q               &        &  18200  \\
Average Acceleration Gradient & MV/m  &  3  \\
Peak rf Power     &  MW  &  6.3  \\
Average Power (15 Hz) & KW & 18.2  \\
Stored Energy   & J  &  609  \\
\hline
\end{tabular}
\end{table}
 It has a diameter of approximately 2 m, only about one third of that of a
conventional pill-box  cavity. To keep its cost down, it would be made of aluminum.
Multipactoring would probably be suppressed by stray fields from the 5 T
focusing coils, but could also be controlled by an internal coating of titanium
nitride.

\begin{figure}[htb!]
\centerline{\epsfig{file=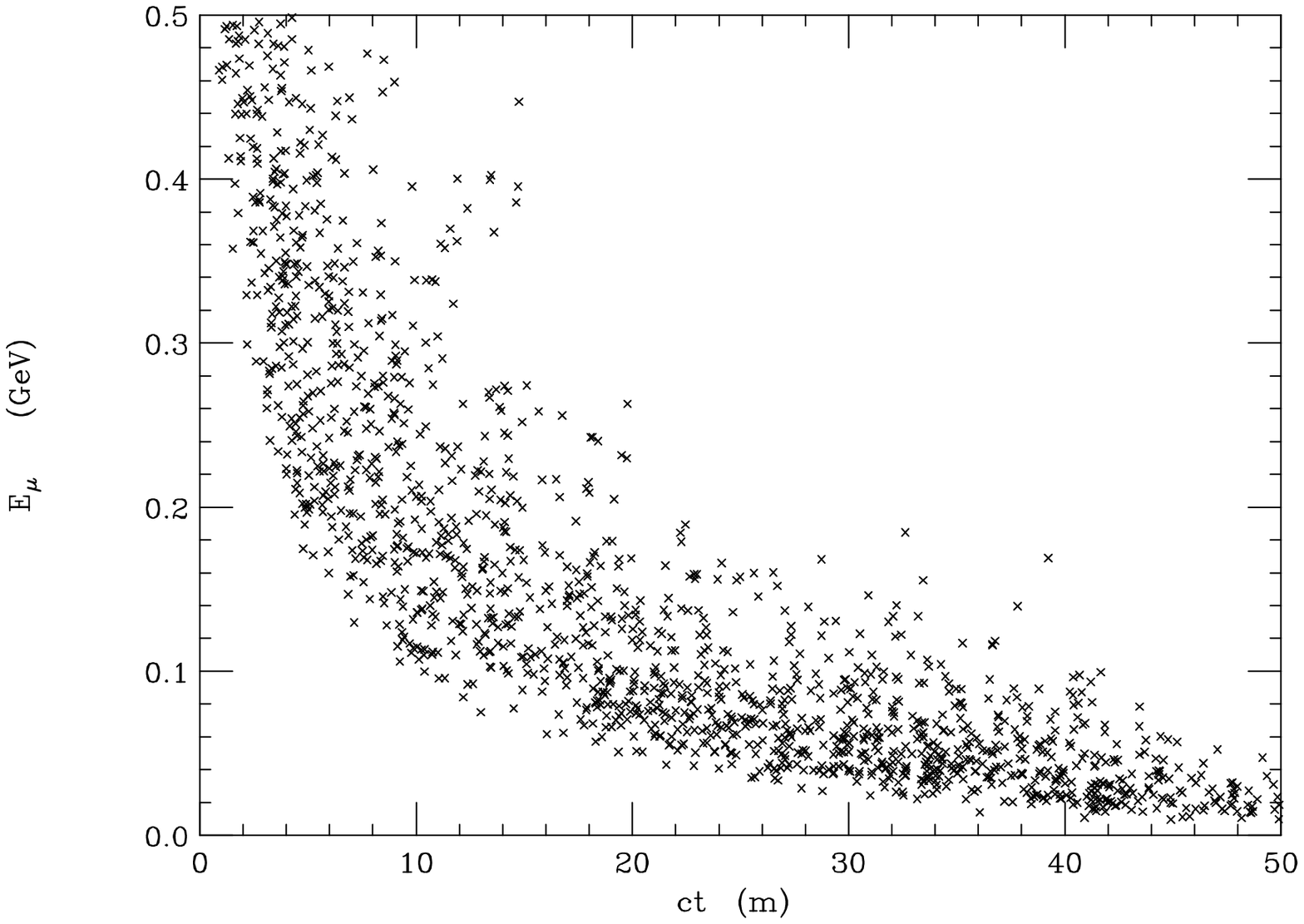,height=4.0in,width=3.5in}}
\caption{Energy vs. ct of Muons at End of Decay Channel without Phase 
Rotation.
\label{Evsct1}}
 \end{figure}

\begin{figure}[htb!]
\centerline{\epsfig{file=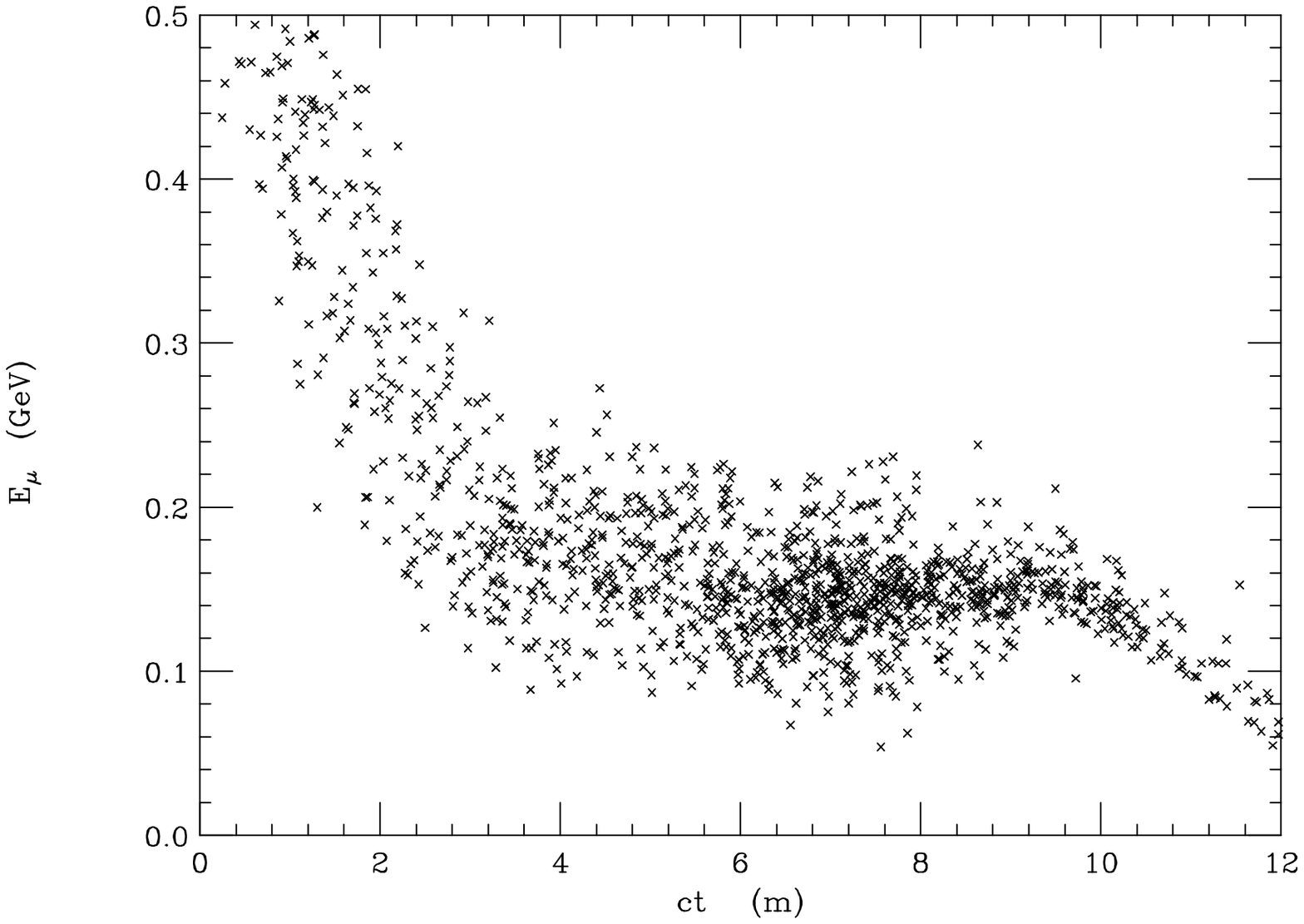,height=4.0in,width=3.5in}}
\caption{Energy vs. ct of Muons at End of Decay Channel with Phase 
Rotation.
 \label{Evsct2}}
 \end{figure}

  Figs.~\ref{Evsct1} and \ref{Evsct2}  show the energy vs. c$\,$t at the end 
of the decay channel with and without phase rotation. Note that the c$\,$t 
scales are very different: the rotation both compacts the energy spread and 
limits the growth of the bunch length. 

After this phase rotation, a bunch can be selected with mean energy 150 MeV,
rms bunch length $1.7\,$m, and rms momentum  spread  $20\,$\% ($95\,$\%,
$\epsilon_{\rm L}= 3.2\,{\rm eV s}$). The number of
muons per initial proton in this selected  bunch is 0.35, about half the
total number of pions initially captured.  As noted above, since the linacs
cannot phase rotate both signs in the same bunch, we need two  bunches: the
phases are set to rotate the $\mu^+$'s of one bunch and the  $\mu^-$'s of the
other. Prior to cooling, the bunch is accelerated to 300 MeV,  in order to
reduce the momentum spread to $10\,\%.$

\subsection{Cooling}
For a collider, the phase-space volume must be reduced within the
$\mu$ lifetime. Cooling by synchrotron radiation, conventional stochastic
cooling and conventional electron cooling are all too slow. Optical stochastic
cooling\cite{ref11a}, electron cooling in a plasma discharge\cite{ref12a} and
cooling in a crystal lattice\cite{ref13a} are being studied, but appear very
difficult. Ionization cooling\cite{ref14a} of muons seems relatively
straightforward.

\subsubsection{Ionization Cooling Theory.}

In ionization cooling, the beam loses both transverse and longitudinal momentum
as it passes through a material medium. Subsequently, the longitudinal
momentum can be restored by coherent reacceleration, leaving a net loss of
transverse momentum. Ionization cooling is not practical for protons and
electrons because of nuclear interactions (p's) and bremsstrahlung (e's), 
but is practical for $\mu$'s because of their low nuclear
cross section and relatively low bremsstrahlung.

The approximate equation for transverse cooling (with energies in GeV)  is:
  \begin{equation}
\frac{d\epsilon_n}{ds}\ \approx \ -\frac{dE_{\mu}}{ds}\ \frac{\epsilon_n}{E_{\mu}}\ +
\ \frac{\beta_{\perp} (0.014)^2}{2\ E_{\mu}m_{\mu}\ L_R},\label{eq1}
  \end{equation}
where $\epsilon_n$ is the normalized emittance, $\beta_{\perp}$ is the betatron
function at the absorber, $dE_{\mu}/ds$ is the energy loss, and $L_R$  is the
radiation length of the material.  The first term in this equation is the coherent
cooling term, and the second is the heating due to multiple scattering.
This heating term is minimized if $\beta_{\perp}$ is small (strong-focusing)
and $L_R$ is large (a low-Z absorber). From Eq.~\ref{eq1} we find a limit to transverse cooling, which occurs when
heating due to multiple scattering  balances cooling due to energy loss. The
limits are $\epsilon_n\approx\ 0.6\ 10^{-2}\ \beta_{\perp}$ for Li, and
$\epsilon_n\approx\ 0.8\  10^{-2}\ \beta_{\perp}$ for Be.

The equation for energy spread  (longitudinal emittance) is:
 \begin{equation}
{\frac{d(\Delta E)^2}{ds}}\ \approx \ -2\ {\frac{d\left( {\frac{dE_\mu}{ds}} \right)} {dE_\mu}}
\ <(\Delta E_{\mu})^2 >\ +\
{\frac {d(\Delta E_{\mu})^2_{{\rm straggling}}}{ds}}\label{eq2}
 \end{equation}
where the first term is the cooling (or heating) due to energy loss, 
and the second term is the heating due to straggling.

Cooling requires that  ${d(dE_{\mu}/ds)\over dE_{\mu}} > 0.$ But at energies
below about 200 MeV, the energy loss function for muons, $dE_{\mu}/ds$, is
decreasing with energy and there is thus heating of the beam.
Above 400 MeV the energy loss function increases gently, giving some
cooling, but not sufficient for our application.

Energy spread can also be reduced by artificially increasing
${d(dE_\mu/ds)\over dE_{\mu}}$ by placing a transverse variation in absorber
density or thickness at a location where position is energy dependent, i.e. where there is
dispersion. The use of such wedges can reduce energy spread, but it
simultaneously increases transverse emittance in the direction of the
dispersion. Six dimensional phase space is not reduced, but it does allow the
exchange of emittance between the longitudinal and transverse directions.

In the long-path-length Gaussian-distribution limit, the heating term (energy
straggling)  is given by\cite{ref15c}
 \begin{equation}
\frac{d(\Delta E_{\mu})^2_{{\rm straggling}}}{ds}\ =\
4\pi\ (r_em_ec^2)^2\ N_o\ \frac{Z}{A}\ \rho\gamma^2\left(1-
\frac{\beta^2}{2}\right),
 \end{equation}
where $N_o$ is Avogadro's number and $\rho$ is the density. Since the energy
straggling increases as $\gamma^2$, and the cooling system size scales as
$\gamma$, cooling at low energies is desired.

\subsection{Low $\beta_{\perp}$ Lattices for Cooling}

We have seen from the above that for a low equilibrium emittance we require 
energy loss in a strong focusing (low $\beta_{\perp}$) region. Three sources 
of strong focusing have been studied:
\subsubsection{Solenoid.}
   The simplest solution would appear to be the use of a long
high field solenoid in which both acceleration and energy loss material could 
be contained. There is, however, a problem: 
when particles enter a solenoid other than on the axis, they are 
given angular momentum  by the radial field components that they must pass. 
This initial angular momentum is proportional to the solenoid field strength, and to 
the particles' radius. In the absence of material, this extra angular momentum 
is maintained proportional to the tracks' radius as they pass along the 
solenoid until they are exactly corrected by the radial fields at the exit.  But if material is introduced, all transverse momenta are ``cooled", including 
the extra angular momentum given by these radial fields. When the cooled 
particles now leave the solenoid, then the end fields overcorrect them, 
leaving the particles with a finite added angular momentum.
In practice, this angular momentum is equivalent to a significant heating term 
that limits the maximum emittance reduction to a quite small factor. The 
problem can only be averted if the direction of the solenoid field is 
periodically reversed.

\subsubsection{Alternating Solenoid (FOFO) Lattice.}

An interesting case of such periodic solenoid field reversals is a lattice 
with rapid reversal that, for example, might approximate sinusoidal 
variations. We describe such a lattice as FOFO (focus focus) in analogy with 
quadrupole lattices that are FODO (focus defocus). Not only do such lattices 
avoid the angular momentum problems of a long solenoid, but they can, if the 
phase advance per cell approaches $\pi$, provide  $\beta_{\perp}$'s at the zero field 
points, that are less than the same field would provide in the long solenoid 
case. 

But as noted above, for cooling to be effective, the ratio of emittance to 
$\beta_{\perp}$ must remain above a given value. This implies that the angular 
amplitude of the particles has to be relatively large (typically greater than 
0.1 radians rms). When tracking of such distributions was performed on 
realistic lattices three apparent problems were observed: 

 \begin{enumerate}
 \item Particles entering with large amplitude (radius or angle) were 
found\cite{neufferandy} to be
lost or reflected by the fringe fields of the lenses. The basic problem is that 
there are strong non-linear effects that focus the large angle particles more 
strongly than those at small angles (this is known as a second order tune 
shift). The stronger focus causes an 
increase in the phase advance per cell resulting in resonant behavior, 
emittance growth and particle loss.
 \item  A bunch, even when monoenergetic, passing along such a lattice would 
be seen\cite{fernow} to
rapidly grow in length because the larger amplitude particles, traveling 
longer orbits, would fall behind the small amplitude ones.
 \item With material present, the energy spread of a bunch grew because the 
high amplitude particles were passing through more material than the low 
amplitude ones.
 \end{enumerate}

Surprisingly however, none of these turns out to be a real problem. If the 
particles are matched, as they must be, into rf buckets, then all particles at 
the centers of these buckets must be traveling with the same average forward 
velocity. If this were not so then they would be arriving at the next rf 
cavity with different phases and would not be at the center of the bucket. It 
follows that large amplitude particles (whose trajectories are longer) must 
have higher momenta than those with lower amplitude. The generation of this 
correlation is part of the matching requirement, and would be 
naturally generated if an adiabatic application of FOFO strength were 
introduced. It could also be generated by a suitable 
gradation of the average radial absorber density.

Since higher amplitude particles will thus have higher momenta, they will, as a 
result, be less strongly focused: an effect of the opposite sign to the second 
order tune shift natural to the lattice. Can the effects cancel ? In practice 
they are found to cancel almost exactly at a specific momentum: close to 100 
MeV/c for a continuous sinusoidal FOFO lattice (the exact momentum will depend 
on the lattice).

A second, but only partial, cancelation also occurs: the higher amplitude, 
and now higher momentum, particles lose less energy in the absorber because of 
the natural energy dependence of the energy loss. This difference of energy 
loss, at 100 MeV/c, actually overcorrects the difference in energy loss from 
the difference in trajectories in the material. But this too is no problem. 
The natural bucket center for large amplitude particles will be displaced not 
only up in energy, but also over in phase, so as to be in a different 
accelerating field, and thus maintain their energy. Again, this would
occur naturally if the lattice is introduced adiabatically and can also be generated by a combination of radially graded absorbers and drifts. Particles of differing momentum or phase will, as in normal synchrotron 
oscillation, gyrate about their bucket centers, but now each amplitude has a 
different center.

\begin{figure}[bht!] 
\centerline{\epsfig{file=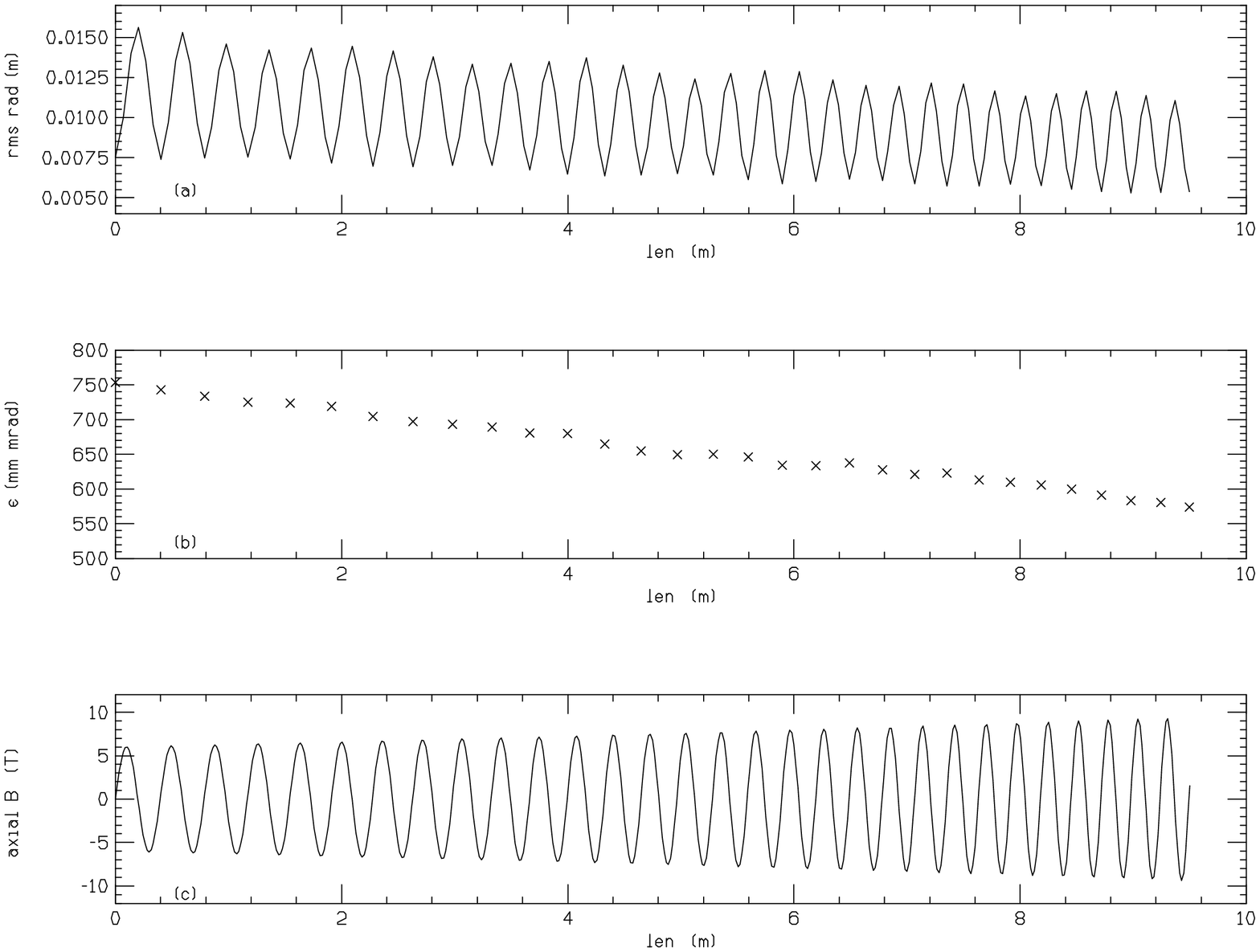,height=5.0in,width=6.in, angle=90}}
\caption{Cooling in a FOFO Lattice: (a) rms radius; (b) normalized emittance; and (c) 
axial magnetic fields; all plotted vs axial length.
 \label{fofo}}
 \end{figure}

Using particles so matched, a simulation using fully Maxwellian sinusoidal 
field has been shown to give continuous transverse cooling without significant 
particle loss (see Fig.~\ref{fofo}). In this simulation, the axial field has been gradually increased,
and its period decreased, so as to maintain a constant rms angular spread  
as the emittance falls. The peak rf accelerating fields were 10 MeV/m, 
their frequency 750 MHz, the absorbing material was lithium, placed at the 
zero magnetic field positions, with lengths such that they occupied 5 \% of the length. The mean momentum was 110 MeV/c, and rms width 2 \%. 500 particles 
were tracked; none were lost.

\subsubsection{Lithium Rods.}
The third method of providing strong focusing and energy loss is to pass the 
particles along a current carrying lithium rod (a long lithium lens). 
The rod serves simultaneously to maintain the low
$\beta_{\perp}$, and attenuate the beam momenta. Similar lithium rods, with
surface fields of $10\,$T , were developed at Novosibirsk\cite{ref16a} and have been used as
focusing elements at FNAL\cite{fermilab_li} and CERN\cite{cern_li}. At the repetition rates 
required here, cooling of a solid rod will not be possible, and circulating 
liquid columns will have to be used. A small lens using such liquid cooling 
has also been tested at Novosibirsk. It is also hoped\cite{20Tli} that because 
of the higher compressibility of the liquid, surface field 
up to 20 T may be possible. 

Lithium lenses will permit smaller 
$\beta_{\perp}$ and therefore cooling to lower emittances than in a practicable 
FOFO lattice, and such rods are thus preferred for the final cooling stages. But 
they are pulsed devices and consequently they are likely to have significant life time problems, 
and are thus not preferred for the earlier stages where they are not absolutely 
needed.                         

Such rods do not avoid the second order tune shift complications
discussed above for the FOFO lattices. The rods must be alternated with 
acceleration sections and thus the particles must periodically be focused 
into and out of the rods. All three of the nonlinear effects enumerated above 
will be encountered. It is reasonable to believe that they can be 
controlled by the same mechanisms, but a full simulation of this has not 
yet been done. 
\subsubsection{Emittance Exchange Wedges.}
Emittance exchange in wedges to reduce the longitudinal emittance has been 
modeled with Monte Carlo calculations and works as theoretically predicted. 
But the lattices needed to generate the required dispersions and focus the 
particles onto the wedges have yet to be designed. The nonlinear complications 
discussed above will again have to be studied and corrected. 

Emittance exchange in a bent current carrying rod has also been studied, both 
for a rod of uniform density\cite{fredmills} (in which the longer path 
length on the outside of the helix plays the role of a wedge; 
and where the average rod density 
is made greater on the outside of the bends by the use of wedges of a more dense material\cite{sasha}.
\subsubsection{Reverse Emittance Exchange.}
   At the end of a sequence of a cooling elements, the transverse emittance 
may not be as low as required, while the  longitudinal emittance, has been
cooled to a value less than is required. The
additional reduction of transverse emittance can then be obtained by  a
reverse exchange of transverse and longitudinal phase-spaces. This can be 
done in one of several ways:
 \begin{enumerate}
 \item by the use of wedged absorbers in dispersive regions between solenoid
elements.
 \item by the use of septa that subdivide the transverse beam size,
acceleration that shifts the energies of the parts, and bending to recombine 
the parts\cite{sasha}.
 \item by the use of lithium lenses at very low energy: at very low energies 
the $\beta_{\perp}$'s, and thus equilibrium emittances, can be made arbitrarily low; 
but the energy spread is blown up by the steep rise in dE/dx. If this blow up 
of dE/dx is left uncorrected, then the effect can be close to an emittance 
exchange. 
 \end{enumerate}
\subsection{Model Cooling System}
We require a reduction of the normalized transverse emittance by almost three
orders of magnitude (from $1\times 10^{-2}$ to $5\times 10^{-5}\,$m-rad), and a
reduction of the longitudinal emittance by one order of magnitude.

A {\it model example} has been generated that uses no recirculating 
loops, and it is assumed for simplicity that the beams of each charge are 
cooled in separate channels (it may be possible to design a system 
with both charges in the same channel). 
The cooling is obtained in a series of cooling stages. In the early stages, 
they each have two components: 
 \begin{enumerate}
 \item FOFO lattice consisting of spaced axial solenoids with alternating 
field directions and lithium hydride absorbers placed at the centers of the 
spaces between them where the $\beta_{\perp}$'s are minimum. RF cavities are introduced between the absorbers along the 
entire length of the lattice.
 \item A lattice consisting of more widely separated alternating solenoids,
and bending magnets between them
to generate dispersion. At the location of maximum
dispersion, wedges of lithium hydride are introduced to interchange
longitudinal and transverse emittance. 
 \end{enumerate}

In the last stages, reverse emittance exchange is achieved using current carrying 
lithium rods. The energy is allowed to fall to 15 MeV, thus increasing the 
focussing strength and lowering  $\beta_{\perp}$.

The design is based on analytic calculations.
The phase advance in each cell of the FOFO lattice is made as close to $\pi$ as
possible in order to minimize the $\beta_{\perp}$'s at the location of the absorber.
The following effects are included: space charge transverse defocusing and longitudinal space charge forces; a 
$3\, \sigma$ fluctuation of momentum and $3\, \sigma$ fluctuations in 
amplitude.

\begin{figure}[bht!] 
\centerline{\epsfig{file=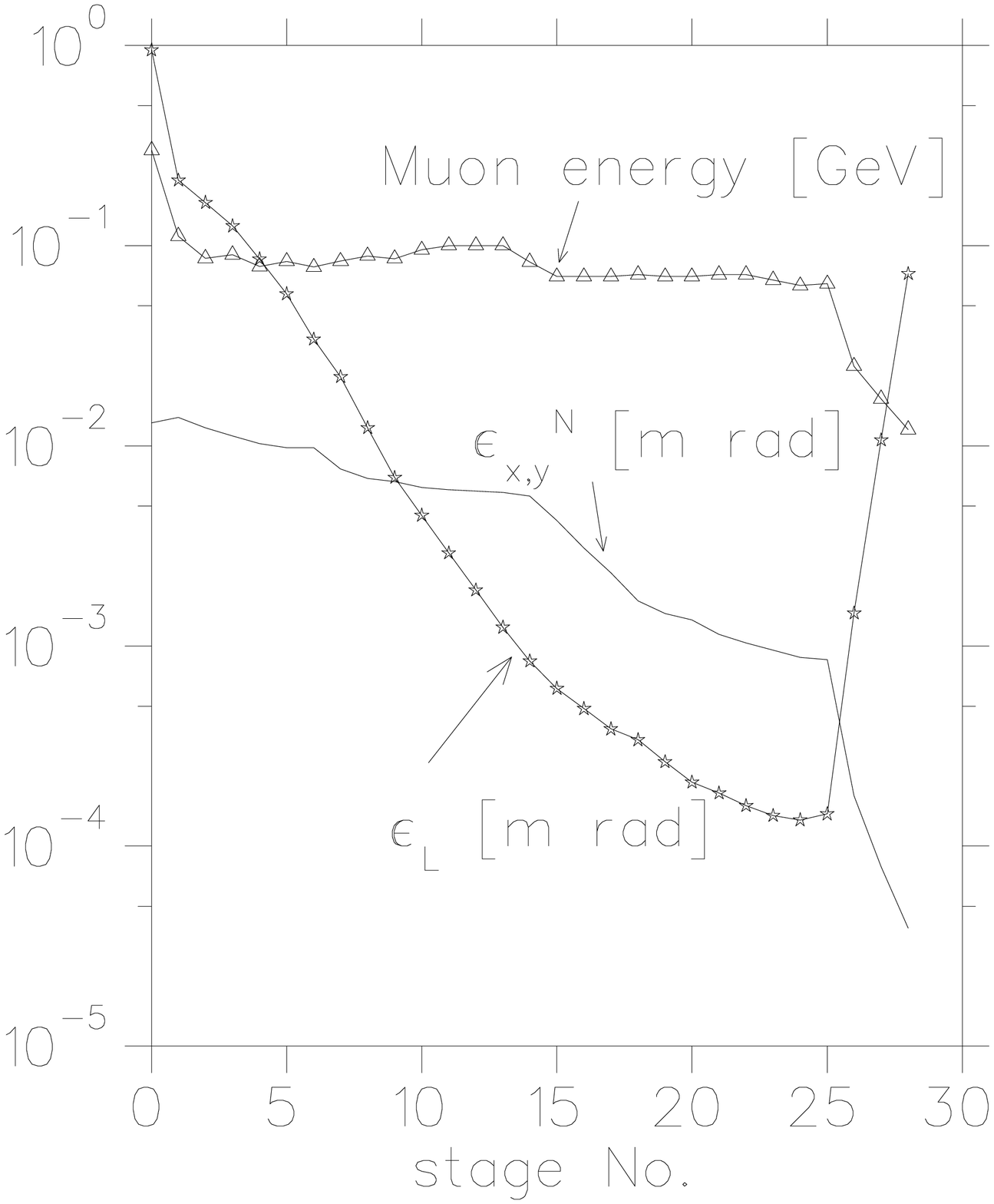,height=4.0in,width=3.5in}}
\caption{$\epsilon_{\perp}$, ${\epsilon_{L}\,c\over \left< {\rm E}_{\mu}\right>} $ and E$_{\mu}$ [GeV] vs. stage number in the cooling sequence.
 \label{cooling}}
 \end{figure}

The emittances, transverse and longitudinal, as a function of stage number, 
are shown in Fig.\ref{cooling}, together with the beam energy. In the first 15 
stages, relatively strong wedges are used to rapidly reduce the longitudinal 
emittance, while the transverse emittance is reduced relatively slowly. The 
objective is to reduce the bunch length, thus allowing the use of higher 
frequency and higher gradient rf in the reacceleration linacs. In the next 10 
stages, the emittances are reduced close to their asymptotic limits. In the final three stages, lithium rods are used to produce an effective emittance exchange, as described above.

Individual components of the lattices have been defined, but a complete 
lattice has not yet been specified, and no complete Monte Carlo study of its 
performance has yet been performed. Wake fields, resistive wall effects, 
second order rf effects and some higher order focus effects are not yet  
included in this design of the system.                                 

   The total length of the system is 750 m, and the 
total acceleration used is
4.7 GeV. The fraction of muons that have not decayed and 
are available for acceleration is calculated to be 
$55\,$\%.

It would be desirable, though not necessarily practical, to economize on linac 
sections by forming groups of stages into recirculating loops. 
\subsection{Acceleration}
Following cooling and initial bunch compression the beams must be rapidly 
accelerated
to full energy (2 TeV, or 250 GeV). A sequence of linacs would work, but would
be expensive. Conventional synchrotrons cannot be used because the muons would
decay before reaching the required energy. The conservative solution is to use
a sequence of recirculating accelerators (similar to that used at CEBAF). A
more economical solution would be to use fast rise time pulsed magnets in
synchrotrons, or synchrotrons with rapidly rotating permanent magnets
interspersed with high field fixed magnets.
\subsubsection{Recirculating Acceleration.}
Tb.~\ref{acceleration} gives an example of a possible sequence of recirculating 
accelerators. After initial linacs, there are two conventional rf 
recirculating accelerators taking the muons up to 75 GeV, then two 
superconducting recirculators going up to 2000 GeV. 
 \begin{table}[hbt!]
\centering
 \caption{Parameters of Recirculating Accelerators
\label{acceleration}} 
 \begin{tabular}{llccccc}
\hline 
		     &         & Linac &  \#1   &  \#2   &   \#3  &   \#4  \\
\hline
initial energy       & GeV     &   0.20&     1 &     8 &    75 &   250 \\
final energy         & GeV     &     1 &     8 &    75 &   250 &  2000 \\
nloop                &         &     1 &    12 &    18 &    18 &    18 \\
freq.                & MHz     &   100 &   100 &   400 &  1300 &  2000 \\
linac V              & GV      &   0.80&   0.58&   3.72&   9.72&  97.20 \\
grad                 &         &     5 &     5 &    10 &    15 &    20  \\
dp/p  initial        & \%      &    12 &   2.70&   1.50&     1 &     1  \\
dp/p  final          & \%      &   2.70&   1.50&     1 &     1 &   0.20 \\
$\sigma_z$ initial   & mm      &   341 &   333 &  82.52&  14.52&   4.79 \\
$\sigma_z$ final     & mm      &   303 &  75.02&  13.20&   4.36&   3.00 \\
$\eta$               & \%      &   1.04&   0.95&   1.74&   3.64&   4.01 \\
$N_\mu$              & $10^{12}$ &   2.59&   2.35&   2.17&   2.09&   2    \\
$\tau_{fill}$        & $\mu$s  &  87.17&  87.17&  10.90&  s.c. &  s.c.  \\
beam t               & $\mu$s  &   0.58&   6.55&  49.25&   103 &   805   \\
decay survival       &         &   0.94&   0.91&   0.92&   0.97&   0.95  \\
linac len            & km      &   0.16&   0.12&   0.37&   0.65&   4.86  \\
arc len              & km      &   0.01&   0.05&   0.45&   1.07&   8.55  \\
tot circ             & km      &   0.17&   0.16&   0.82&   1.72&  13.41  \\
phase slip           & deg     &     0 &  38.37&   7.69&   0.50&   0.51  \\
\hline
\end{tabular}
\end{table}

\begin{figure}[thb!] 
\centerline{\epsfig{file=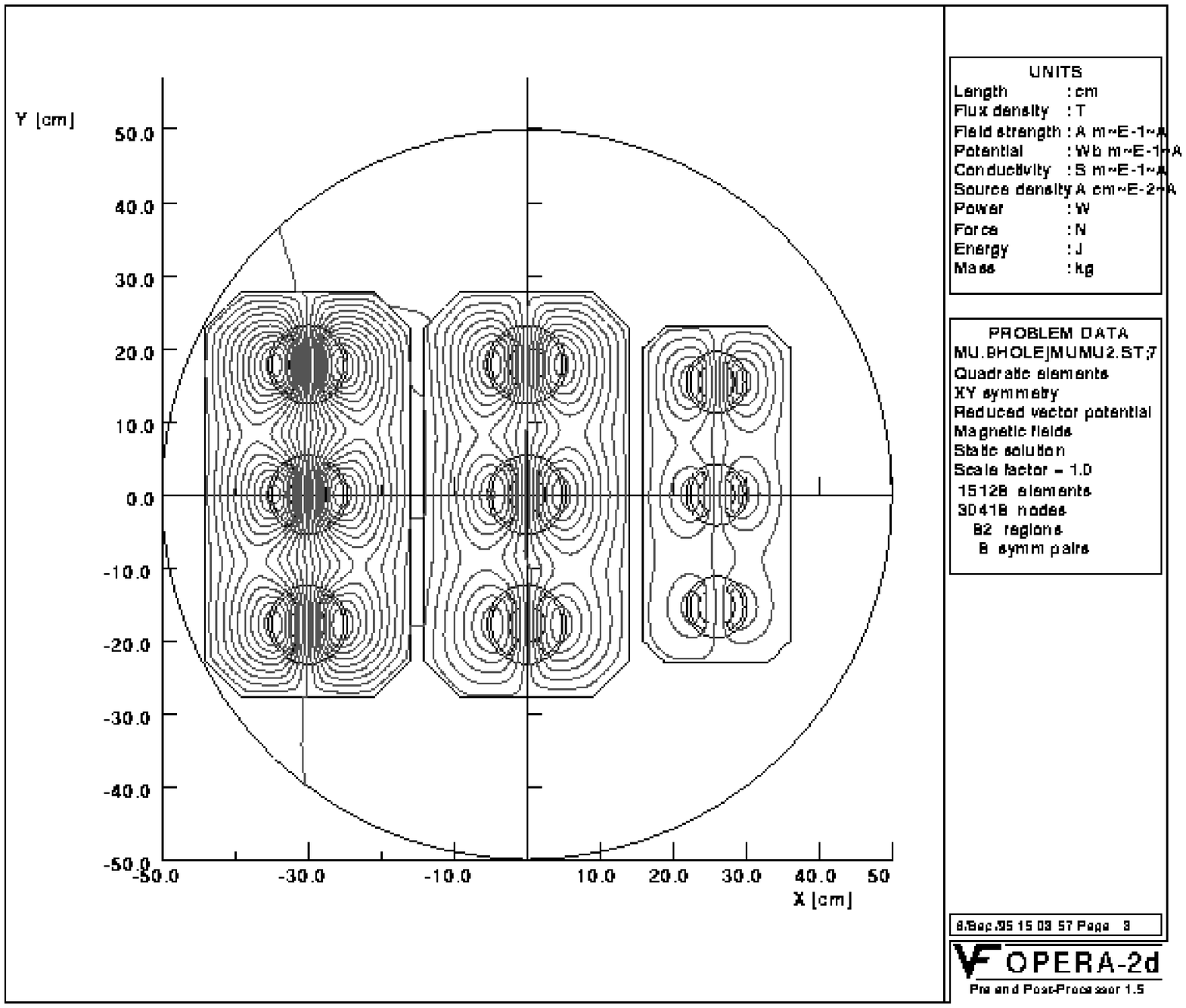,height=4.0in,width=4.5in}}
\caption{A cross section of a 9 aperture sc magnet.  
 \label{9hole}}
 \end{figure}

   Criteria that must be considered in picking the parameters of such
accelerators are:
\begin{itemize}
\item The wavelengths of rf should be chosen to limit the loading, $\eta$, (it
is restricted to below 4 \% in this example) to avoid excessive longitudinal
wakefields and the resultant emittance growth.
 \item  The wavelength should also be sufficiently large compared to the bunch
length to avoid excessive  second order effects (in this example: 10 times).
 \item  For power efficiency, the cavity fill time should be
long compared to the acceleration time. When conventional cavities cannot
satisfy this condition, superconducting cavities are specified.
 \item  In order to minimize muon decay during acceleration (in this example
73\% of the muons are accelerated without decay), the number of
recirculations at each stage should be kept low, and the rf acceleration
voltage correspondingly high. For minimum cost, the number of
recirculations appears to be of the order of 18. In
order to avoid a large number of separate magnets, multiple aperture magnets
can be designed (see Fig.\ref{9hole}).
 \end{itemize}
Note that the linacs see two bunches of opposite signs, passing through
in opposite directions. In the final accelerator in the 2 TeV case, each bunch
passes through the linac 18 times. The total loading
is then $4\times 18\times \eta = 288 \%.$ With this loading, assuming 60\%
klystron efficiencies and reasonable cryogenic loads, one could probably
achieve 35\% wall to beam power efficiency, giving a wall power
consumption for the rf in this ring of 108 MW.

A recent study\cite{neufferacc} tracked particles through a
similar sequence of recirculating accelerators and found a dilution of
longitudinal phase space of the order of 15\% and negligible particle loss.

\subsubsection{Pulsed Magnet Acceleration.}
An alternative to recirculating accelerators for stages \#2 and \#3 would be 
to use pulsed magnet synchrotrons with rf systems consisting of significant 
lengths of superconducting linac.  

\begin{figure}[bth!]
\centerline{\epsfig{file=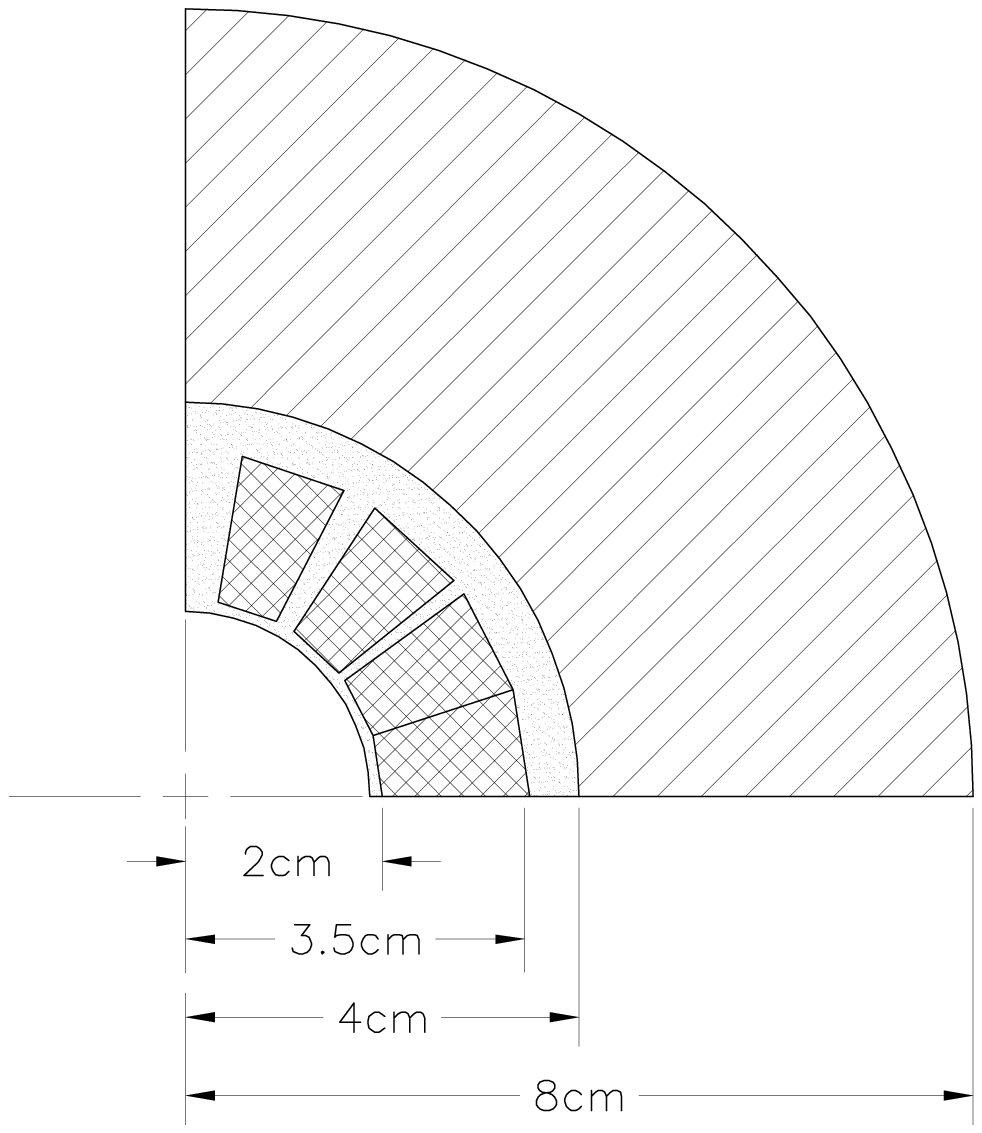,height=4.1in,width=2.9in}}
\caption{Cross section of pulsed magnet for use in the
acceleration to 250 GeV. 
 \label{pulse}}
 \end{figure}

The cross section of a pulsed magnet for this
purpose is shown in Fig.~\ref{pulse}. If desired, the number of recirculations
could be higher in this case, and the needed rf voltage correspondingly lower,
but the loss of particles from decay would be somewhat more. The cost for a
pulsed magnet system appears to be significantly less than that of a multi-hole
recirculating magnet system, and the power consumption is moderate for energies 
up to 250 GeV.  Unfortunately, the power consumption is impractical at
energies above 500 GeV.

\subsubsection{Pulsed and Superconducting Hybrid.}

For the final acceleration to 2 TeV in the high energy machine, the power 
consumed by a ring using only pulsed magnets would be excessive,
but a hybrid ring with alternating 
pulsed warm magnets and fixed superconducting magnets\cite{rotating}\cite{summers2}  should be a good alternative. 

Tb.~\ref{acceler} gives an example of a possible sequence of       such 
accelerators. Fig.~\ref{accpict} used a layout of this sequence. The 
first two rings use pulsed cosine theta magnets with peak fields of $3\,$T and 
$4\,$T. Then follow two hybrid magnet rings with $8\,$T fixed magnets 
alternating with $\pm2\,$T iron yoke pulsed magnets. The latter two rings 
share the same tunnel, and might share the same linac too. The survival from 
decay after all four rings is $67\,\%.$ Phase space dilution should be similar to that determined for the recirculating accelerator design above.
\begin{table}[htb!]
\centering
 \caption{Parameters of Pulsed Accelerators
\label{acceler}}
\begin{tabular}{llcccc}
\hline 
   & Ring     &             1   & 2 & 3 & 4 \\ \hline
$E_{init}$ &(GeV)  &    2.5  &  25  &  250  &  1350  \\
$E_{final}$& (GeV) &  25  &  250  &  1350  &  2000  \\
fract pulsed  & \% &  100   &  100   &   73 & 44 \\   
$B_{pulsed}$ &(T) &  3  &  4  &  $\pm 2$  &  $\pm 2$  \\
Acc/turn &(GeV)  &  1  &  7  &  40  &  40  \\
Acc Grad &(MV/m)  &  10  &  12  &  20  &  20  \\
RF Freq &(MHz)  &  100  &  400  &  1300  &  1300  \\
circumference &(km)  &    0.4  &    2.5  &   12.8  &   
12.8  \\
turns & &  22  &  32  &  27  &  16  \\
acc. time& ($\mu s$)  &  26  &  263  &  1174  &  691  \\
ramp freq& (kHz)  &   12.5  &    1.3  &    0.3  &    0.5 
 \\
loss &(\%)  &   13.4  &   13.2  &    9.0  &    2.2  \\
\hline
\end{tabular}
\end{table} 
\subsection{Collider Storage Ring} 
After acceleration, the $\mu^+$ 
and $\mu^-$ bunches are injected into a storage ring that is separate from 
the accelerator. The highest possible average bending field is desirable, to 
maximize the number of revolutions before decay, and thus maximize the 
luminosity. Collisions would occur in one, or perhaps two, very low-$\beta^*$ 
interaction areas. Parameters of the ring were given earlier in  Tb.\ref{sum}. 

\subsubsection{Bending Magnet Design.}

   The magnet design is complicated by the fact that the $\mu$'s decay within
the rings ($\mu^-\ \rightarrow\ e^-\overline{\nu_e}\nu_{\mu}$), producing
electrons whose mean energy is approximately $0.35$ that of the muons. These
electrons travel toward the inside of the ring dipoles, radiating a 
fraction of their energy as synchrotron radiation towards the outside of the
ring, and depositing the rest on the inside. The total beam power, in the 4 
TeV machine, is 38 MW.  The total power deposited in the ring is 13 MW, yet 
the maximum power that can reasonably be taken from the magnet coils at 4$\,$K 
is only of the order of 40 KW.  Shielding is required.

\begin{figure}[htb!]
\centerline{\epsfig{file=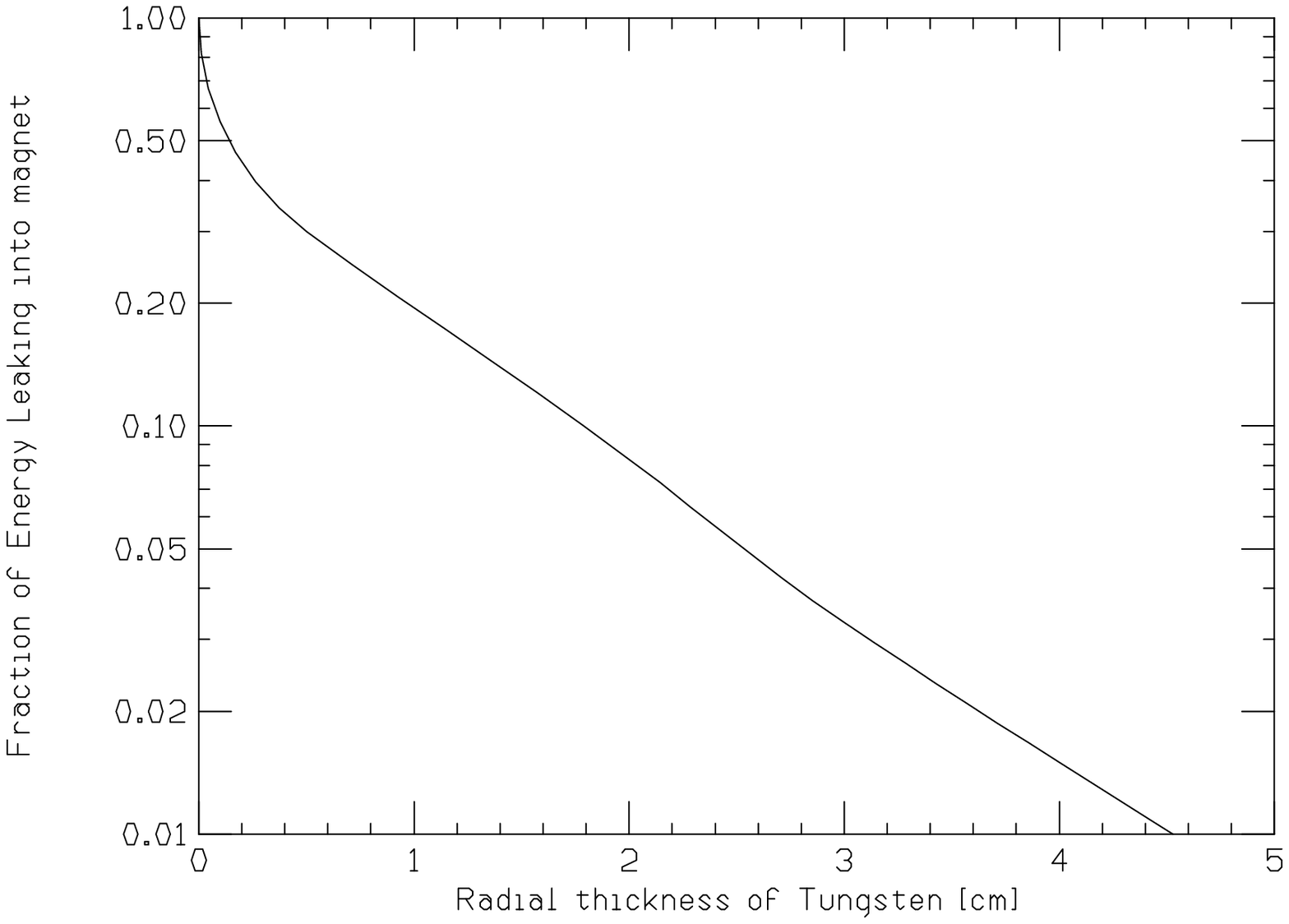,height=4.0in,width=3.5in}}
\caption{Energy attenuation vs. the thickness of a tungsten liner. 
 \label{shield}}
 \end{figure}

The beam is surrounded by a thick warm shield, located inside a large aperture 
magnet. Fig.\ref{shield} shows the attenuation of the heating produced as a 
function of the thickness of a warm tungsten liner\cite{Iuliupipe}. If 
conventional superconductor is used, then the thicknesses required in the 
two cases would be as given in Tb.\ref{linert}. If high Tc superconductors 
could be used, then these thicknesses could probably be halved. 

 \begin{table}[hbt!]  
\caption{Thickness of Shielding for Cos Theta Collider Magnets.
\label{linert}}
 \begin{tabular}{llcc}
\hline
		  &        &  2TeV   &   0.5 TeV Demo   \\
\hline
Unshielded Power  &   MW   &  13          &    .26   \\
Liner inside rad &    cm   &  2            &   2     \\
Liner thickness  &   cm   &   6           &     2    \\
Coil inside rad  &   cm   &   9    &       5     \\ 
Attenuation      &        &   400  &     12   \\
Power leakage     &   KW   &  32   &       20   \\
Wall power for $4\,K$ &  MW   &   26      &   16    \\
\hline
\end{tabular}
\end{table}

\begin{figure}[hbt!]
\centerline{\epsfig{file=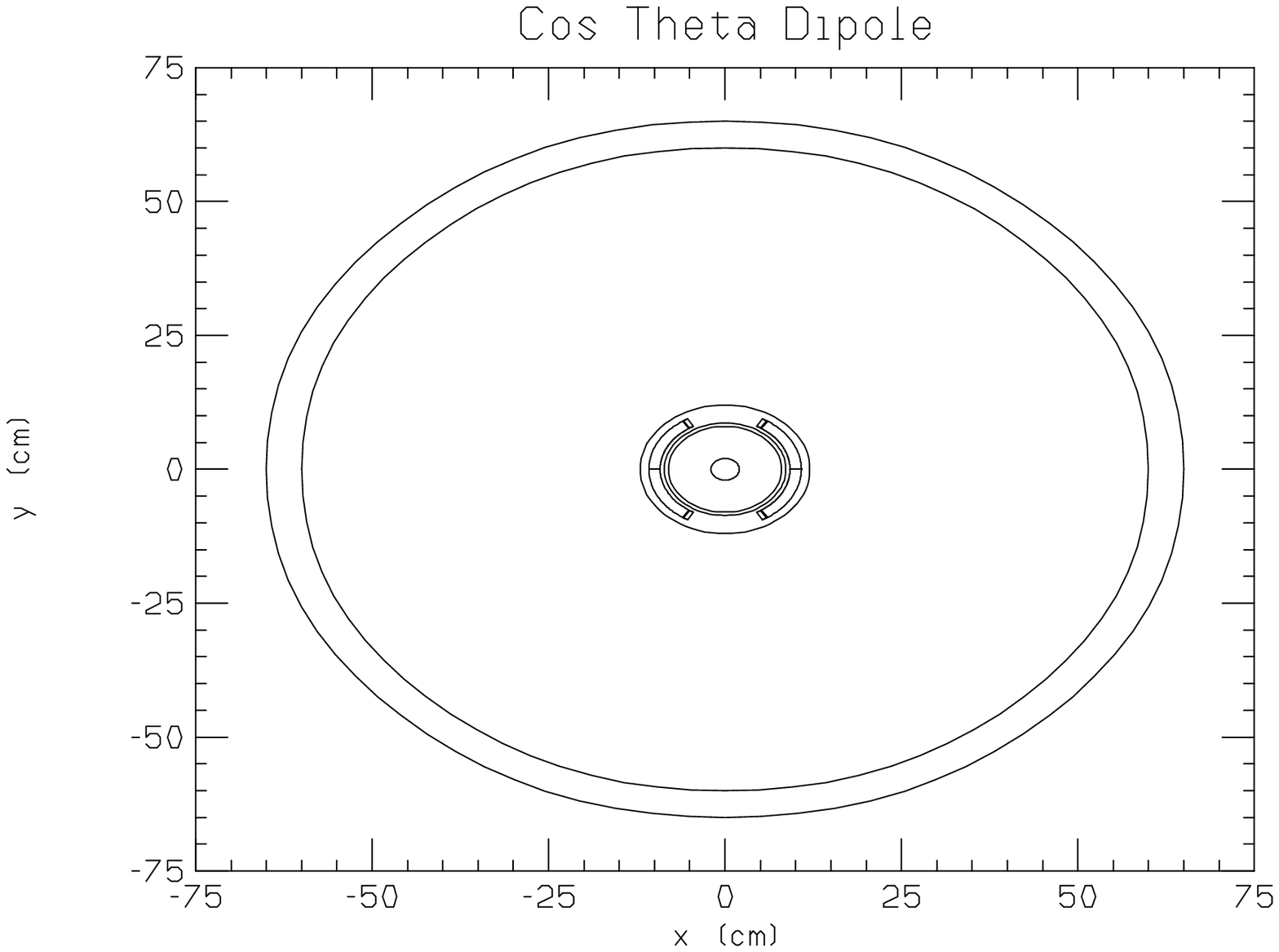,height=4.0in,width=4.0in}} 
\caption{Cos Theta Arc Bending Magnet
 \label{costhetamag}}
 \end{figure}

The magnet could be a conventional cosine-theta magnet 
(see Fig.\ref{costhetamag}), or, in order to reduce the compressive forces on 
the coil midplane, a rectangular block design.

The power deposited could be further reduced if the 
beams are kicked out of the ring prior to their their complete decay. Since 
the luminosity goes as the square of the number of muons, a significant power 
reduction can be obtained for a small luminosity loss.
\subsubsection{Quadrupoles.}
The quadrupoles could have warm iron
poles  placed as close to the beam as practical. The coils could be either 
superconducting or warm, as dictated by cost considerations. If an elliptical 
vacuum chamber were used, and the poles were at 1 cm radius, then gradients of 
150 T/m should be possible. 
\subsubsection{Lattice Design.}

\begin{enumerate}
\item{{\bf Arcs:} In a conventional 2 TeV FODO lattice the tune would be of the order of 
200 and the momentum compaction $\alpha$ around $2 \times 10^{-3}$. In this 
case, in order to maintain a bunch with rms length 3 mm, 45 GeV of S-band rf 
would be required. This would be excessive. It is thus proposed to use an 
approximately isochronous lattice of the dispersion wave type\cite{ref17a}. 
Ideally one would like an $\alpha$ of the order of $10^{-7}$. In this case 
the machine would behave more like a linear beam 
transport and rf would be needed only to correct energy spread introduced by 
wake effects. 

It appears easy to set the zero'th order slip 
factor $\eta_0$ to zero, but if nothing is done, there is a relatively large 
first order slip factor $\eta_1$ yielding a minimum $\alpha$ of the order of 
$10^{-5}$. The use of sextupoles appears able to correct this $\eta_1$ 
yielding a minimum $\alpha$ of the order of $10^{-6}$. With octupoles it may 
be possible to correct $\eta_2$, but this remains to be seen. But even with an 
$\alpha$ of the order of $10^{-6}$ very little rf is needed.

It had been feared that amplitude dependent anisochronisity generated in the 
insertion would cause bunch growth in an otherwise purely isochronous design. 
It has, however, been pointed out\cite{oide} that if chromaticity is 
corrected in the ring, then amplitude dependent anisochronisity is 
automatically removed.} 
 \item{{\bf Low $\beta$ Insertion:} In order to 
obtain the desired luminosity we require a very low beta at the intersection 
point: $\beta^*=3\,{\rm mm}$ for 4 TeV, $\beta^*=8\,{\rm mm}$ for the .5 TeV 
design. An initial final focusing quadruplet design 
used $6.4\,$T
maximum fields at $4\,\sigma.$
This would allow a radiation shield of the order of 5 cm, while 
keeping the peak fields at the conductors less than 10 T, which should be 
possible using ${\rm Nb_3Sn}$ conductor. 
The maximum beta's in both x and y were of the order of 400 
km in the 4 TeV case, and 14 km in the 0.5 TeV machine. The chromaticities 
$(1/4\pi\int \beta dk)$ are approximately 6000 for the 4 TeV case, and 600 for 
the .5 TeV machine. A later design\cite{ref35a} has lowered these chromaticities somewhat, but in either case the chromaticities are too large to correct within the 
rest of a conventional ring and therefore require local correction\cite{ref18}\cite{ref19}. 

It is clear that there is a great advantage in using very powerful final 
focus quadrupoles. The use of niobium tin or even more exotic materials should 
be pursued.}
\item{{\bf Model Designs:} Initially, two lattices were generated\cite{ff}\cite{ref29},\cite{oide1}, one of 
which\cite{oide1}, with the application of octupole and 
decapole correctors, had an adequate calculated dynamic 
aperture. More recently, 
a new lattice and IR section has been generated\cite{ref35a} with 
much more desirable properties than those in the 
previously reported versions. Stronger final focusing quadrupoles were 
employed to reduce the maximum $\beta$'s and chromaticity, the dispersion was 
increased in the chromatic correction regions, and the sextupole strengths 
reduced. It was also discovered that, by adding dipoles near the intersection 
point, the background in the detector could be reduced.\cite{ref35a} 
}
\end{enumerate}
\subsubsection{Instabilities.}
Studies\cite{stability} of the  resistive wall impedance instabilities indicate 
that the required muon bunches (eg. for $2\,$TeV: $\sigma_z=3\ mm,\ N_{\mu}= 
2\times 10^{12}$) would be unstable in a conventional ring. In any case, the 
rf requirements to maintain such bunches would be excessive. 

If one can obtain momentum-compaction factor  $\alpha \leq 10^{-7}$, then the 
synchrotron oscillation period is longer than the effective storage time, and 
the beam dynamics in the collider behave like that in a linear beam 
transport\cite{ngstab}\cite{chengstab}. In this case, beam breakup 
instabilities are the most important collective effects.  Even with an aluminum
beam pipe of  radius  $b=2.5$~cm, the resistive wall effect will cause the tail
amplitude of  the bunch to double in about 500~turns.  For a broad-band
impedance of $Q=1$  and $Z_\parallel/n=1$~Ohm, the doubling time in the same
beam pipe is only  about 130 turns; which is clearly unacceptable. But both
these instabilities can easily be  stabilized using BNS\cite{bns} damping. For
instance, to stabilize the resistive wall instability, the required tune
spread, calculated\cite{ngstab} using the two particle
model approximation, is (for Al pipe)
\begin{equation}
{\Delta \nu_{\beta} \over \nu_{\beta}}=\left\{ \begin{array}{ll}
1.58\,10^{-4} & b=1.0\,{\rm cm}\\
1.07\,10^{-5} & b=2.5\,{\rm cm}\\ 
1.26\,10^{-6} & b=5.0\,{\rm cm}
\end{array}
\right.
\end{equation} 

This application of the BNS damping to a quasi-isochronous ring, where
there  are other head-tail instabilities due to the  chromaticities $\xi$ and $
\eta_1$, needs more careful study. 

If it is not possible to obtain an $\alpha$ less than $10^{-7}$, then 
rf must be introduced and synchrotron oscillations will occur. The above 
instabilities are then somewhat stabilized because of the interchanging of head 
and tail, but the impedance of the rf now adds to the problem and simple 
BNS damping is no longer possible. 

If, for example, a momentum-compaction  factor $|\alpha|\approx1.5\times10^{-
5}$ is obtained, then rf of $\sim 1.5$~GV is  needed which gives a synchrotron 
oscillation period of 150 turns.  Three  different impedance models: 
resonator,  resistive wall, and a  SLAC-like or a  CEBAF-like rf accelerating 
structure have been used in the estimation for  three sets of design 
parameters. The impedance of the ring is dominated by  the rf cavities, and 
the microwave instability is well beyond threshold. Two  approaches are being 
considered to control these instabilities: 1) BNS damping  applied by rf 
quadrupoles as suggested  by Chao\cite{chaobook}; and 2) applying an 
oscillating perturbation on the chromaticity\cite{cheng2}. 

When the ring is nearly isochronous, a longitudinal head-tail (LHT) instability 
may occur because the nonlinear slip factor $\eta_1$ becomes more important 
than the first order $\eta_0$. The growth time for the rf impedance when 
$\eta \simeq 10^{-5}$ is about $0.125\, b\, \eta_0/\eta_1$~s, where $b$ is the 
pipe radius in cm.  This would be longer than the storage time of $\sim 41$~ms 
if $\eta_1 \sim \eta_0$. However, if $\eta_1 \sim \eta_0/\delta$, with 
$\delta \sim 10^{-3}$, then the growth time is about $0.125 b $~ms, which 
is much shorter than the storage time. More study is needed.

\section{BACKGROUND AND DETECTOR}
\vskip -1pc
\subsection{Design of the Intersection Region}
\begin{figure}[htb!]
\centerline{\epsfig{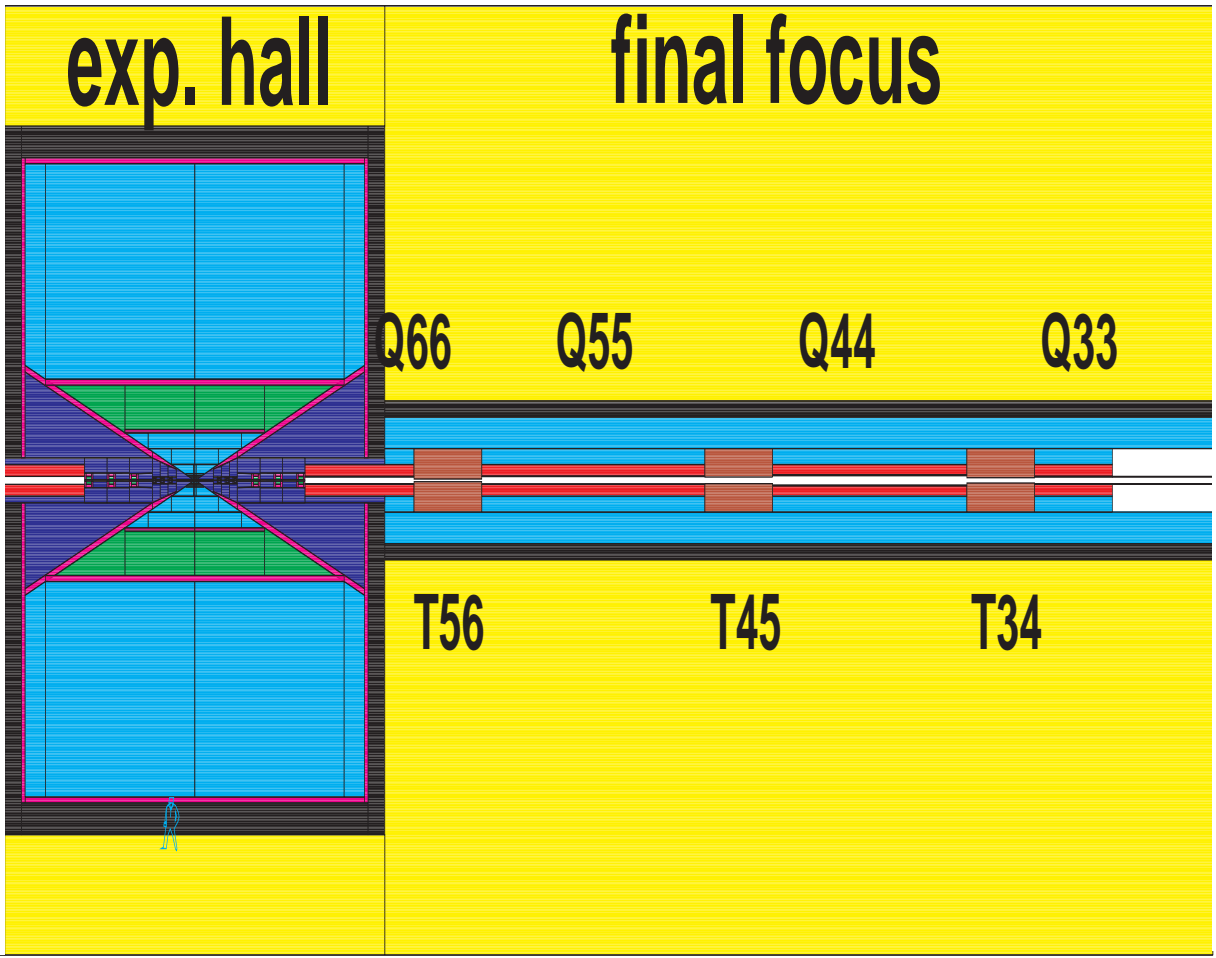}}
\caption[Region around the
 Intersection Region modeled in GEANT. ]
{Region around the
 Intersection Region modeled in GEANT.  The black regions represent
 tungsten shielding.  The final quadrupoles (Q) and toroids
(T) on one
 side of the detector enclosure are shown.  The shaded areas around
 the intersection point represent the various detector volumes used in
 calculating particle fluences.
 \label{fg.iuliu1}}
\end{figure}

\begin{figure}
\Twofigs{\epsfysize=3in\epsfbox{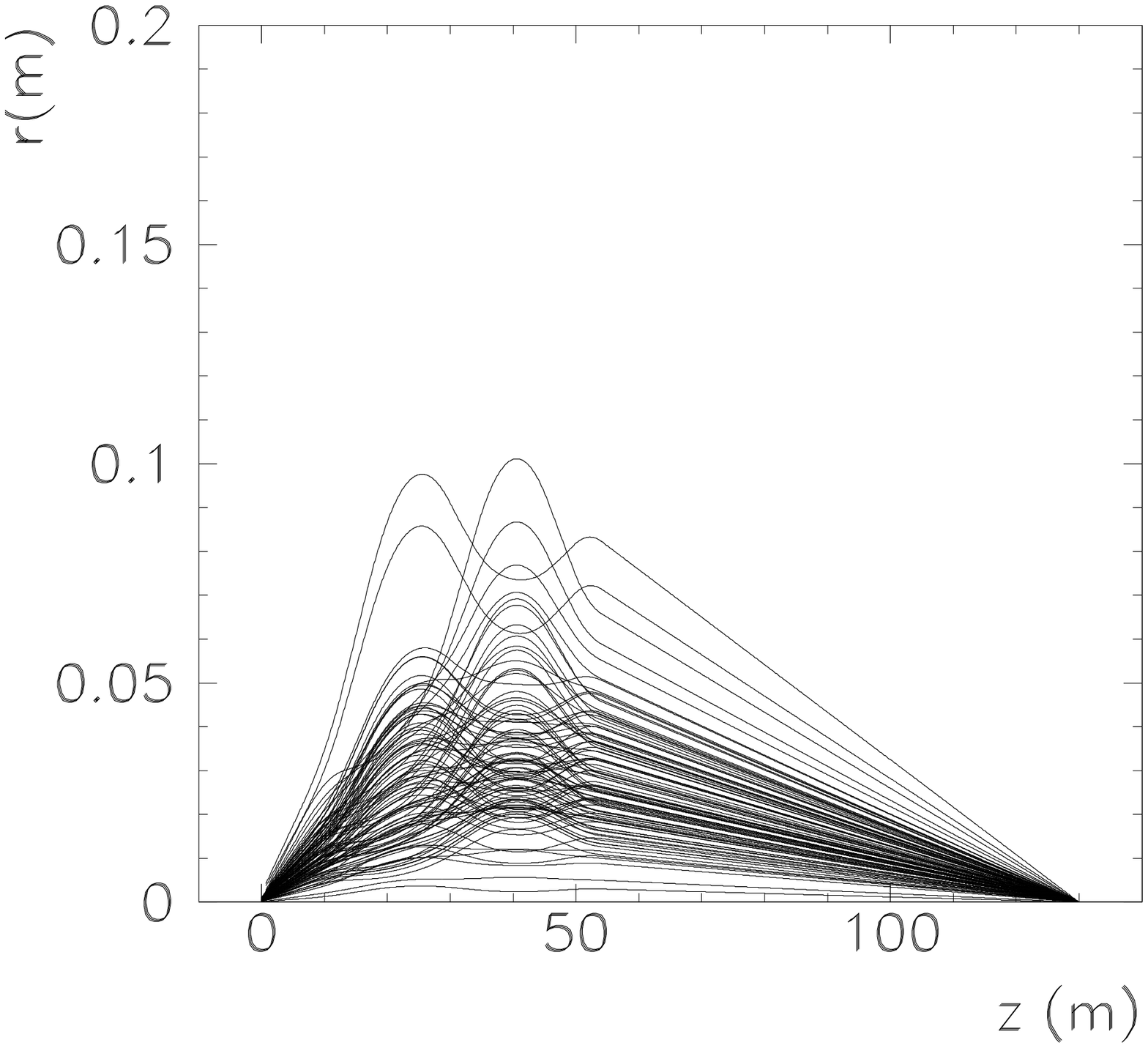}}{\epsfysize=3in\epsfbox{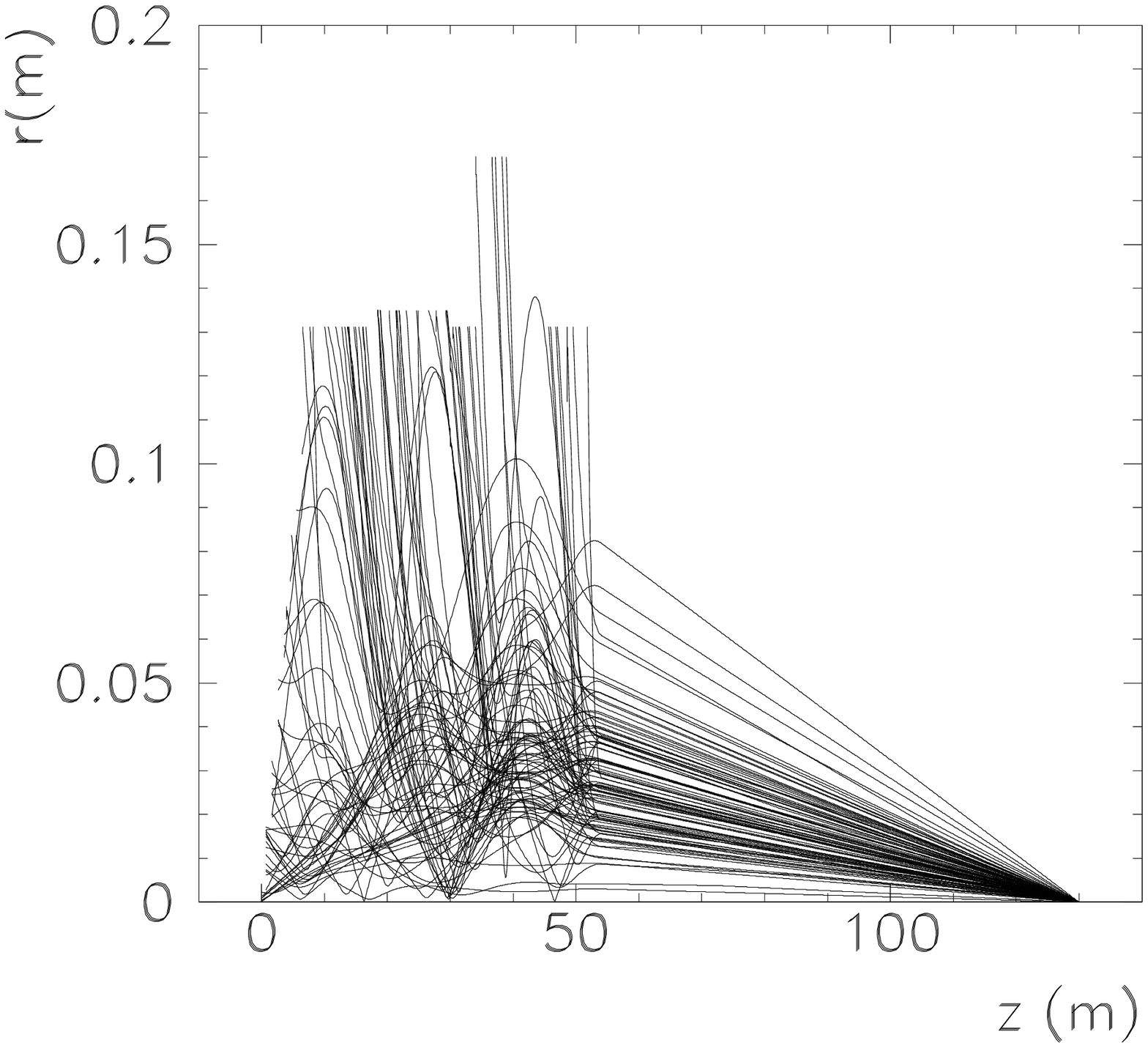}} 
\Lcaption[Trajectories in Final Focus Region.]{Trajectories
in the Final Focus Region with Muon decays turned
off\label{fg.iuliu6}} 
\Rcaption[Muon Decay Trajectories in the Final Focus Region.]{Trajectories in the Final Focus Region with Muon Decays allowed.  The
decay electrons are tracked until they reach either a magnet or
shielding.\label{fg.iuliu7}} 
\end{figure}

The design of the Intersection Region\cite{istumer} is driven by the
    desire to reduce the background from muon decays in the detector
    as much as possible.  For this study a 130 m final focus
    section (\Fig{iuliu1}) which included four final quadrupoles,
    three toroids, a 2 T solenoidal field for the detector and the
    connecting beam pipe and shielding was modeled in GEANT
     with all the appropriate magnetic fields and
    shielding materials. The parameters used were taken from \cite{ff}\cite{ref29}. Trajectories of particles with and without decay are shown later in Figs.\ref{fg.iuliu6} and \ref{fg.iuliu7}. Studies of the effects of high
    energy electrons hitting specific edges and surfaces were carried
    out and the shielding adjusted or
 augmented to mitigate the
    apparent effects of particular background problems.  Effects due to
    electrons, photons, neutrons and charged hadrons and muons were
    considered in turn to try to optimized the design.  While the
    current design is
not fully optimized, it is a marked improvement
    over a much simpler design which had been used in the past.  More
    importantly, it helped develop the tools and strategy to do such an
    optimization as the lattice is further developed. A second study\cite{mokhovov_dete} using a somewhat different final focus design and selecting shielding parameters has given results that are of the same order of magnitude as those that will be discussed in detail here.

\begin{figure}
\Twofigs{\epsfysize=3in\epsfbox{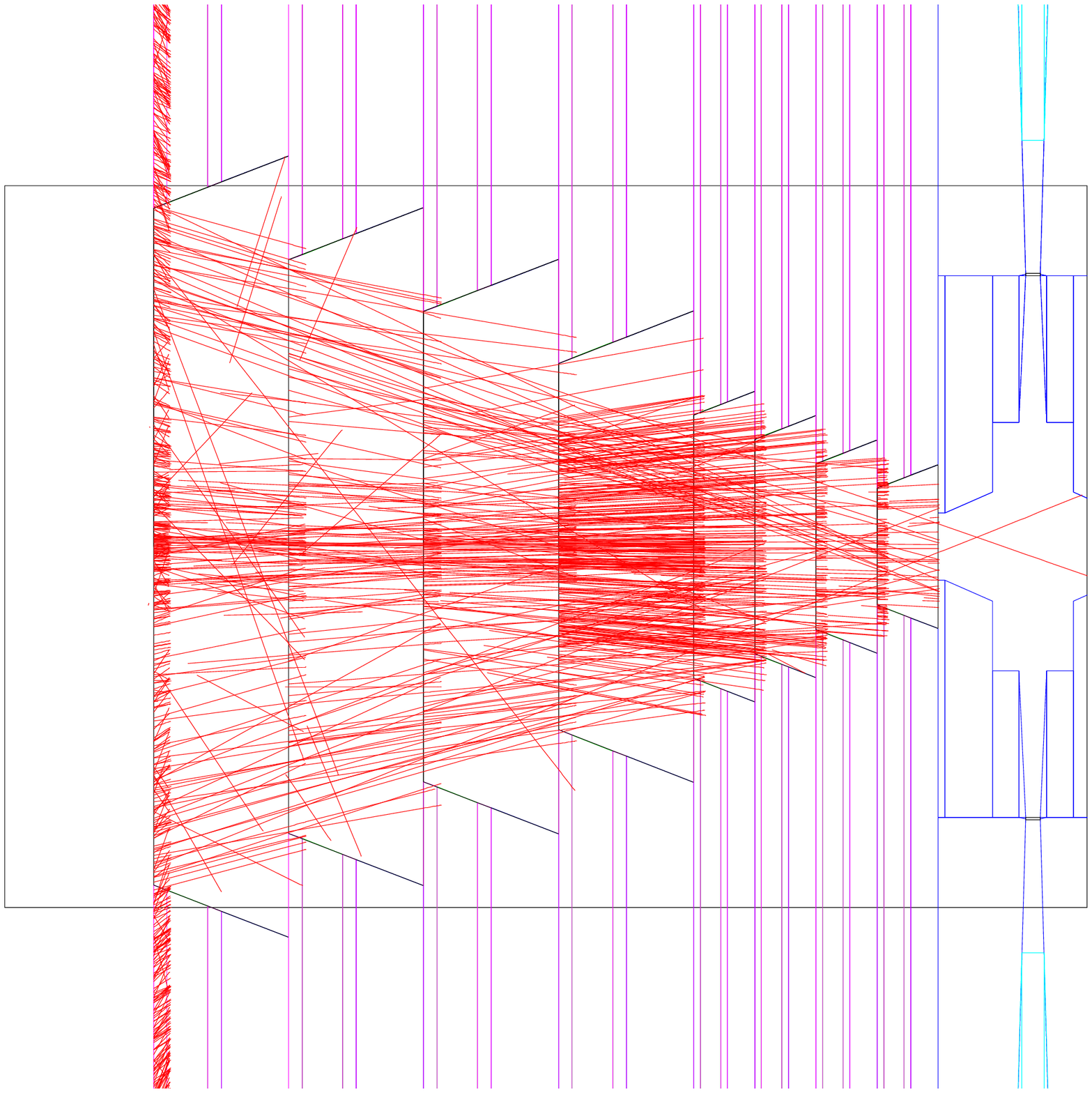}}{\epsfysize=3in\epsfbox{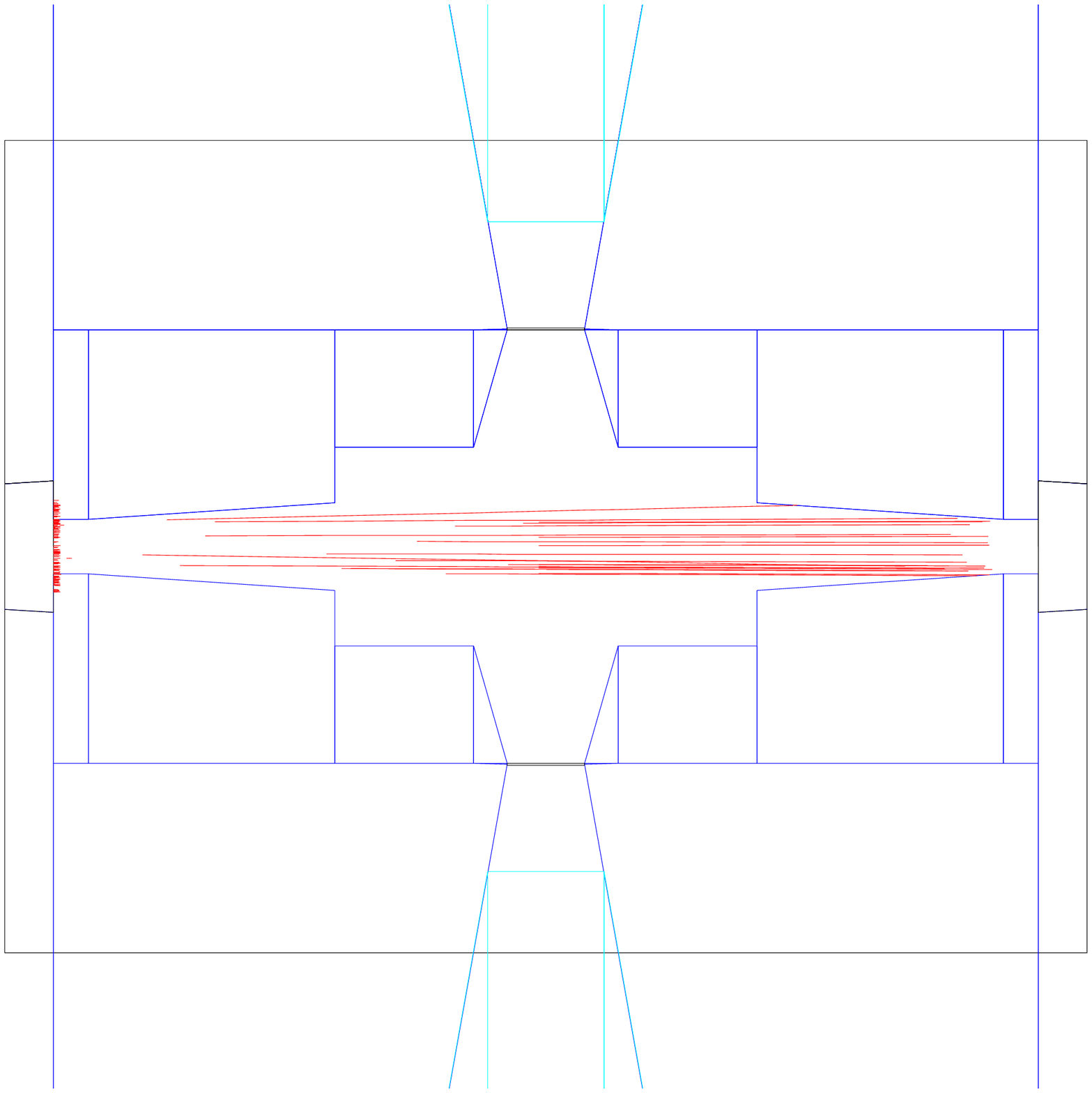}}
\Lcaption[Expanded view near Intersection Region.]{Expanded View of
Region (2) near the Intersection point.  The lines represent electrons
from a random sample of muon decays.\label{fg.iuliu4}} 
\Rcaption[Detailed view of Intersection Region.]{Detailed View of Region
(1), the Intersection Region.  The lines represent electrons from a
random sample of muon decays.\label{fg.iuliu5}}
\end{figure}

    The final focus may be thought to be composed of 3 separate
    regions.  The longest of these, from 130 m to approximately 6.5 m
    contains the quadrupole magnets which bring the beam to the final
    focus in the intersection region.  The space available between the
    four quadrupoles was used to install toroids. They fulfill a
    double role: first they are used as scrapers for the
    electromagnetic debris; secondly, they serve as magnetic
    deflectors for the Bethe--Heitler(BH) muons generated upstream.  The
    effect of the toroids on the BH muons will be discussed later.  In
    order to optimize the inner aperture of the toroids, the
   $\sigma_x$ and $\sigma_y$ envelope of the muon bunch at every exit
    of the quadrupoles has been estimated.  The inner aperture of each
    toroid was chosen to match the 4 $\sigma$ ellipse of the muon
    bunch at that point.  
    The second region, from 6.5 m to 1.1 m
    contains tungsten plus additional shielding boxes to help contain
    neutrons produced by photons in the electromagnetic showers.
    A shielding box consists of a block of Cu
    surrounded by polyboron.  The shielding here is designed with
    inverted cones to reduce the probability of electrons hitting the
    edges of collimators or glancing off shielding surfaces
    (\Fig{iuliu4}).  The beam aperture at the entrance to this section
    is reduced to 2.5 cm and by the exit of the section to 4.5 mm.
    This profile follows the beam envelope as the particles approach
    the intersection region.  The intersection region itself
    (\Fig{iuliu5}) is designed as an inverse cone to prevent electrons
    which reach this region from hitting any shielding as this region
    is directly viewed by the detector.  
    A $20^\circ$ tungsten cone around the intersection region is
    required for the reduction of the electromagnetic component of the
    background.  The cone is lined, except very near the
    intersection region with polyboron to reduce the slow neutron
    flux. In the shielding calculations it is also assumed that there
    is a polyboron layer before the calorimeter and surrounding the
    muon system.  In earlier designs this cone was only $9^\circ$.
    Whether or not the full $20^\circ$ is required is still under
    study and work is ongoing to evaluate the physics impact of this
    choice of the shielding cone angle.  It is likely that, after
    optimization is completed, the cone angle will be reduced.

\subsubsection{Muon Decay Backgrounds.}

\subsubsection{Results using GEANT Simulation.}
 The backgrounds in the detector are defined as the fluence of
    particles (number of particles per cm$^{2}$ per beam crossing)
    across surfaces which are representative of the various kinds of
    detectors which might be considered. For this study the
    calorimeter was assumed to be a composition of copper and liquid
    argon in equal parts by volume which represents a good resolution
    calorimeter with approximately 20\% sampling fraction.  The other
    volumes of the detector were vacuum.  The calorimeter starts from
    a radius of 150 cm and is 150 cm deep.  The tracker volume is
    defined from 50 to 150 cm .  An array of horizontal and vertical
    planes were placed in the detector volumes. These planes
    were used for flux calculations; their positions are evident in the
    tables of rates below.

\begin{table}[ht!]
\centering
\caption[Longitudinal Particle Fluences from Muon Decays and Interactions
from the GEANT Calculation. ]
{Longitudinal Particle Fluences from Muon Decays and Interactions
from the GEANT Calculation.
Fluence = particles/cm$^2$/crossing for two bunches of 2 10$^{12}\mu$'s each.}
\label{fluencesL}
\begin{tabular}{lccccccc} \hline
Detector&Radius(cm)&$\gamma$'s& neutrons&e$^{\pm}$&$\pi^{\pm}$&protons&$\mu^{\pm}$ \\ \hline
Vertex&5-10&15800&2200&138&
28.8&1.6&3.0 \\
&10-15&6200&2400&&7.4&0.1&1.0 \\
&15-20&3200&2000&&9.2&8.0&4.6 \\
Tracker&20-50&900&1740&&1.6&7.8&0.6 \\
&50-100&240&1040&&0.2&4.4&0.12\\
&100-150&260&660&&0.006&0.8&0.02 \\
Calorimeter& 160-310&&&&&&0.004 \\
Muon&310-10000&&&&&&0.0004 \\ \hline
\end{tabular}
\end{table}

\begin{table}[ht!]
\centering
\caption[Radial Particle Fluences from the GEANT Calculation.]{Radial Particle Fluences from Muon Decays and Interactions
from the GEANT Calculation. 
Fluence = particles/cm$^2$/crossing for two bunches of 2 10$^{12}\mu$'s each.}
\label{fluencesR}
\begin{tabular}{lccccccc} \hline
Detector&Radius(cm)&$\gamma$'s&
neutrons&e$^{\pm}$&$\pi^{\pm}$&protons&$\mu^{\pm}$ \\ \hline
Vertex&5&34000&3200&168.0&19&3.4&.7 \\
&10&9600&3400&19 &9&2.8&0.86 \\
&15&4400&3400&4.2&4.2&2.2&0.66 \\
&20&2500&3400&&2.6&3.8&0.40 \\
Tracker&50&880&3000&&0.44&8.4&0.064\\
&100&320&720&&0.08&1.6&0.016 \\ \hline
\end{tabular}
\end{table}

\begin{table}[hbt!]
\centering
\caption[Mean kinetic energies and momenta of part
icles as calculated by GEANT.]{Mean kinetic
energies and momenta of particles as calculated by GEANT.}
\label{mean_E}


\begin{tabular}{ccccccc} \hline
Detector  &Radius &$\mu$ &$\gamma$ &$p$ & $\pi^{\pm}$ &$n$ \\
	  &       & GeV  & MeV & MeV& MeV& MeV\\ \hline
Vertex&10-20&24&1 &30 &240&10  \\ \hline
Tracker&50-100&66&" & " & " & "\\
&100-150&31&" & " & " & "\\ \hline
Calorimeter&160-310&19&" & " & " & "\\ \hline
\end{tabular}

\end{table}
\subsection{ Detector Specifications and Design}
The physics requirements of a muon collider detector are similar to
those
of an electron collider. The main difference has to do with the
machine related backgrounds and the added shielding that is needed
near the beam pipe.

At this time little detailed work has been done on the design of a complete 
detector.  Most of the discussion has centered around the types of
detector elements which might function well in this environment.  The
background levels detailed in the previous section are much higher
than the comparable levels calculated for the SSC detectors and appear
to be somewhat in excess of the levels expected at the LHC. Clearly
segmentation is the key to successfully dealing with this environment.
One major advantage of this muon collider over high energy hadron
colliders is the long time between beam crossings; the LHC will have
crossings every 25 ns compared to the 10 $\mu$s expected for the
4 TeV $\mu$-collider.  Much of the detector discussion has focused on
ways to exploit this time between crossings to increase the
segmentation while holding the number of readout elements to manageable
levels.

The real impact of the backgrounds will be felt in the inner tracking
and vertex systems.  One attractive possibility for a tracking system is
a Time Projection Chamber (TPC)\cite{Nygren}.  This is an example of a low
density, high precision device which takes advantage of the long time
between crossings to provide low background and high segmentation with
credible readout capability.

Silicon, if it can withstand damage from the neutron fluxes, 
appears to be an adequate option for vertex detection.  Again,
because of the time between beam crossings, an attractive option 
is the Silicon Drift Detector\cite{Gatti1984}. 
Short drift TPC's with microstrip\cite{unknown} readout could also be considered 
for vertex detection.

An interesting question which has yet to be addressed is whether or
not it is possible
to tag high energy muons which penetrate the
tungsten shielding which, in the present design, extends to $20^\circ$
from the beam axis.  For example, in the case of
$\mu\mu\rightarrow\nu\nu W^+ W^-$ the primary physics background is due
to $\mu\mu\rightarrow\mu\mu W^+ W^-$. To reduce the background, in
addition to a high $p_T$ cut on the WW pair, in might be advantageous
to tag forward going muons.  These $\mu$'s would penetrate the
shielding.
\subsection{Strawman Detector}

\begin{table}[htb!]
\centering
\caption[Detector Performance Requirements.]{Detector Performance Requirements.}
\label{detper}
\begin{tabular}{ll}
\hline
Detector Component & Minimum Resolution/Characteristics \\ \hline
Magnetic Field&Solenoid; B$\geq$2 T\\ \hline \hline
Vertex Detector&b-tagging, small pixels\\ \hline
Tracking&$\Delta$p/p$^2\sim 1\cross 10^{-3}$(GeV)$^{-1}$ at large p\\
&High granularity\\ \hline
EM Calorimeter & $\Delta$E/E$\sim 10\%/\sqrt{\rm E}\oplus0.7\%$\\
&Granularity: longitudinal and transverse \\
&Active depth: 24 X$_0$\\ \hline
Hadron Calorimeter&$\Delta$ E/E$\sim 50\%/\sqrt{\rm E}\oplus2$\%\\
&Granularity: longitudinal and
 transverse\\
&Total depth (EM + HAD)$\sim 7 \lambda$\\ \hline
Muon Spectrometer&$\Delta$p/p $\sim 20\%$\@ 1 TeV\\ \hline
\end{tabular}
\end{table}

The detector performance criteria that are used
for the design of the detector are summarized in Tb.~\ref{detper}.
 The object of this present exercise is to see if
 a
detector can be built using state-of-the-art
(or not far beyond) technology to satisfy the physics needs of the
muon collider.

\begin{figure}[tbh!]
\centerline{\epsfig{file=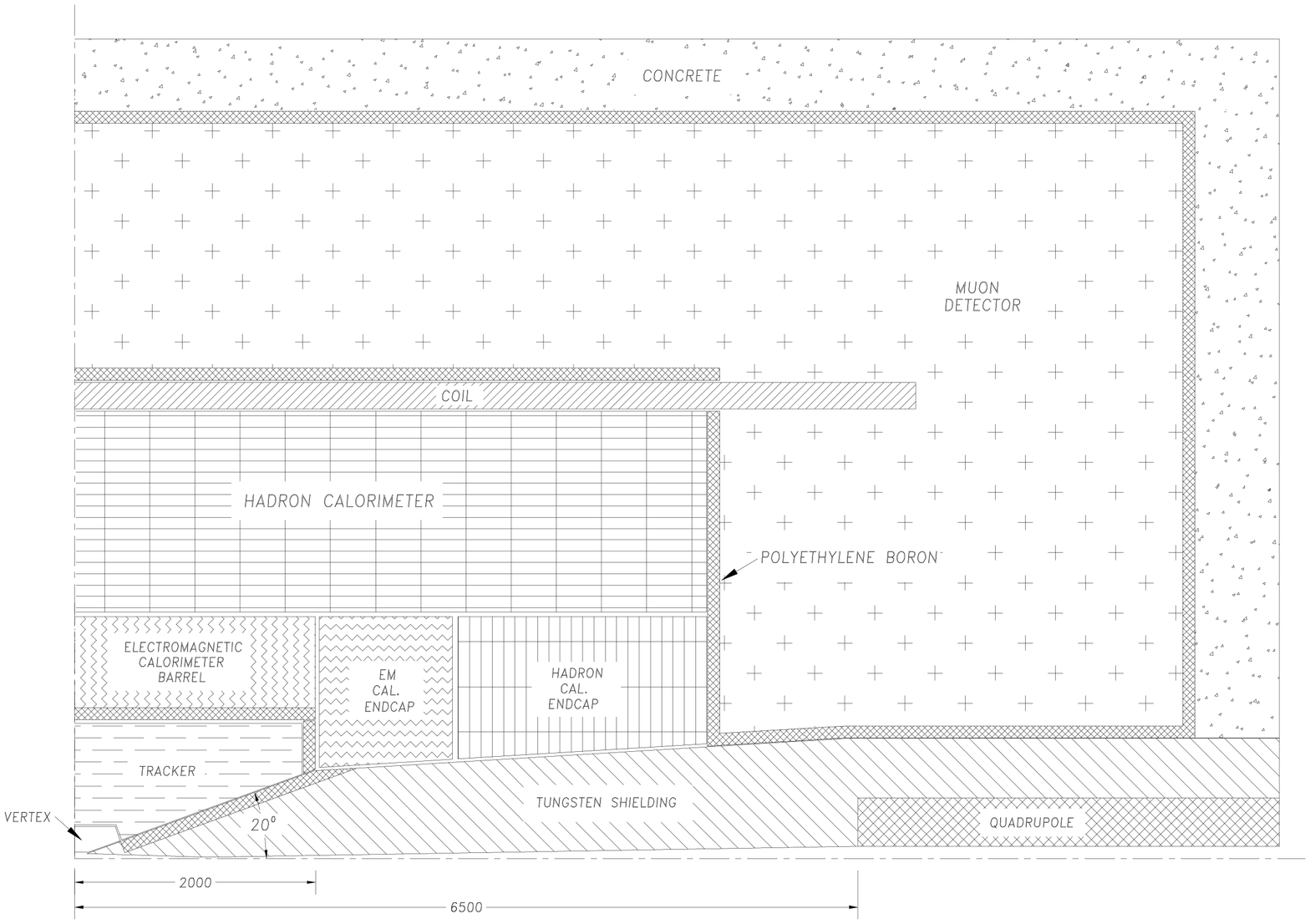,width=5in,height=5in,angle=90}}
\caption[Strawman Detector.]{Strawman Detector\label{fg.detector}}
\end{figure}

A layout of the strawman detector is shown in \Fig{detector}.   A large cone ($20^\circ$) that is probably not
instrumented and is used to shield the detector from the machine
induced background. 

The main features of the detector are:The element nearest to the
intersection region is the vertex detector located at as small a
radius as possible.  A number of technologies including Silicon Drift
Detectors(SDD), Silicon Pixels\cite{ATLAS TDR}, and CCD detectors
have been considered, as well as short drift TPC's.  
SDD seem especially attractive
because of the reduced number of readout channels and potentially
easier construction. A micro TPC would have lower occupancy and greater 
radiation resistance.  Tracking technologies considered were
cathode pad chambers, silicon strips and TPCs. The use of a TPC is
interesting as the amount of material is minimized and thus the
detector
does not suffer as much from low energy photon and neutron
backgrounds.

For the calorimeter system there are many options.  A liquid argon
accordion calorimeter for the EM part\cite{ATLAS TDR},\cite{GEM TDR}
and a scintillator hadronic calorimeter appear appropriate. This
combination gives a hadronic resolution that is of order 50\%/$\sqrt E$ 
which may be good enough. The high granularity of the EM section
allows good electron identification which will be of help in tagging
b-jets. In addition the longitudinal
 and transverse granularity allow
for corrections on an event by event basis to compensate for the fact
that the calorimeter is non-compensating as well as to allow the
identification of catastrophic muon bremsstrahlung.

There is a single solenoid magnet, with a
 field of 2 T in the tracking
region.  The magnet is located behind the hadron calorimeter.

The muon system is a stand-alone system.
The chambers in the muon system are Cathode Strip Chambers(CSC) that
can be used for both a two dimensional readout
as well as a
trigger. These chambers have good timing resolution and relatively
short drift time which minimizes neutron background problems. We now consider these components in detail:
\subsection{Silicon Drift Vertex Detector}
The best silicon detectors are capable of achieving a position resolution
of 4 $\mu$m with a detector 300 $\mu$m thick. However,
these results were obtained only
with  normally incident  charged particles. For other angles of incidence,
the resolution can be degraded by the fluctuations in the density
of the ionization (Landau) in the silicon.
The presence of a magnetic field modifies the trajectories of electrons
in a \sdc\ and normal incidence is no longer the ideal one. There is, however,
a proper incidence angle which does not give any degradation of the
position resolution.

	The granularity of the proposed detector seems be reasonable
for the rates of background particles. Based on rates in
 Tbs.~\ref{fluencesL},\ref{fluencesR},\ref{mean_E},  
 a layer located 10 cm from the beam would have
22 hits per
$cm^2$ from the interaction of neutral particles and 32 from
crossings of charged particles. 
For a pixel size of $316\times 316~\mu m^2$ the number of pixel per $cm^2$ is 1000.  In this case the
occupancy of background hits is less than 6\%. 

	The damage due to the radiation dose may be a serious
problem. 
Only about 1/3 of the neutrons have 
energies above 
100 KeV and contribute to this damage, but the 
integrated flux per 
year for this example would still be 
${\rm 3\ 10^{14}\ n's\ cm^{-2}year^{-1}}$.
If detectors are produced from an n--type silicon with a
bulk doping level of $1.5\times 10^{12}/cm^3$ the detectors would have to be
replaced after a year of operation. The use of p--type material seems
to be more appropriate for this application. P--type silicon drift
detectors are being developed in LBNL. These detectors are supposed to
be much more
radiation resistant. Some R\&D may be required.
\subsection{Time Projection Chamber (TPC)}
An interesting candidate for tracking at a muon collider is a Time
Projection Chamber (TPC).  This device has good track reconstruction
capabilities in a low density environment, good 3-dimensional imaging
and provides excellent momentum resolution and track pointing to the
vertex region.  It is perhaps particularly well suited to this
environment as the long time between bunch crossings ($\sim 10~\mu$s)
permits drifts of $\sim$1 m and the average density of the device is
low compared to more conventional trackers which helps to reduce the
measured background rates in the device.  In the present detector
considerations the TPC would occupy the region between the conic
tungsten absorber and electromagnetic
calorimeter in the region from 35 cm to 120 cm, divided into two parts, 
each 1 m long.

To reduce background
gamma and neutron interactions in the detector volume, a low density
gas mixture should be chosen as the detection medium of the
TPC. Another important parameter is the electron drift velocity. Since
the time between beam crossings is fixed (10 $\mu$s in the present
design) the drift velocity should be high enough to collect all the
ionization deposited in the drift region.
Finally the detection medium should not contain low atomic number gases
to help reduce the transfer energy to the recoil nucleus and in this way to
reduce its range in the gas. The gas mixture 90\% He + 10\% CF$_4$
satisfies all these
requirements and  could be an excellent  candidate. It does not
contain hydrogen which would cause a deleterious effect from the
neutrons, has a density $1.2$ mg/cm$^3$ and a drift velocity of
9.4 cm/$\mu$s. The single electron longitudinal diffusion for this gas is
\begin{equation}
\sigma_{l} = 0.15\,{\rm mm}/\sqrt{\rm cm}.
\end{equation}
 The transverse diffusion, which is
strongly suppressed by the 2 T magnetic field  is given by,

\begin{equation}
\sigma_{t}=\frac{\sigma_{t}(B=0)}{\sqrt{1+(\omega\tau)^2}}=0.03\,{ \rm
mm}/{\sqrt{\rm cm}}
\end{equation}

Each time slice will contain about 25 ionization electrons, and the
expected precision in r- $\phi$ and z coordinates is,

\begin{equation}
\sigma_{\phi} = \sqrt{\frac{Z\sigma_{t}^2}{25}+(50)^2}\approx 100\, ({\rm \mu m})
\qquad
\sigma_{Z} = \sqrt{\frac{Z\sigma_{l}^2}{25}+(150)^2}\approx 300\,({\rm \mu m})
\end{equation}
   where    Z, the drift length is 1 m.
The precision of r-coordinate is defined by the anode wire pitch - 3 mm.

\subsubsection{Occupancy from Photons.}

Low energy photons, neutrons and charged particles produce the main
backgrounds in the tracker.  Photons in the MeV region interact
with matter mainly by Compton scattering. For a 1 MeV photon the
probability of producing a Compton electron in 1 cm of gas is
$\xi_{\gamma}=4.5\times 10^{-5}$.  For an average photon fluence
$h_{\gamma} = 200$ cm$^{-2}$ about $N_{\gamma} = 8\times 10^4$ electron
tracks are created in the chamber volume.  Because the transverse
momentum of Compton electrons is rather small the electrons are
strongly curled by
 the magnetic field and move along the magnetic
field lines.  Most of the electrons have a radius less than one
millimeter and their projection on the readout plane covers not more
than one readout pitch, $0.3\times 0.4$ cm$^2$.  The average length of
the
 Compton electron tracks in the TPC is 0.5 meter and therefore, the
volume occupied by electron tracks is $v_{comp.e}=4.8\times 10^5$
cm$^3$.  Since the total chamber volume is $10^{7}$ cm$^{3}$, the
average occupancy due to background photon interactions
is equal to,
\begin{equation}
\  <occupancy>_{\gamma} = \frac{V_{comp.e}}{V_{total}} = 4.4\times 10^{-2}
\end{equation}
and could be further reduced by subdividing the chamber in length, thus
shortening the drift distances. Indeed this may be required to reduce 
an excessive space charge distortion from the accumulation of ions.

\begin{figure}[hbt!]
\centerline{\epsfig{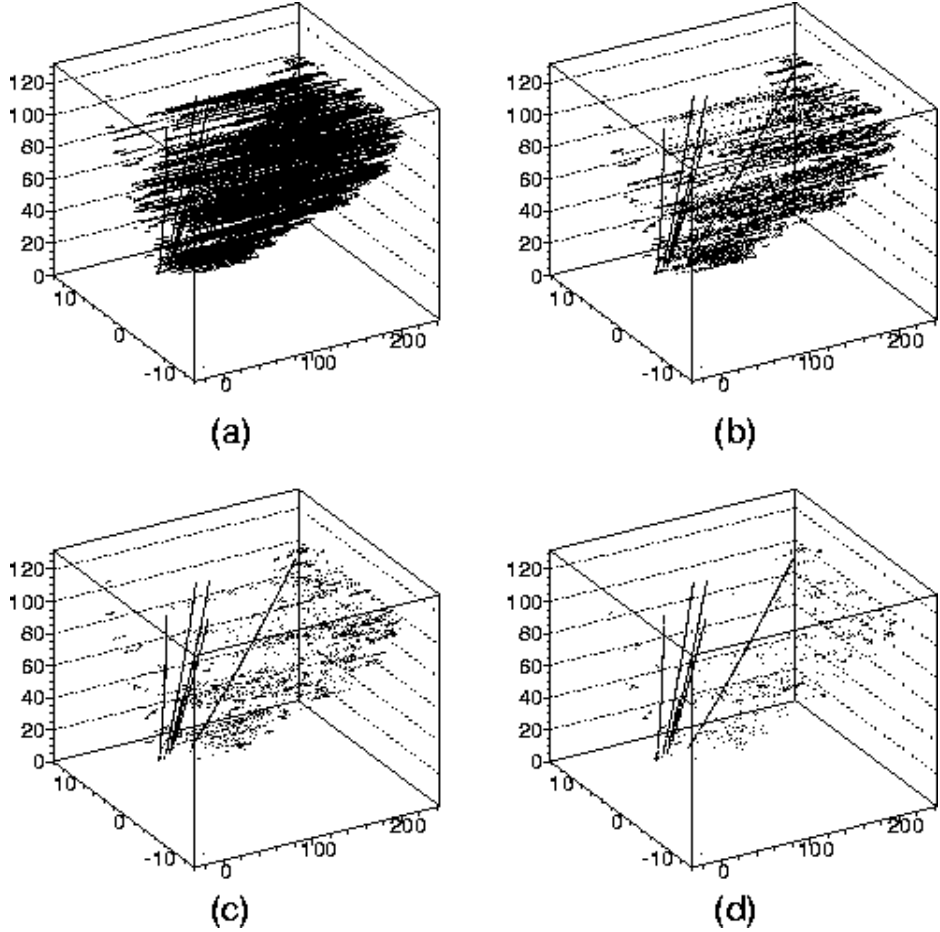}}
\caption[TPC signals with threshold cuts]{Charge distribution from a
signal event and background Compton electrons with threshold
cuts; (a) has no cut applied while (d) has the most stringent cut.\label{fg.tchern3}}
\end{figure}

These Compton tracks can easily be identified and removed.
Because almost all points of a Compton track
lie along the $z$-axis most of them will be projected into one cell and
therefore the
 number of points in this cell will be very different from hit
cells from non-background tracks. 
To remove low momentum electron tracks, all cells containing more than
some threshold number of points should be excluded. Applying this
procedure a few percent of volume is lost but the quality of the high
momentum tracks is not substantially changed.  This is
illustrated in \Fig{tchern3} where one sector of the TPC is shown
after the application of different value threshold cuts.
\subsubsection{Occupancy from Neutrons.}
For neutrons in the MeV region the primary interaction with matter is
elastic collisions. 
In this case the energy transfer to the nucleus has a flat distribution and the maximum transfer energy is given by
$4E_nA/(A+1)^2$ or $4E_n/A$ when $A\gg1$.  

The calculated mean energy of background neutrons is $E_n=27$
$MeV$. In this case, for hydrogen, 
their mean range in the gas is
several meters, but for the gas
chosen, the mean length of the recoil nucleus tracks will only be a
few millimeters.  

The calculated neutron fluence is $<n> = 2\times10^3$
${cm^{-2}}$.  The track of the recoil nucleus
occupies, typically, not more than one volume cell of the TPC,
$v_{n}=0.3\cross0.4\times 1.0$ cm$^3$. The probability of a background
neutron interacting in 1 cm of the gas is $\xi_{n}=2\times 10^{-5}$, the
number of recoil tracks $N_{n}=<n>\cdot\xi_{n}\cdot V_{total}=4\times
10^5$ and therefore the
neutron occupancy is,
\begin{equation}
\   <occupancy>_{n} = \frac{N_{n}\cdot v_{n}}{V_{total}} = 0.48\times 10^{-2}
\end{equation}

It is easy to clean out these recoil tracks owing to their large
ionization density per cell. Only a simple cut to remove
all volume cells which contain a charge in excess of some preset
threshold is required. This cut will only eliminate about 1\% of the
TPC volume.

\subsection{Micro TPC for vertex detection}

TPCs with very short drift distances (3 cm) and microstrip readout (0.2 x 2 mm
pads) might be an interesting alternative to silicon drift chambers for 
vertex detection. The resolution of such chambers would be somewhat worse
(of the order of $30\,\mu$m, compared with $4\,\mu$m), but the greater number of points
would compensate for this to some extent. Such chambers would have less 
occupancy and much greater radiation resistance than silicon devices.

\subsection{Electromagnetic Calorimeter}

	An accordion liquid argon calorimeter is being developed for the
ATLAS collaboration\cite{ATLAS TDR}.  A similar calorimeter designed
for the GEM Collaboration at the SSC.

From the GEANT background calculations, the total energy deposited from the
electromagnetic debris is $\sim$ 13 TeV but relatively uniformly distributed.   If one divides the
calorimeter into $\sim 2\cross 10^5$ cells, the mean energy would be about 65
MeV/cell.  Certainly, energetic electromagnetic showers from $\gamma$'s or
electrons or the core of jets will stand out above this uniform noise.  Since
the readout is every 10\ $\mu$s, multiplexing is possible to reduce costs
compared to the LHC where collisions occur every 25 ns.
\subsection{Hadron Calorimeter}

	A good choice for the hadron calorimeter is a scintillator
tile device being designed for ATLAS\cite{ATLAS TDR}.  It uses a novel
approach where the tiles are arranged perpendicular to the beam
direction to allow easy coupling to wave-length shifting
fibers\cite{Guildermeister}.
With a tile calorimeter of
the type discussed here it should be possible to achieve a resolution
of $\Delta$ E/E$\sim 50\%/\sqrt{\rm E},$ satisfying the requirements in Tb.~\ref{detper}.

From the GEANT background calculations, the total energy
 deposited in
the calorimeter from electromagnetic and hadronic showers and muons is
about 200 TeV.  Again, this is rather uniform with 
and if subdivided into $10^5$ towers would 
introduce 2 GeV pedestals with 300 
MeV 
fluctuations: also acceptable. But the muons, arising 
from 
Bethe-Heitler pair production in EM showers or from a 
halo in the 
machine, though modest in number, have high average 
energies. They 
would not be a problem in the tracking detectors, but in 
the 
calorimeters, they would occasionally induce deeply 
inelastic 
interactions, depositing clumps of energy deep in the 
absorbers. If 
a calorimeter is not able to recognize the direction of 
such 
interactions (they will be pointing along the beam axis) 
then they 
would produce unacceptable fluctuations in hadron energy 
determination. Segmenting the 
calorimetry in depth should allow these interactions to 
be 
subtracted.  We are studying various solutions, including the use of fast time digitizing\cite{jsand} to provide such segmentation, but 
ultimately 
there will have to be some hardware tests to verify the 
MC study.

\subsection{Muon Spectrometer}

Triggering is probably the most difficult aspect of muon spectrometers
in large, 4$\pi$ detectors in both lepton and hadron colliders. In
addition, a muon system should be able to cope with the larger than
usual muon backgrounds that would be encountered in a muon
collider. Segmentation is, again, the key to handling these high
background rates.  Cathode Strip Chambers (CSC) are an example of a
detector that could be used in the muon system of a muon collider
experiment.
This detector performs all
functions necessary for a muon system:
\begin{itemize}
\item Precision coordinate(50 to 70 $\mu $m)
\item Transverse coordinate(of order mm or coarser as needed)
\item Timing (to a few ns)
\item Trigger primitives
\end{itemize}
In addition, the cathodes can be lithographically segmented almost
arbitrarily resulting in pixel detectors the size of which is limited
only by the density and signal routing of the readout electronics.
\subsection{Halo Background}
   There could be a very serious background from the 
presence of even 
a very small halo of near full energy muons in the 
circulating 
beam. The beam will need careful preparation 
before 
injection into the collider, and a collimation system 
will have to be 
designed to be located on the opposite side of the ring 
from the 
detector. 
\subsection{Pair Production}
\begin{figure}[hbt!]
\centerline{\epsfig{file=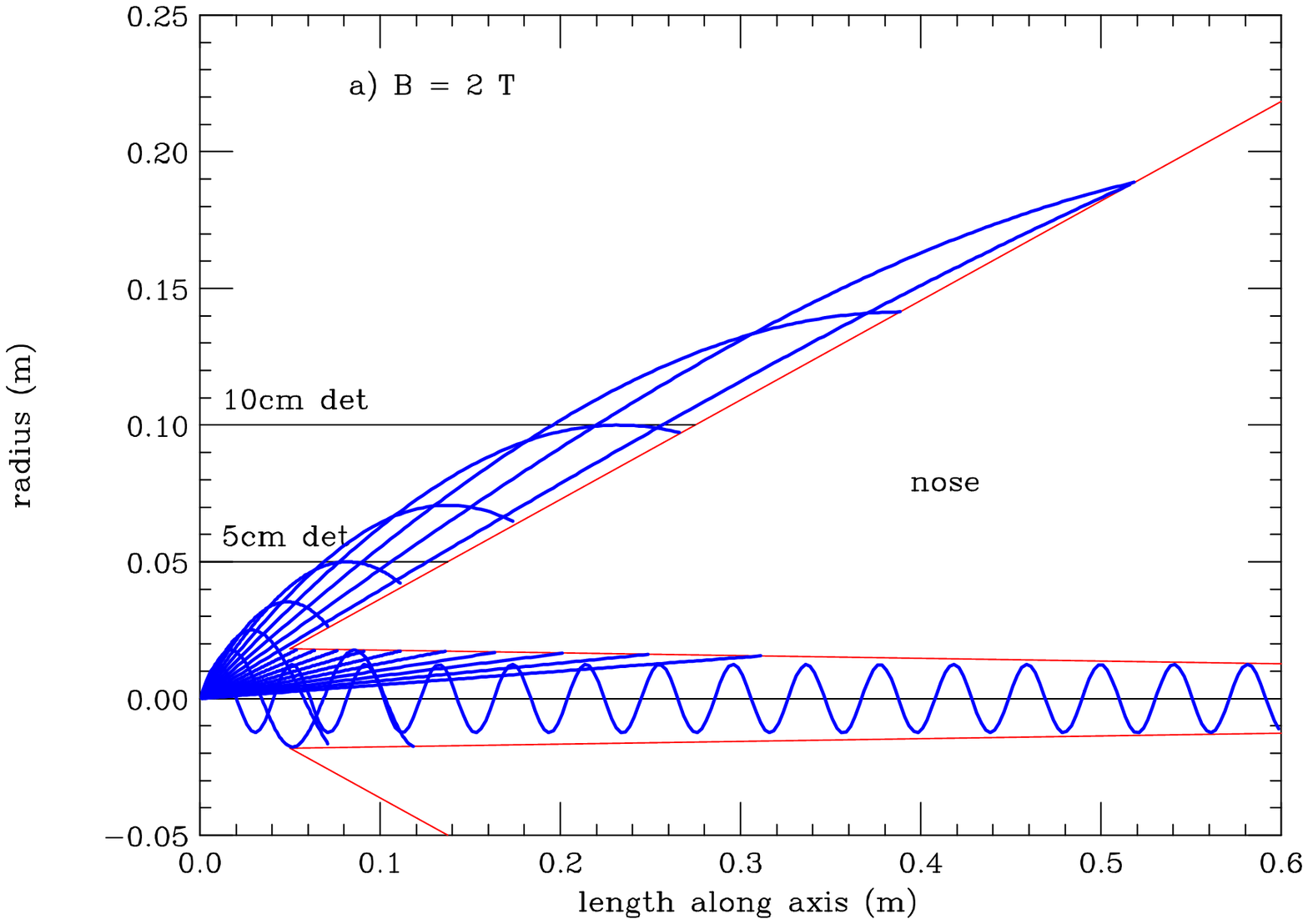,width=3in,height=3in}}
\centerline{\epsfig{file=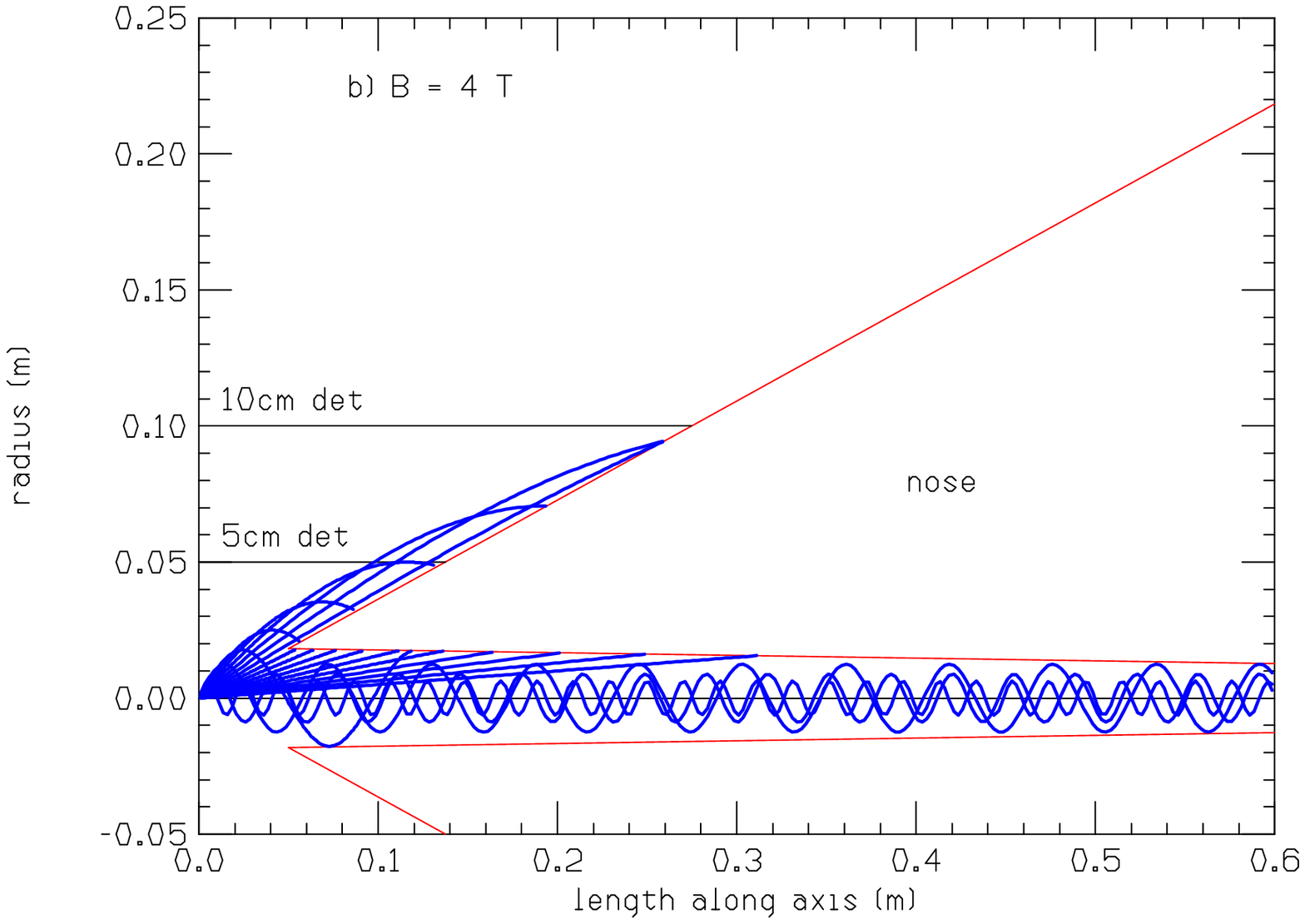,width=3in,height=3in}}
\caption[Radius vs. length of electron pair tracks ]
{Radius vs. length of electron pair tracks for
initial momenta from 3.8 to 3000 MeV 
in geometric steps of $\sqrt{2};$ (a) for
a solenoid field of 2 T, (b) for 4 T.
\label{pair}}
\end{figure}
     
   There is also a background from incoherent 
(i.e. \mumu 
$\rightarrow$ \ee) pair production in the 4 TeV Collider 
case\cite{ref37}. The cross section is estimated to be 
$10\,$mb\cite{ginzburg}, which would give rise to a 
background of 
$\approx 3\times 10^4$  electron pairs per bunch crossing. 
The electrons at production, do not have significant 
transverse momentum, but the
fields of the oncoming 3 $\mu$m bunch can deflect them towards the detector.
A simple program was written to track electrons from close to the axis 
(the worst case) as they are deflected away from the bunch center. 
Once clear of the opposing bunch, the tracks spiral under the 
influence of the experimental solenoid field.
Fig.~\ref{pair} (a) shows the radii vs, length of these electron tracks for
initial momenta from 3.8 to 3000 MeV in geometric steps of $\sqrt{2}$ and
a solenoid field of 2 T, see Fig.\ref{pair}a ( Fig.~\ref{pair}b it is 4 T). In the 
2 T case, tracks with initial energy below 30 MeV do not make it out 
to a detector at 10 cm, while those above 100 MeV have too small an 
initial angle and remain within the shield. Approximately $10\ \% (3000)$ of 
these are in this energy range and pass through a detector at 10 cm. The track fluence at the ends of the detector are less 
than 10 tracks per cm$^2$ which should not present a serious problem. At 5 cm, there are 4500 tracks giving a fluence of 30 per cm$^2$, which is also probably acceptable. If 
the detector solenoid field is raised to 4 T then no electrons reach 10 cm and the flux at 5 cm is reduced by a factor of 2.

   There remains some question about the coherent pair production generated by 
virtual photons interacting with  the coherent electromagnetic fields of 
the entire oncoming bunch. A simple Weizs\"acker-Williams 
calculation\cite{ref37} yields a background that  would 
consume the entire beam at a rate comparable with its decay. However, I. 
Ginzburg\cite{ginzburg} and others have argued that the integration must be 
cut off due to the finite size of the final electrons. If this is true, then 
the background becomes negligible.  A more detailed study of this problem is 
now underway\cite{chen}\cite{pisin}. 

   If the coherent pair production problem is confirmed, then there are two 
possible solutions: 

   1) one could design a two ring, four beam machine (a $\mu^+$ and a
$\mu^-$ bunch coming from each side of the collision region, at the same
time). In this case the coherent electromagnetic fields at the intersection
are canceled and the pair production becomes negligible.

   2) plasma could be introduced at the intersection point to cancel the beam
electromagnetic fields\cite{plasma}.

\subsection{Detector and Background Conclusions.}
	Two independent background calculations have been used for a
preliminary study of the expected background level at a 4 TeV muon
collider.  The optimization of the intersection region is still at its
infancy, but the results of both studies show that the level of
background while still large, can be managed with proper design of the
intersection region and choice of detector technologies.  This is in
large
 part due to the fact that the background is composed of many
very soft particles which behave like a pedestal shift in the
calorimeter.  The tracking and vertexing systems will have to be
highly segmented to handle this flux of background particles.

A large amount of work is still needed in order to optimize the
intersection region and the final focus.  In particular a better
understanding of  the trade off between the different backgrounds is required.
 The strawman detector is meant only to show
that the muon collider detector has unique problems and advantages.
An optimized detector needs to be developed taking these problems into
consideration. 

	Some preliminary calculations for machine related backgrounds
for a lower energy collider (250 GeV x 250 GeV) have also been carried
out.  It appears that the backgrounds in this case are
comparable to those at the 4 TeV machine.  Since little attention 
has yet been paid to the details of the final focus for this lower
energy machine it is possible that reductions in the machine related
backgrounds will be achievable in the future.
\section{OPTIONS}
\vskip -1pc
\subsection{Introduction}

   Up to this point, this report has concentrated on the design of a muon collider 
with 

1) beam energies of 2 + 2 TeV 

2) operating at its maximum energy

3) with a fixed rms energy spread of 0.12 %

4) with no attention to maximizing polarization

In this section we discuss modifications to enhance the muon polarization's, 
operating parameters with very small momentum spreads, operations at energies 
other than the maximum for which a machine is designed, and designs of 
machines for different maximum energies.
\subsection{Polarization}
\vskip -1pc
\subsubsection{Polarized Muon Production.}
The specifications and components in the baseline design have not been  
optimized for polarization. Nevertheless, simple manipulations of parameters 
and the addition of momentum selection after phase rotation does generate 
significant polarization with relatively modest loss of luminosity. The only 
other significant changes required to give polarization at the interaction point are 
rotators in the transfer lines, and a chicane snake in the collider opposite the 
IP.

      \begin{figure}[bht!] 
\centerline{\epsfig{file=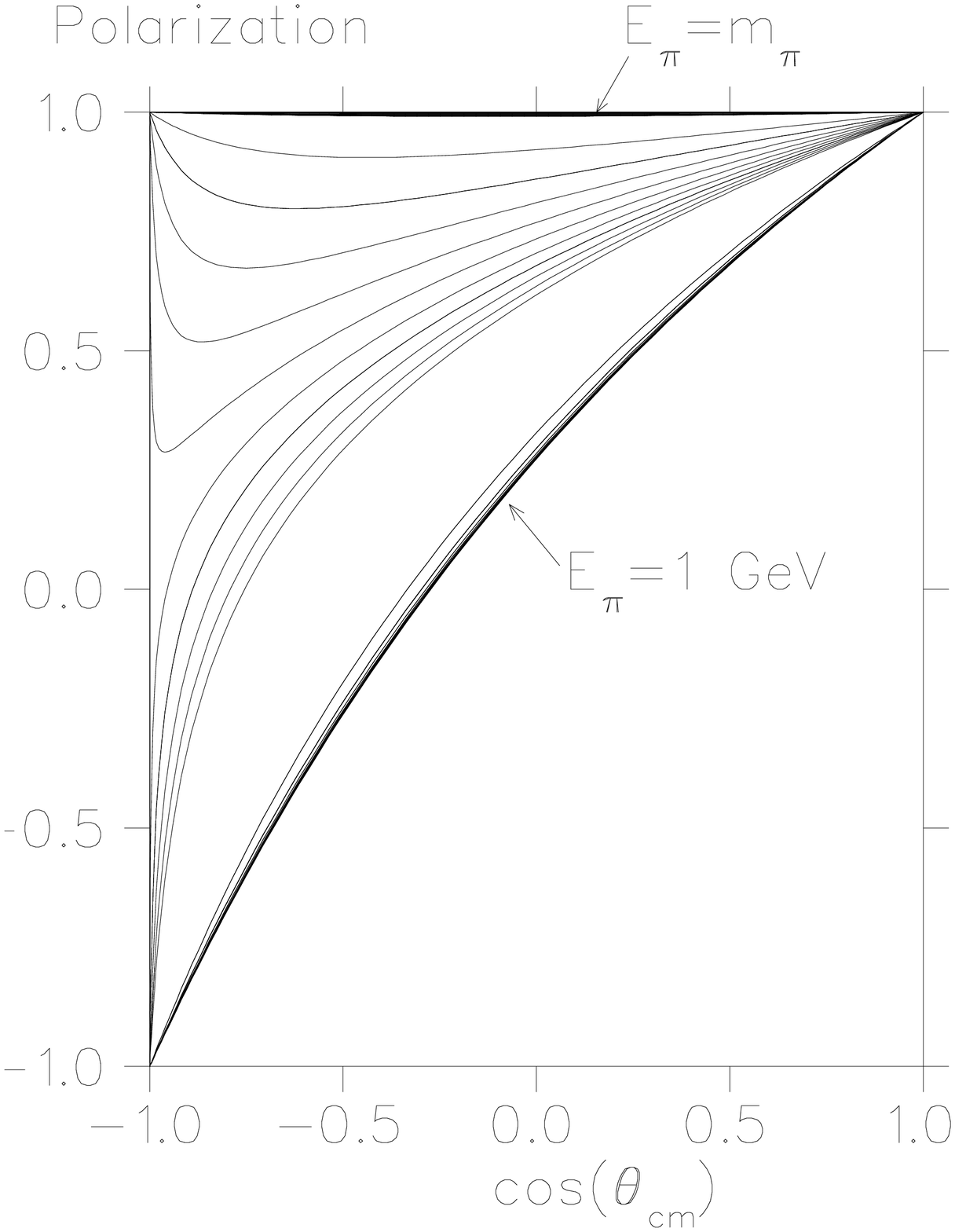,height=4.0in,width=3.5in}}
\caption{Polarization in the lab. frame vs. the cosine of the center of mass decay angle, 
for a number of pion energies.
 \label{polar}}
 \end{figure}

   In the center of mass of a decaying pion, the outgoing muon is fully 
polarized (-1 for $\mu^+$ and +1 for $\mu^-$). In the lab system the 
polarization depends\cite{decaypolar} on the decay angle $\theta_d$ and 
initial pion energy. Figure \ref{polar} shows this polarization as a function 
of the cosine of the center of mass decay angle, for a number of pion 
energies. It is seen that for pion kinetic energy larger than the pion mass, 
the dependence on pion energy becomes negligible and the polarization is given 
approximately by: 
 \b
P_{\mu^-} \approx \cos{\theta_d} + 0.28(1 -\cos^2{\theta_d})
 \e
The average value of this is about 0.19. A Monte Carlo calculation\cite{mc} 
of the capture, decay and phase rotation discussed above gave 
muon polarization of approximately 0.22. The slight difference of this value 
from the average comes from an accidental bias towards forward decay muons.

If higher polarization is required, some deliberate selection of muons from 
forward pion decays  $(\cos{\theta_d} \rightarrow 1)$ is required. This could 
be done by selecting pions within a narrow energy range and then selecting 
only those muons with energy close to that of the selected pions. But such a 
procedure would collect a very small fraction of all possible muons and would 
yield a very small luminosity. Instead we wish, as in the unpolarized case, to 
capture pions over a wide energy range, allow them to decay, and to use rf to 
phase rotate the resulting distribution. 

\begin{figure}[bht!] 
\centerline{\epsfig{file=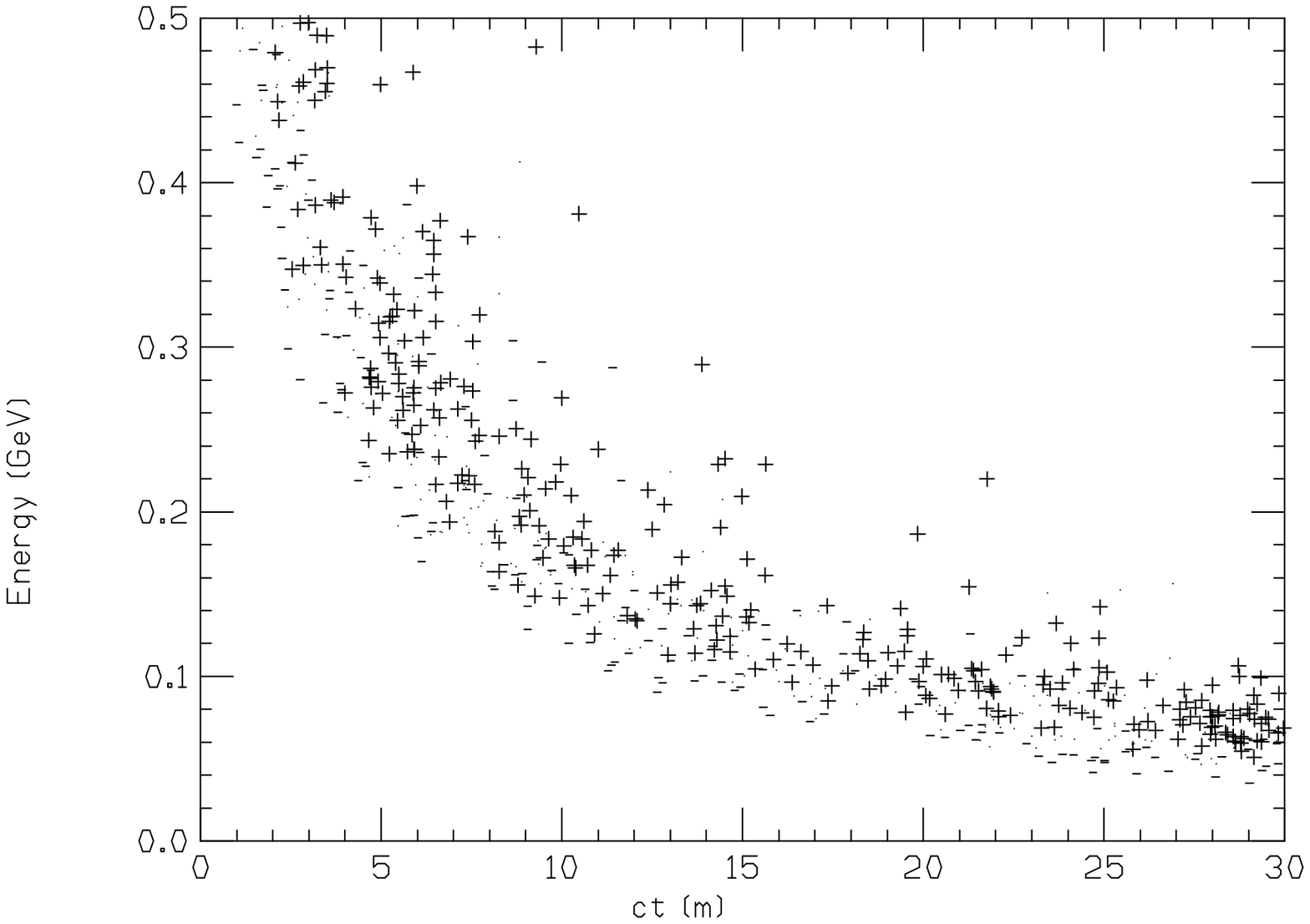,height=4.0in,width=3.5in}}
\caption[Energy vs. ct of $\mu$'s  at end of decay channel without phase 
rotation ]
{Energy vs. ct of $\mu$'s at end of decay channel without phase 
rotation; muons
with polarization P$>{1\over 3}$, $-{1\over 3}< P<{1\over 3}$, and P$<-{
1\over 3}$ are marked by the symbols `+', `.' and `-' respectively. 
 \label{Evsctpol1}}
 \end{figure}

   Consider the distributions in velocity  vs. ct at the end of a decay  
channel. If the source bunch of protons is very short and if the pions were  
generated in the forward direction, then the pions, if they did not decay, 
would all be found  on a single curved line. Muons from forward decays would 
have gained velocity  and would lie above that line. Muons from backward 
decays would have lost  velocity and would fall below the line. A real 
distribution will be diluted  by the length of the proton bunch, and by 
differences in forward velocity due to the finite angles of particles 
propagating in the solenoid fields. In order to  reduce the latter, it is 
found desirable to lower  the solenoid field in the  decay channel from 5 to 3 
Tesla. When this is done, and in the absence of phase rotation, one obtains 
the distribution shown in Fig.~\ref{Evsctpol1}, where the polarization 
P$>{1\over 3}$, $-{1\over 3}< P<{1\over 3}$, and P$<-{1\over 3}$ is marked by 
the symbols `+', `.' and `-' respectively. One sees that the +'s are high, and 
the -'s are low, all along the distribution. 

\begin{figure}[hbt] 
\centerline{\epsfig{file=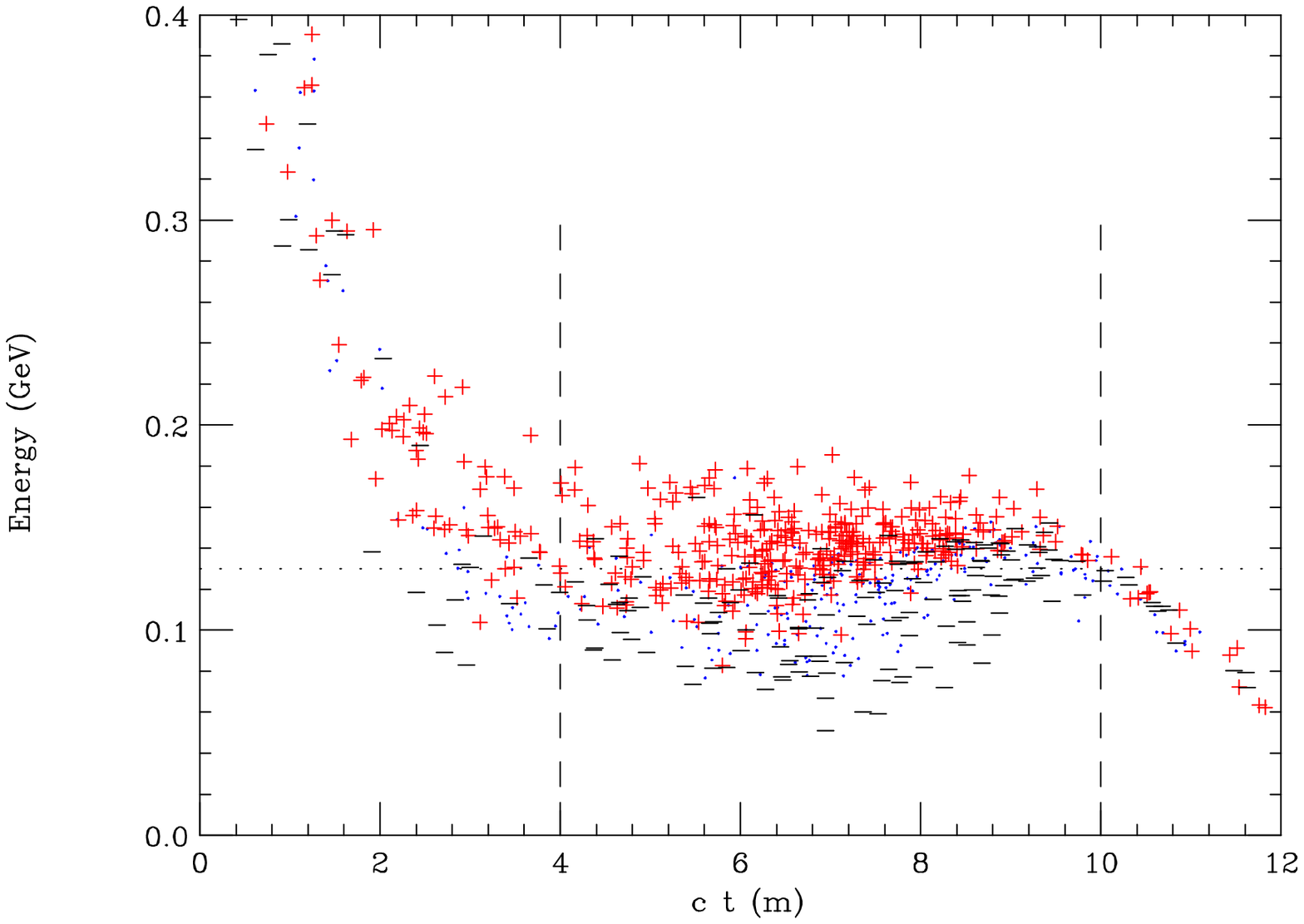,height=4.0in,width=3.5in}}
\caption[Energy vs. ct of $\mu$'s at end of decay channel with phase 
rotation ]
{Energy vs. ct of $\mu$'s at end of decay channel with phase 
rotation; muons with 
polarization P$>{1\over 3}$, ${-1\over 3}< P<{1\over 3}$, and P$<{-
1\over 3}$ are marked by the symbols `+', `.' and `-' respectively.  
 \label{Evsctpol2}}
 \end{figure}

It is found that phase rotation does not remove this correlation: see 
Fig.~\ref{Evsctpol2}. Now, after  time cuts to eliminate decays from high and low  energy pions, there is a simple correlation of polarization with the energy of the muons. 
  If a selection is made on the minimum energy of the muons, then net polarization is obtained. The higher the cut 
on energy, the greater the polarization, but the less the fraction $F_{loss}$ 
of muons that are selected. The cut in time can probably be obtained from the 
phasing of the rf used to capture the bunch. Alternatively, it could be 
provided by a second energy cut applied after a 90 degree longitudinal phase 
rotation.

   In order to provide the required cut on energy, one needs to generate 
dispersion that is significantly larger than the beam size. Collimation from  
one side can then select the higher energy muons. After collimation, the 
remaining dispersion should be removed. The generation of sufficient 
dispersion, in the presence of the very large emittance, is non-trivial. The 
only practical method appears to be the use of a bent solenoid (as discussed 
above, in the section of Muon Collider Components). First the solenoid is bent one way to generate the dispersion; 
the collimator is introduced; then the solenoid is bend the other way to 
remove the dispersion. The complete system thus looks like an ``S" or ``snake". 
   Particles with momentum $p_{\mu}$ in a magnetic field $B$ have a 
bending radius of $R_B$, given by:
 \b
R_B={\({ep_\mu / mc}\)\over c \ B}.
 \e
If the particles are trapped in a solenoid with this field, and the solenoid 
is bent with a radius $R_{bend}$, where $R_{bend} >> R_B,$ 
 then those particles, besides their normal helical motion in the solenoid, 
will drift in a direction ($z$) perpendicular to the bend, with a drift 
angle ($\theta_{drift}=dz/ds$) given by: 
 \b
\theta_{drift}\ \approx \ {R_B \over R_{bend}}
 \e
The integrated displacement in $z$, ie. the dispersion $D$, is then:
 \begin{equation}
D  \= \theta_{drift}\ s \ \approx \ \phi\ R_B,
 \end{equation}
where $\phi$ is the total angle of solenoid bend. 

\begin{figure}[hbt!] 
\centerline{\epsfig{file=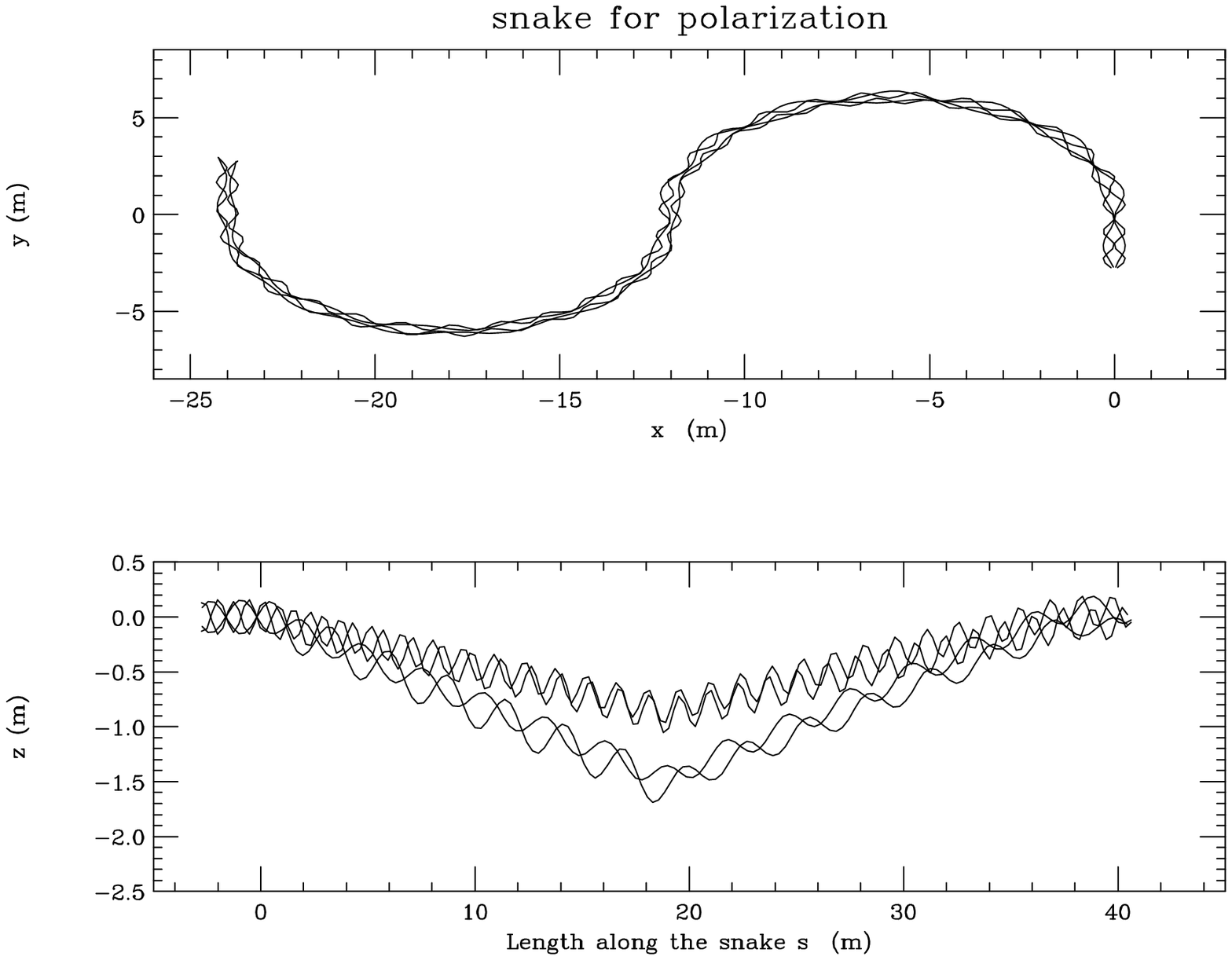,height=4.0in,width=3.5in}}
\caption{Dispersion Snake:  trajectories as seen from the $z$ direction (top); vertical (z) particle positions vs. length (s) along the snake (bottom).
 \label{snakes}}
 \end{figure}

   As an example, we have traced typical particles with momenta of 150 and 300 
MeV/c through a snake with $B=1\, T$, $R_{bend}=6\ m,$ with a first band with $\phi=\pi$ followed by a reverse bend $\phi=-\pi.$  
Fig.~\ref{snakes} shows the trajectories of muons as viewed from the z 
direction. No significant dispersion is seen. 
The two momenta are seen to be dispersed during the 
right hand turn and recombined by the left hand turn.

\begin{figure}[bht!] 
\centerline{\epsfig{file=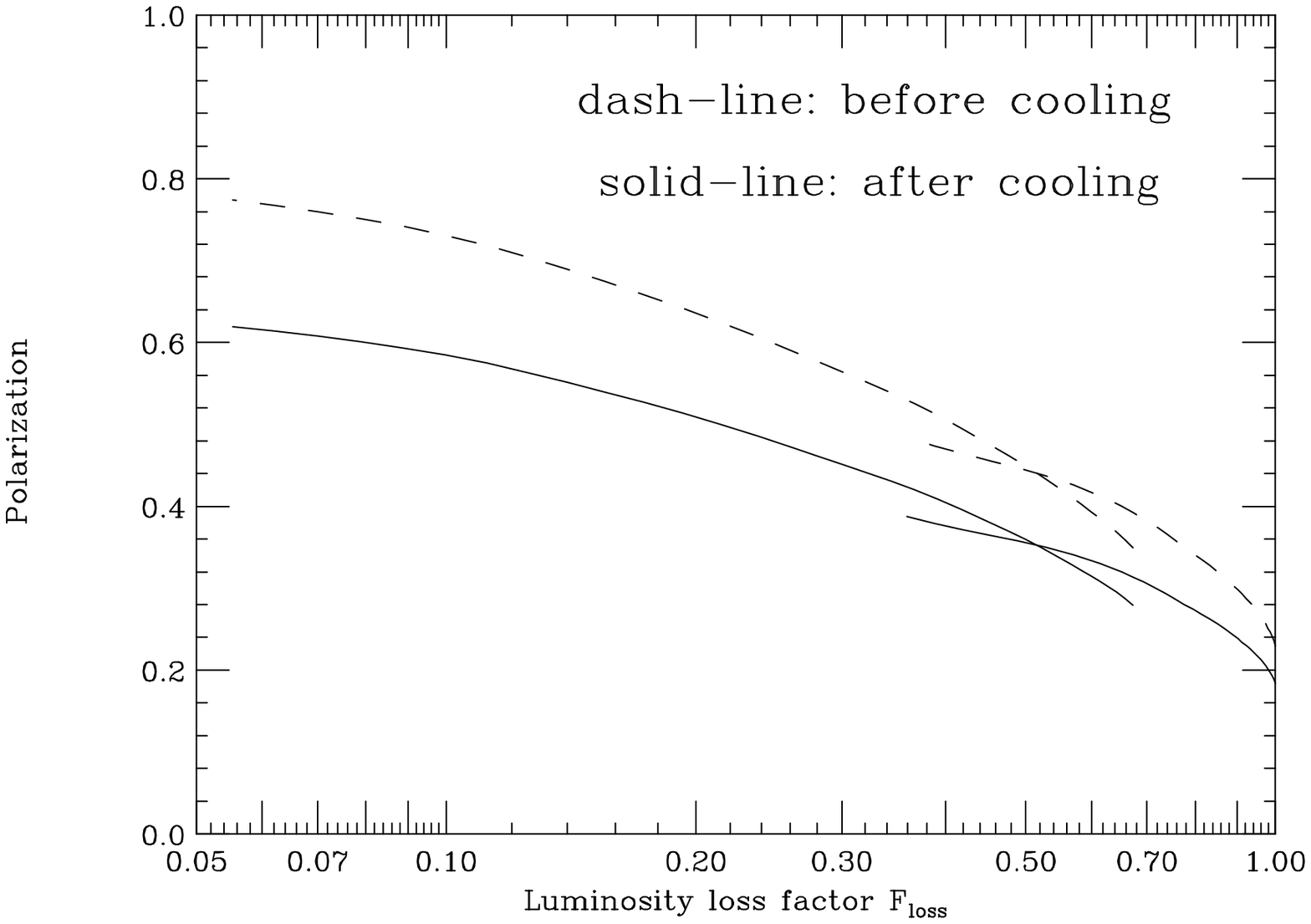,height=4.0in,width=3.5in}}
\caption[Polarization vs. $F_{loss}$ of muons accepted ]
{Polarization vs. $F_{loss}$ of muons accepted; the dashes show 
polarization as selected, the line gives polarization after cooling.
 \label{polvscut}}
 \end{figure}

   Fig.~\ref{polvscut} and Tb.~\ref{poltable} give the results of a Monte 
Carlo study\cite{mc} in which dispersion is introduced, and progressive cuts are applied, 
to the muons at the end of the phase rotation. In this calculation, in order 
to calculate more rapidly, the trajectories were not actually traced through a 
snake. Instead, the particles were propagated through 20 m of straight 
solenoid, followed by the application of dispersion equal to 6 times the 
momentum in GeV/c. A snake that would give such dispersion could have the 
parameters:  solenoid field of 3 T (giving $R_B=0.25\ m$ at the average momentum of 230 MeV/c), diameter of snake bends greater than 5 m and  bend angles of 
$320^{\circ},$ (which would require some variations 
in bend curvature to avoid the solenoid crossing itself), 

Tb.~\ref{poltable} gives results for two fields in the decay channel 
solenoids: 5 T, the field in the point design; and 3 Tesla, chosen to increase 
the polarization. It is seen that for weak cuts and small polarization, it is 
better to avoid the loss of muons from the lower, 3 T, 
field, but with stronger cuts, the lower field gives greater polarization.  In Fig.~\ref{polvscut}, and 
subsequent plots, only data from the preferred fields are shown beyond the 
cross over. 

\begin{table}[htb!]
\centering \protect
\caption[Production Polarization vs. Position]
{Production Polarization vs. Position. B, decay channel field; cut, the position of the colimator; $P_{init}$ initial polarization; $P_{final}$ polarization after dilution in the cooling section; $P_{vec}$ effective vector polarization (see Eq.\ref{vector}); $R_{v/s}$ vector to scalar ratio (see Eq.\ref{vec_sca}); $L_{vec}$ luminosity enhancement for vector state; $E_{ave}$ average final muon energy; $\Delta$E {\it rms} final energy spread
\label{poltable}}
\begin{tabular}{cccccccccc}
\hline
B & cut &$F_{loss}$& $P_{init}$&$P_{final}$ &$P_{vec}$& $R_{v/s}$  &$L_{vec}$&$E_{ave}$&$\Delta$E \\
T & m & & & & & & & MeV & MeV \\
\hline
5 & 0.00& 1.000 & 0.23 & 0.18 & 0.36 & 1.45 & 1.18 & 130  & 23 \\
5 & 1.00& 0.960 & 0.27 & 0.21 & 0.41 & 1.54 & 1.21 & 144  & 23 \\
5 & 1.12& 0.890 & 0.30 & 0.24 & 0.46 & 1.64 & 1.24 & 147  & 20 \\
5 & 1.24& 0.759 & 0.36 & 0.29 & 0.53 & 1.80 & 1.29 & 151  & 18 \\
5 & 1.30& 0.614 & 0.41 & 0.33 & 0.60 & 1.99 & 1.33 & 157  & 17 \\
5 & 1.40& 0.360 & 0.48 & 0.39 & 0.67 & 2.26 & 1.39 & 166  & 15 \\
5 & 1.50& 0.163 & 0.56 & 0.45 & 0.75 & 2.64 & 1.75 & 177  & 15 \\
\hline
3 & 0.00& 0.801 & 0.22 & 0.18 & 0.34 & 1.43 & 1.18 & 130  & 22 \\
3 & 1.06& 0.735 & 0.29 & 0.23 & 0.44 & 1.61 & 1.23 & 133  & 22 \\
3 & 1.16& 0.673 & 0.35 & 0.28 & 0.52 & 1.77 & 1.28 & 137  & 19 \\
3 & 1.26& 0.568 & 0.41 & 0.33 & 0.59 & 1.98 & 1.33 & 141  & 17 \\
3 & 1.32& 0.417 & 0.50 & 0.40 & 0.69 & 2.32 & 1.40 & 147  & 15 \\
3 & 1.40& 0.264 & 0.59 & 0.47 & 0.77 & 2.78 & 1.47 & 151  & 13 \\
3 & 1.48& 0.126 & 0.70 & 0.56 & 0.86 & 3.58 & 1.56 & 159  & 13 \\
3 & 1.56& 0.055 & 0.77 & 0.62 & 0.90 & 4.25 & 1.62 & 168  & 12 \\
\hline
\end{tabular}

\end{table}

\begin{figure}[bht!] 
\centerline{\epsfig{file=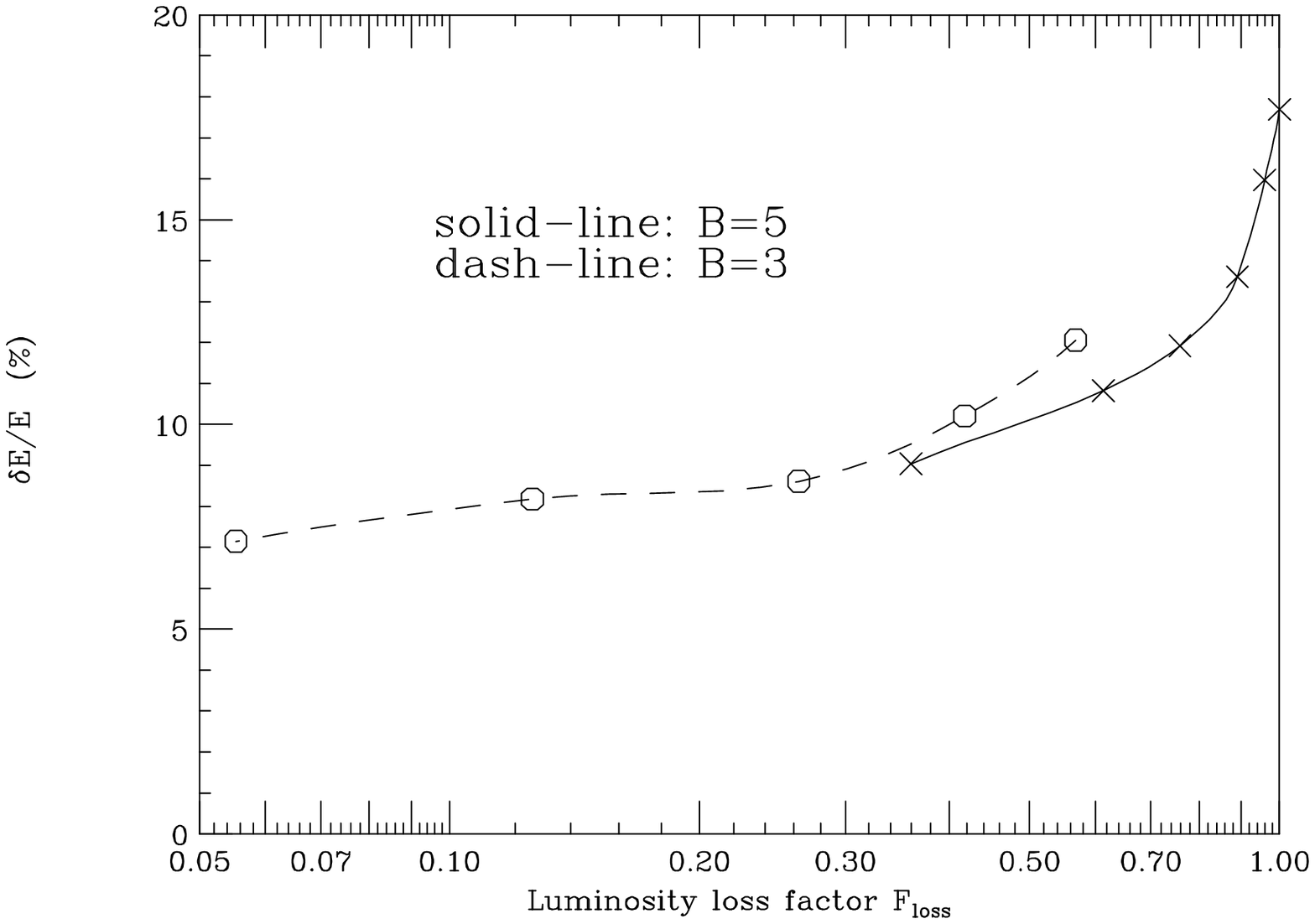,height=4.0in,width=3.5in}}
\caption{the fractional energy spread $\Delta E/E$ 
is plotted against the loss factor $F_{\rm loss}$.
 \label{dEvsF}}
 \end{figure}

   It is seen from Tb.~\ref{poltable} that the energy cut not only 
increases the polarization, but also decreases the energy spread $\Delta E$ of 
the remaining muons. In Fig.~\ref{dEvsF} the fractional energy spread $\Delta E/E$ 
is plotted against the loss factor $F_{loss}$. The energy spread is reduced almost a 
factor of two for reasonable collimator positions. This reduction in energy 
spread would eliminate the need for the first stage of emittance cooling. 

\begin{figure}[bht!] 
\centerline{\epsfig{file=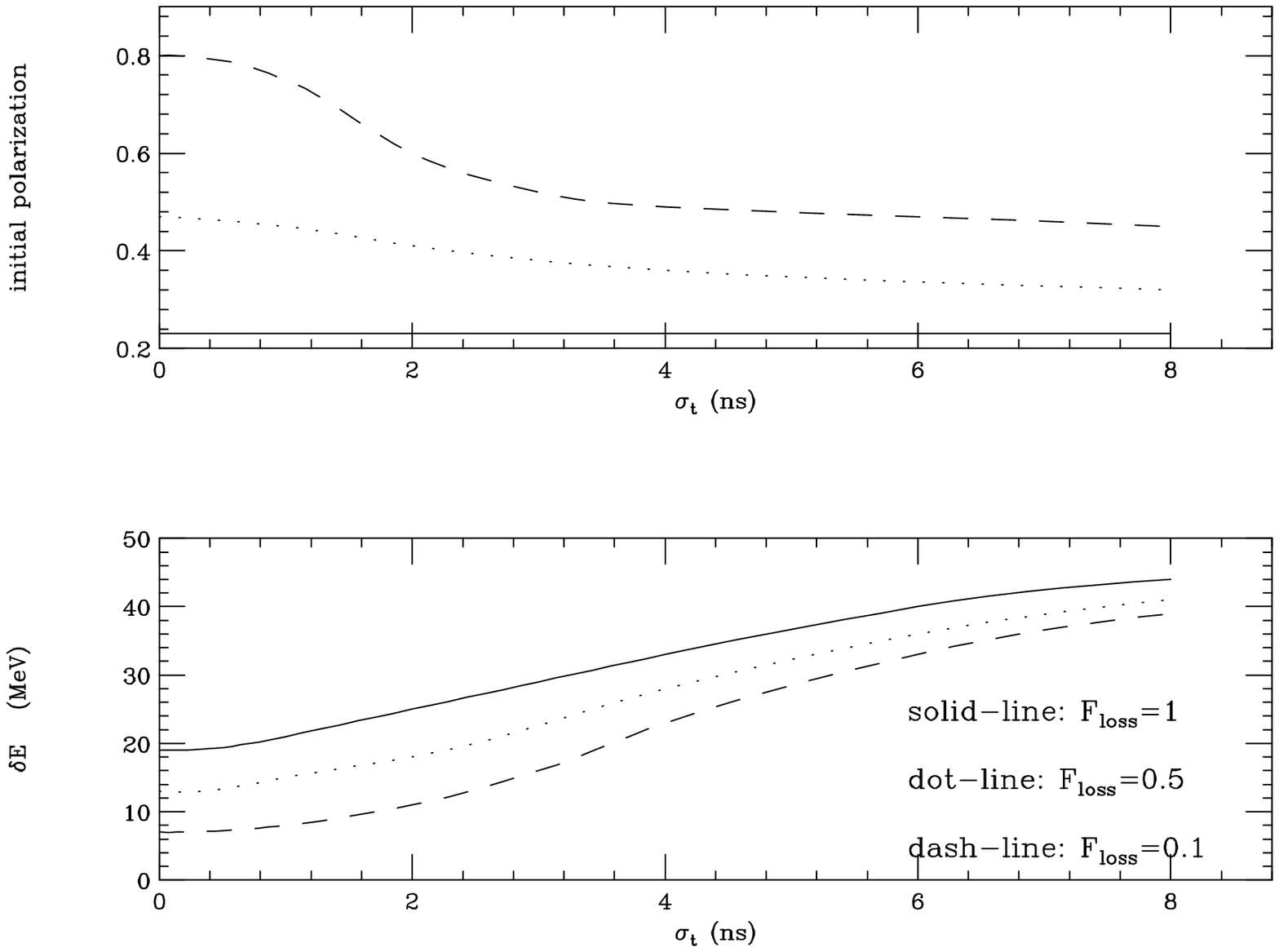,height=5in,width=4in}}
\caption[Polarization vs. $\sigma_t,$ the proton bunch length (upper plot). 
Muon {\it rms} energy spread vs $\sigma_t$ (lower plot)]
{Polarization vs $\sigma_t,$ the proton bunch length (upper plot). Muon
{\it rms} energy spread vs. $\sigma_t$ (lower plot). Both plots for three values
of the loss factor $F_{\rm loss}.$  }         
 \label{sigmat}
 \end{figure}

   A Monte Carlo study has also been done on the effect of variations of the 
proton bunch length $\sigma_t$. Fig.~\ref{sigmat}a shows the polarization 
before cooling as a function of $\sigma_t$ for three values of the loss factor 
$F_{loss}$. It is seen that serious loss of polarization can occurs when the rms width is 
more than 1 nsec. Fig.~\ref{sigmat}b shows the muon rms energy spread after 
the polarization cut. Again it is shown as a function of $\sigma_t$ for three 
values of the loss factor $F_{loss}$. With no cut, the rise in energy spread would be 
serious ($\Delta E > 20\,{\rm MeV}$ is difficult to cool) for an rms width more than 1 ns. 
But with polarization cuts, the energy spread is so reduced that a larger 
 proton bunch length would not be a problem. 

\subsection{Polarization Preservation}
   A recent paper\cite{rossmanith} has discussed the preservation of muon 
polarization in some detail. During the ionization cooling process the muons 
lose energy in material and have a spin flip probability ${\cal P},$ where 
 \b
{\cal P}\approx \int {{m_e}\over {m_{\mu}}}\beta_v^2\ {\Delta E\over E }
 \e
where $\beta_v$ is the muon velocity divided by c, and $\Delta $E/E is the fractional 
loss of energy due to ionization. In our case the integrated energy loss 
is approximately 3 GeV and the typical energy is 150 MeV, so the integrated 
spin flip probability is close to 10\%. The change in polarization $dP/P$ is 
twice the spin flip probability, so the reduction in polarization is 
approximately 20~\%. This dilution is included in the ``$P_{final}$" column in 
Tb.~\ref{poltable} and is plotted as the line in Fig.~\ref{polvscut}.

   During circulation in any ring, the muon spins, if initially longitudinal, 
will precess by $(g-2)/2 \gamma$ turns per revolution in the ring; 
where $(g-2)/2$ is $1.166\times 10^{-3}$. An energy spread $d\gamma/\gamma$ will 
introduce variations in these precessions and cause dilution of the 
polarization. But if the particles remain in the ring for an exact integer 
number of synchrotron oscillations, then their individual average $\gamma$'s 
will be the same and no dilution will occur. It appears reasonable to use this 
``synchrotron spin matching"\cite{rossmanith} to avoid dilution during 
acceleration. 

   In the collider, however, the synchrotron frequency will be too slow to use 
``synchrotron spin matching'', so one of two methods must be used. 

\begin{itemize}
 \item{ Bending can be performed with the spin orientation in the vertical 
direction, and the spin rotated into the longitudinal direction only for the 
interaction region. The design of such spin rotators appears relatively 
straightforward. The example given in the above reference would only add 120 m 
of additional arc length, but no design has yet been incorporated into the 
lattice.} 
 \item{ The alternative is to install a Siberian Snake\cite{siberian}
at a location exactly opposite to the intersection point.  Such a snake 
reverses the sign of the horizontal polarization and generates a cancelation 
of the precession in the two halves of the ring.} 
 \end{itemize} 

   Provision must also be made to allow changes in the relative spins of the 
two opposing  bunches. This could be done, prior to acceleration, by switching 
one of the two beams into one or the other of two alternative injection lines. 

\subsection{Benefits of Polarization of Both Beams}

   We consider two examples of the general advantage of having polarization 
in both beams. Individual physics experiments would have to be considered to 
determine how important such advantages are.

\begin{figure}[bht!] 
\centerline{\epsfig{file=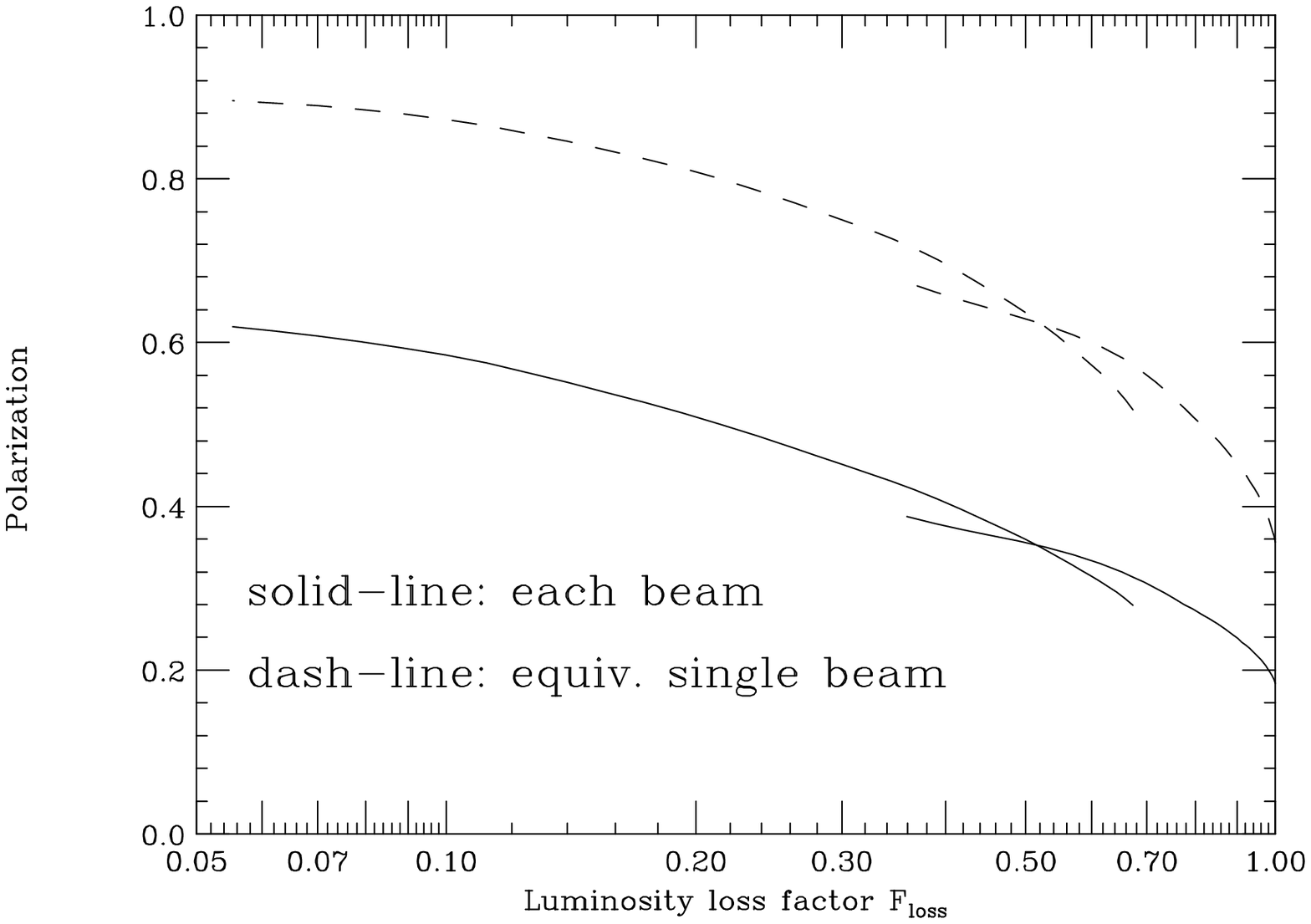,height=4.0in,width=3.5in}}
\caption{ Polarization of each beam $P$, and the resulting polarization of a 
vector state $P_{vec}$ vs. the loss factor $F_{loss}$.
 \label{pol2}}
 \end{figure}

   Consider the polarization 
of a vector spin state generated by the annihilation of the two muons. 
 \b
P_{vec}\={F^{++}-F^{--}\over F^{++}+F^{--}}
 \e
When only one beam has polarization $P_1$, then $P_{vec}=P_1$. But if both 
beams have polarization $P$ in the same direction (ie. with opposite 
helicities), then 
 \b
P_{vec}\= {(P+1)^2-(P-1)^2 \over (P+1)^2+(P-1)^2}\label{vector}
 \e
In Fig.~\ref{pol2} both the polarization of each beam $P$, and the resulting 
polarization of a vector state $P_{vector}$ are plotted against the loss factor $F_{loss}$.                          
    
\begin{figure}[bht!] 
\centerline{\epsfig{file=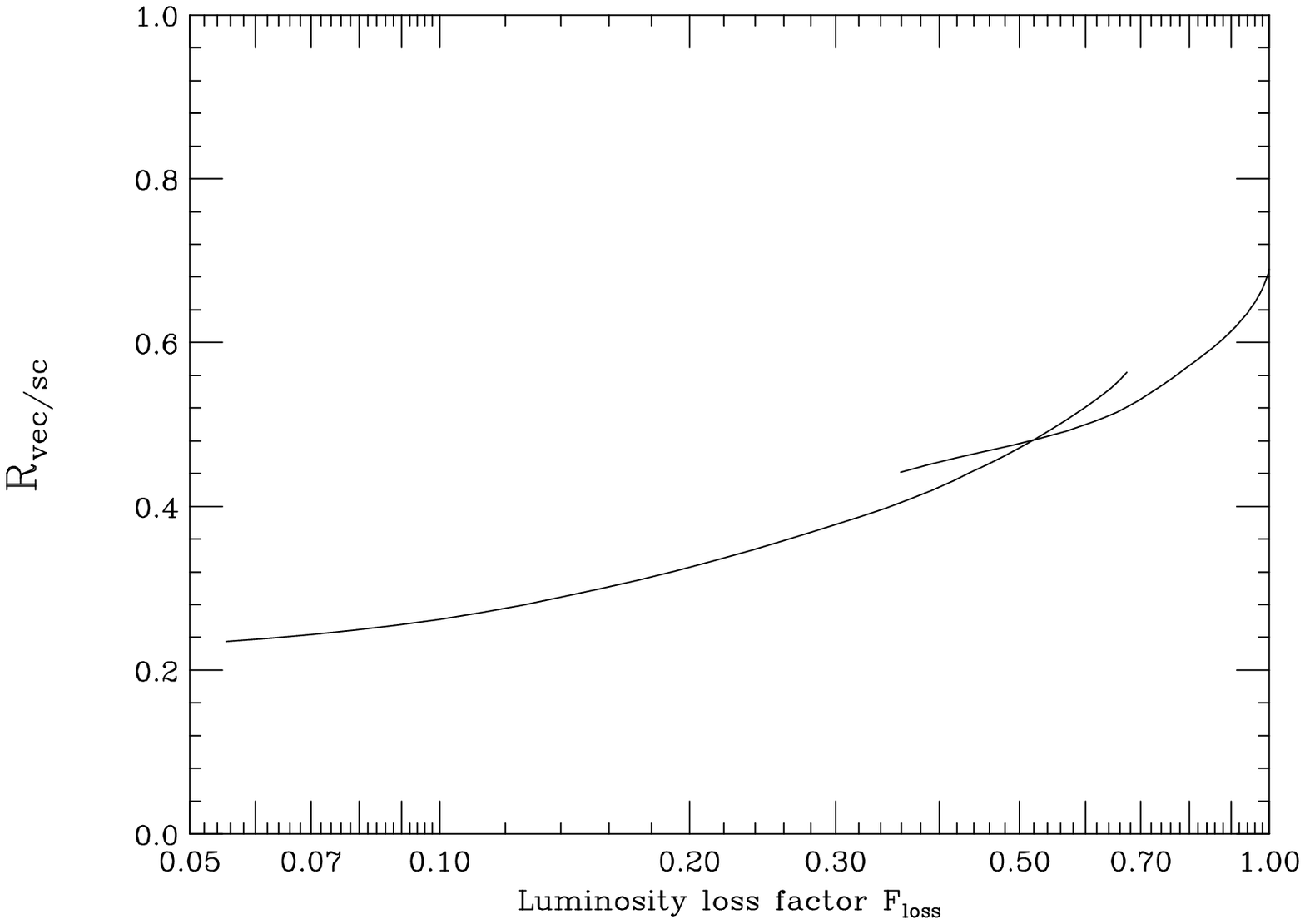,height=4.0in,width=3.5in}}
\caption{ Ratio of vector to scalar states, $R_{v/s}$ vs. the 
loss factor $F_{loss}.$
 \label{scalarvector}}
 \end{figure}

   A second advantage is that the ratio $R_{v/s}$ of vector to 
scalar luminosity can be manipulated to enhance either the vector or the 
scalar state. If the polarization directions have been chosen to enhance the 
ratio of vector to scalar states, then: 
 \b
R_{v/s} \= {1+P \over 1-P}\label{vec_sca}.                             
 \e
Tb.~\ref{poltable} and Fig.~\ref{scalarvector} show this ratio as a 
function of the loss factor $F_{loss}$.
    
Tb.~\ref{poltable} also shows that the fraction of total luminosity in a 
given state can be enhanced. If polarizations are chosen to enhance the 
vector state, then the fraction of vector luminosity is increased from 1/2 to 
$(1+P)/2$, giving a gain factor of
\begin{equation}
{\cal L}_{vec}=1+P
\end{equation}

\subsection{Luminosity loss}
 If nothing else is done, then the luminosity, which is proportional to $n_{\mu},$ will drop as $F_{loss}^2$; where $F_{loss}$ 
is the fraction muons lost by the muon momentum cut. At the same time, 
however, the space charge, wakefield, and loading during the cooling and 
acceleration will all be reduced; as will the beam beam tune shift in the 
collider. Clearly, the machine parameters should be reoptimized and some part of the lost luminosity recovered. 

   One way to recover the luminosity would be to increase the 
proton bunch intensity by the factor $F_{loss}$. If this were done, then the original
number of muons per bunch would be generated; all the wake field, loading and 
space charge effects would be the same; and the luminosity per bunch crossing 
would be the same. If we assume that the total proton current is determined 
by the driver, then such an increase in proton intensity per bunch will 
necessitate a reduction in the number of bunches or repetition rate, by the 
same factor $F_{loss}$. The luminosity will then fall by $F_{loss}$ and not by  $F_{loss}^2$ as before.

   For instance, in the unpolarized case of the $4\,$TeV collider, there were 
two bunches of each sign. If the momentum cut is chosen to give a value of 
$F_{loss}$  = 1/2, and the proton beam is distributed into 2 instead of 4 
initial bunches, then the final number of muons per bunch, the loading, beam 
beam tune shift, etc. would all be the same as in the unpolarized case. The 
luminosity would be down by a factor of only two, for a polarization of 34 \% in 
both beams.

   For higher polarization at good luminosity it would be desirable to have a 
proton source with the option of a lower repetition rate, but even 
larger numbers of 
protons per spill. For example 4 x $10^{14}$ protons per pulse at 4 Hz. It 
should then be possible to extend this method to an operation with 
$F_{loss}$ = 1/8, and polarization of both beams of 57 \%.

   One also notes that the luminosity could be maintained at the full 
unpolarized value if the proton source intensity could be increased. Such an 
increase in proton source intensity in the unpolarized case would be 
impractical because of the resultant excessive high energy muon beam power, 
but this restriction does not apply if the increase is used to offset losses 
in generating polarization. If, for instance, the driver repetition rate were 
increased from 15 to 30 Hz, the fractions $F_{loss}$   set at 0.5, and the 
number of bunches reduced to one, then the full luminosity of $10^{35}\ (cm^{-
2} s^{-1})$ would be maintained with polarization of both beams of 34\%. 
\subsection{Luminosity}
The bunch populations decay exponentially, yielding an integrated luminosity 
equal to its initial value multiplied by an {\it effective} number of turns 
$n_{{\rm eff}}\approx 150\ B,$ where B is the mean bending field in T. 

The luminosity is given by:
 \b
\Ls\={n_{\mu}^2\ n_b f_{rep}\ n_{eff} \gamma\over 4\pi\ \beta_{\perp}^*\ \epsilon_n} 
H(A,D)   \label{lumi}
 \e
where $n_{\mu}$ is the number of muons per bunch, $n_b$ is the number of bunches, $\gamma$ is the normalize energy, $\beta_{\perp}^*$ is the beta function at the IP and $\epsilon_n$ is the transverse normalize emittances (assumed vertical and horizontal to be equal), and the enhancement factor $H(A,D)$ is
 \b
H(A,D)\approx 1+D^{1/4} \left[ {D^3\over 1+D^3} \right] \left\{
\ln{(\sqrt{D}+1)} + 2\ln{({0.8\over A})} \right\},
 \e
 \b
A=\sigma_z / \beta^*,
 \e
 and
 \b
D={\sigma_z n_\mu \over \gamma \sigma_t^2} r_e ({m_e\over m_{\mu}})
 \e

In the cases we are considering\cite{pisin2}: 
A = 1, D $\approx$ .5 and H(A,D) $\approx 1$.
\subsection{Luminosity vs. Energy, for a Given Ring}
  For a fixed collider lattice, operating at energies lower than the design 
value, the luminosity will fall as $\gamma^3.$ One power comes from the 
$\gamma$ in Eq.~\ref{lumi}; a second comes from $n_{eff}$, the effective 
number of turns, that is proportional to ${\gamma}$; the third factor comes from 
$\beta^*,$ which must be increased proportional to $\gamma$ in order to keep 
the beam size constant within the focusing magnets. The bunch length 
$\sigma_z$ must also be increased proportional to $\gamma$ so that the 
required longitudinal phase space is not decreased; so A = $\sigma_z/\beta^*$ 
remains constant. 
\subsection{Scaling for Collider Rings for Different Energies} 
   As noted above, the luminosity in a given ring will fall as the third power 
of the energy at which it is operated. Such a drop is more rapid than the gain 
in typical cross sections, and, as we shall see, it is more rapid than the 
drop in luminosity obtained with rings designed for the lower energies. It 
would thus be reasonable, having invested in a muon source and accelerator, 
to build a sequence of collider rings at spacings of factors of 2-3 in 
maximum energy. 
   We will now derive scaling rules for such collider rings.

The luminosity 
 \b
{\cal L}\ =\ 
{n_{\mu}^2\ n_{\rm eff}\ n_b\ f_{\rm rep} \ \gamma \over 4\ \pi\ \epsilon_n\ 
\beta_{\perp}^*    }
\ \propto\ {n_{\mu}\ I_{\mu}\ \gamma \over \epsilon_n\ \beta^*    }
 \e
where $I_{\mu}=n_{\mu}n_bf_{rep},$ is the muon flux and which, since
$\Delta\nu_{bb}$, the
beam beam tune shift is given by: 

 \b
\Delta\nu_{bb}\ \propto\ {n_{\mu} \over \epsilon_n },
 \e
gives:
 \b
{\cal L}\ \propto \  
{I_{\mu} \ \Delta\nu_{bb} \ \gamma \over  \beta_{\perp}^*    }
 \e

   If a final focus multiplet is scaled keeping the relative component lengths
and the pole tip fields constant, then one obtains:
 \b
 \ell^*     \ \propto \ \sqrt{a_{max}\ \gamma}    \e  \b
 \theta^*\ \propto \ \sqrt{{a_{max}\over \gamma}} 
 \propto\ \sqrt{{\epsilon_n \over \beta_{\perp}^*\ \gamma}} \e \b
 \beta_{\perp}^*\ \propto\ {\epsilon_n \over a_{max}}
 \e
where $\theta^*$ is the rms angle of muons diverging from the focus, $\ell^*$
is the free space from the target to the first quadrupole (proportional to 
all quadrupole lengths in the multiplet), and $a_{max}$ is the maximum 
aperture of any quadrupole (proportional to all apertures in the multiplet).

   The normalized emittance $\epsilon_n$ is constrained by the ionization 
cooling, but since one can exchange transverse and longitudinal emittance,
it is, in principle, the six dimensional emittance $\epsilon_6$ that is 
constrained. Extending the lepton emittance conventions, we define:
 \b
 \epsilon_6\ = \ (\epsilon_n)^2\ {dp\over p} \sigma_z \gamma \beta_v.
 \e
With this definition, the six dimensional phase space 
$ \Phi_6\ =\ \pi^3\ m_{\mu}^3\ \epsilon_6$.
$\sigma_z$~cannot be large compared with the focus parameter $\beta^*$, so, 
taking them to be proportional to one another, and taking the $\beta_v=1$,
then:
 \b
 \epsilon_6\ \propto \ (\epsilon_n)^2\ {dp\over p} \beta^* \gamma 
 \e
and from the above:
 \b
 (\epsilon_n)^3\ \propto\ {\epsilon_6\ a_{max} \over \gamma\  {dp\over p}} 
 \e  \b
 (\beta_{\perp}^*)^3\ \propto {\epsilon_6 \over  \gamma\ {dp\over p}\ a_{max}^2 }
 \e
\subsection{Six Dimensional Emittance dependence on $n_\mu$ and 
$\epsilon_n$}
The six dimensional emittance $\epsilon_6$ obtained from 
the cooling will, because of more detailed constraints, depend 
to some extent  
on the number of muons $n_{\mu}$, and on the final transverse 
emittance $\epsilon_n$. 

The approximate dependence on the number of muons is relatively transparent. As the 
number of muons per bunch rises, the longitudinal space charge forces increase 
and it becomes impossible, without changing the rf gradients, to maintain the 
same bunch lengths. As a result the bunch lengths must be increased
by the square root of the number of muons.

\begin{figure}[bht!] 
\centerline{\epsfig{file=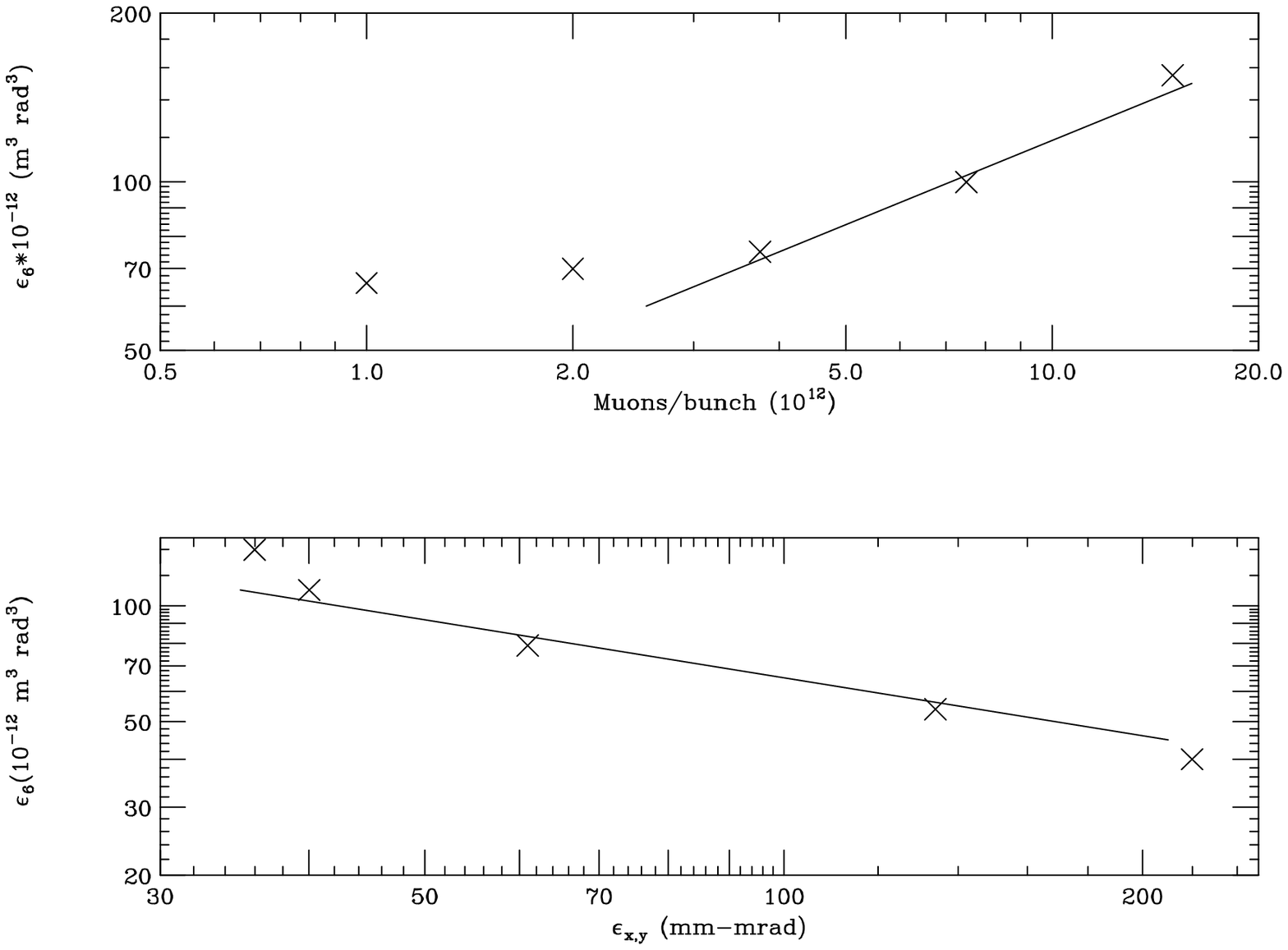,height=4.0in,width=3.5in}}
\caption{Six dimensional emittance $\epsilon_6$ vs. a) muon intensity $n_\mu$ 
entering the cooling, and b) the transverse emittance $\epsilon_n$ at the end 
of the cooling }
 \label{6dim}
 \end{figure}

A study, with the analytic formulae used for the model cooling system discussed before, was used again to derive cooling sequences 
with different final parameters. First, sequences were calculated with
numbers of initial 
muons per bunch of 1, 2, 3.75, 7.5, and 15 x $10^{12}$ (corresponding to muons 
in the collider of 0.1, 0.2, 1, 2, and 4 x $10^{12}$). The final
transverse emittance at the end of the cooling was required to be 
$4\times 10^{-5}$ m, 
(corresponding to an emittance in the collider of $5\times 10^{-5}$ m).
The six dimensional 
emittances obtained are plotted in Fig.~\ref{6dim}a. It is seen that for 
$n_{\mu} > 10^{12}$ the six dimensional emittances are indeed approximately 
proportional to the root of the number of muons (the line shows this 
dependence). 

The study also obtained cooling sequences giving
six dimensional emittances for a range of final 
transverse emittances. The dependence here is more complicated. If emittance 
exchange between longitudinal and transverse emittances could be achieved 
without material then the six dimensional emittance should be independent of 
the final transverse emittance chosen. But the exchange does require material  and Coulomb scattering in this material increases the six dimensional 
emittances; and it 
does so to a greater extent if the transverse emittance is 
small. In Fig.~\ref{6dim}b, we show the six dimensional emittances obtained 
for 5 representative transverse emittances. Over the range of interest the 
dependence of $\epsilon_6$ is approximately the inverse root of $\epsilon_n$ 
(the line shows this dependence).

For the purposes of this study, we may thus assume that:
  \b
\epsilon_6\ \propto\ \sqrt{n_{\mu}\over \epsilon_n}
\label{eps6}
  \e

\subsection{Energy Scaling, allowing the emittances to vary} 
If $n_{\mu}$ is limited by the beam beam tune shift:
 \b
n_{\mu}\ \propto\ \epsilon_n\ \Delta\nu_{bb}
 \e
substituting this in Eq.~\ref{eps6}:
 \b
\epsilon_6\ \propto\  \sqrt{ \Delta\nu_{bb}}
  \e
giving:
 \b
\epsilon_n \propto\ \Delta\nu_{bb}^{1/6}\ \({a_{max}\ \over \gamma\ 
dp}\)^{1/3}
 \e  \b
\beta_{\perp}^* \ \propto\ {\epsilon_n \over a_{max}}                            
 \e  \b 
n_{\mu}\ \propto\ (\Delta\nu_{bb})^{1{1\over 6}}\  \({a_{max}\ \over \gamma\ 
dp}\)^{1/3}
 \e  
so:
 \b  
{\cal L}(\Delta\nu) \ \propto \ 
I_{\mu}\ \gamma^{4/3}\  \Delta\nu_{bb}^{5/6}\  a_{max}^{2/3}\ dp^{1/3} 
 \e

One notes however that as $\gamma$ or $dp$ fall the required number of muons 
$n_{\mu}$ rises, and will at some point become unreasonable. 
If we impose a maximum number of 
muons $n_{max}$, then, when this bound is reached,
 \b
\epsilon_n \propto\ n_{max}^{1/7}\ \({ a_{max}\  \over \gamma\ dp}\)^{2/7}
 \e  \b 
\beta_{\perp}^* \ \propto\ {\epsilon_n \over a_{max}}                       
 \e  
and: 
 \b 
{\cal L}(n) \ \propto \ 
I_{\mu}\ n_{max}^{12/7}\ \gamma^{11/7}\  a_{max}^{3/7}\ dp^{4/7}
 \e

\begin{figure}[hbt!] 
\centerline{\epsfig{file=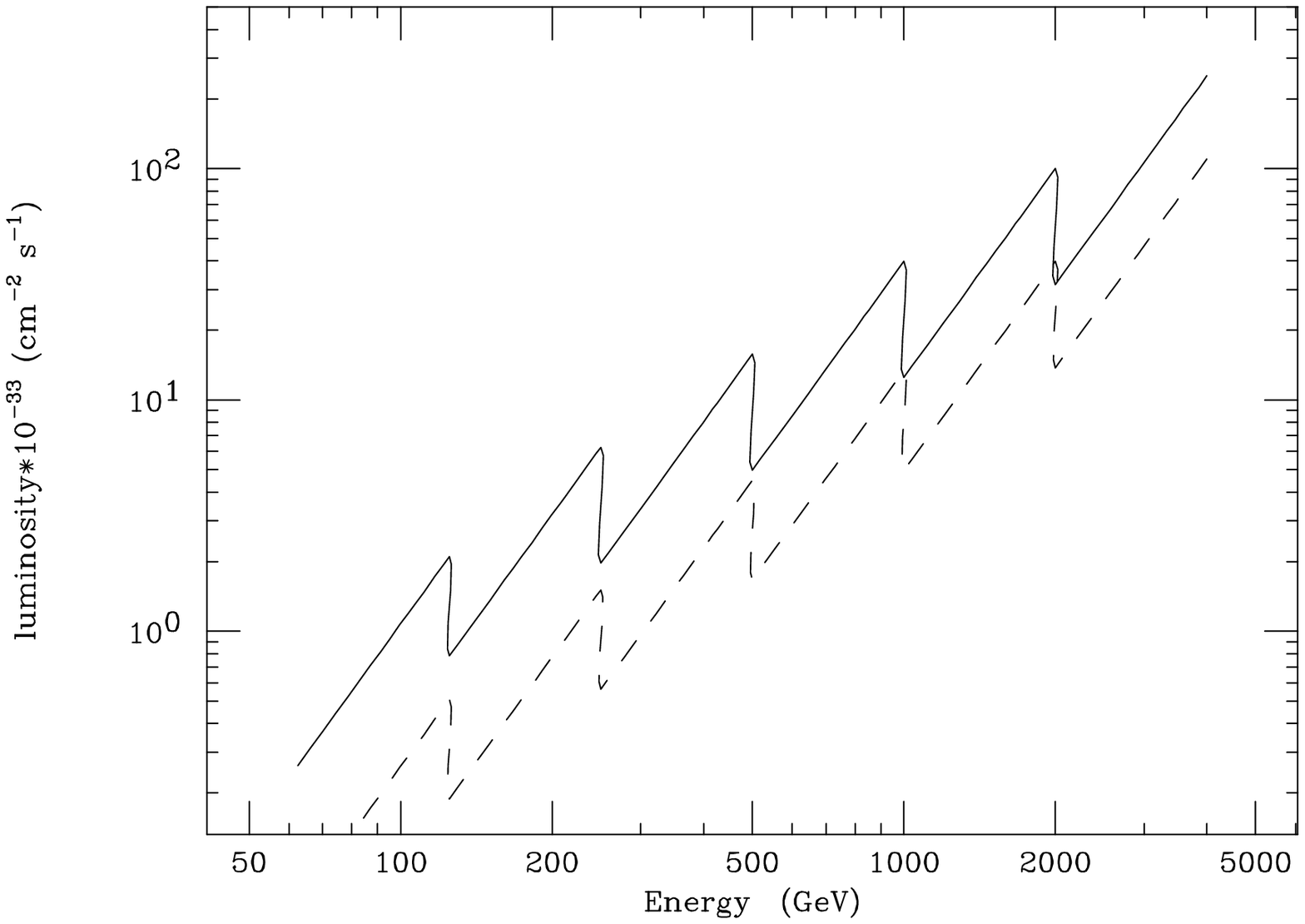,height=4.0in,width=3.5in}}
\caption{Luminosity vs. energy assuming rings spaced by factors of two in 
energy; the line is for $\Delta E/E=0.12 \%,$ the dashed line is for $\Delta E/E=0.01 \%.$
 \label{scaling}}
 \end{figure}

   Using the above relationships. and assuming a constant value of $a_{max}$ we 
obtain the scaled parameters for a sequence of colliding rings given in Tb.~\ref{scalingtbl}. Fig.~\ref{scaling} shows the luminosities that would be available 
at all energies, including those requiring the use of rings  at energies less 
than their maximum. The lines and dashed lines indicate the luminosities with 
a bound on $n_{\mu}$ of $4\times 10^{12}$.
The line gives luminosities for the nominal rms 
$dp/p$ of 0.12\%, while the dashed line is for a $dp/p$ of 0.01\%.

\begin{table}[hbt!]
\centering \protect
\caption{Scaling of Parameters with Energy and Momentum spread.
\label{scalingtbl}}
\begin{tabular}{cccccccccc}
\hline
  E   &Luminosity& emittance & $n_{\mu}$ &$\delta\nu_{bb}$& $\beta_{\perp}^*$ &len$^*$&$\beta_{max}$&chrom&$\Delta$ E/E    \\
 GeV  &$cm^{-2}s^{-1}$&$\pi$ m rad &$10^{12}$&   & mm   &  m   &  km  &   & \%  \\
\hline
 4000 &  2.5E+35 &  4.0E-05 & 1.6 & 0.040&  2.4 &  9.2 &  882 &12829  &0.12 \\
 2000 &  1.0E+35 &  5.0E-05 & 2.0 & 0.040&  3.0 &  6.5 &  350 & 3600  &0.12  \\
 1000 &  4.0E+34 &  6.3E-05 & 2.5 & 0.040&  3.8 &  4.6 &  139 & 1010   &0.12 \\
  500 &  1.6E+34 &  7.9E-05 & 3.2 & 0.040&  4.8 &  3.3 &   55 &  283    &0.12\\
  250 &  6.3E+33 &  1.0E-04 & 4.0 & 0.040&  6.0 &  2.3 &   22 &   80  &0.12  \\
  125 &  2.1E+33 &  1.2E-04 & 4.0 & 0.033&  7.3 &  1.6 &    9 &   23   &0.12 \\
\hline
 4000 &  1.1E+35 &  9.1E-05 & 3.6 & 0.040&  5.5 &  9.2 &  385 & 5604 &0.01  \\
 2000 &  4.0E+34 &  1.1E-04 & 4.0 & 0.036&  6.7 &  6.5 &  156 & 1603 &0.01  \\
 1000 &  1.3E+34 &  1.4E-04 & 4.0 & 0.029&  8.2 &  4.6 &   64 &  465  &0.01 \\
  500 &  4.5E+33 &  1.7E-04 & 4.0 & 0.024& 10.0 &  3.3 &   26 &  135  &0.01 \\
  250 &  1.5E+33 &  2.0E-04 & 4.0 & 0.020& 12.2 &  2.3 &   11 &   39  &0.01 \\
  125 &  5.1E+32 &  2.5E-04 & 4.0 & 0.016& 14.9 &  1.6 &    4 &   11  &0.01 \\

\hline
\end{tabular}
\end{table}
\section{RESEARCH AND DEVELOPMENT PLAN}
In this section we discuss a Research and Development 
plan  
aimed at the operation of a $0.5\,{\rm TeV}$ 
demonstration machine by 
the year 2010, and of the 4 TeV machine by year 2020. It 
assumes 5 years of 
theoretical study, component modeling and critical 
subsystem 
demonstration; followed by 4 years of component 
development and demonstration machine design. 
Construction of the 
demonstration machine would follow and take about 4 
years. 
The high energy machine would come a decade later. 
\subsection{Theoretical Studies}

Much progress has been made during the last year. New 
problems 
continue to be uncovered, but new solutions have been 
found. Much 
work remains to be done: The first object will be to 
define a 
single self consistent set of parameters for the 4 TeV 
collider. 
Items needing study include: 
 \begin{enumerate}
 \item
Define parameters for the proton source, target, 
capture and phase rotation systems. 
 \item
Incorporate operating parameters for the optional 
operation with 
polarized, or very low energy spread, beams.
 \item
Define and simulate a complete cooling scenario.
 \item
Define a preferred acceleration scenario and perform 
complete simulations. Study the required shielding of 
the superconducting 
cavities from muon decay electrons.
 \item
Design a halo scraping system for the collider ring.
 \item
Continue work on the collider lattice, including a study 
of the effect 
of lattice errors, and an investigation of the use of 
higher order multipole 
correctors. Continue the study of the stability of the 
proposed beams and 
design an rf system for BNS damping.
 \item
Continue optimization of the shielding of the detector. 
 \item
Design a ``strawman'' detector with all components 
capable of 
withstanding the backgrounds, and simulate some 
representative 
physics observations. 
 \item
Study safety and radiation exposures both on and off 
site, including the 
hazards from neutrino fluxes. 

 \end{enumerate}

   It is estimated (see Tb.~\ref{effort}) that the 
current 
effort is about 22 full time equivalents, but only a few 
of these 
are funded specifically for such work. Not only should 
the effort 
be legitimized, but, if we are to determine if such 
machines are 
practical, it needs to be expanded. The machine is 
complex 
and unconventional. Many separate systems need study. 
Some 
have hardly been looked at yet. 
\begin{table}[thb!]
\centering
\caption{Required Base Manpower
\label{effort}}  
\begin{tabular}{lcc}
\hline
	& Now   & Required \\
\hline
ANL     & 1     &   2   \\
BNL     & 8     &  16    \\
FNAL    & 7     &  16    \\
LBNL    & 4     & 8     \\
BINP    & 1     & 3     \\
Other US & 1    & 3     \\
	 & --- & --- \\
Total FTE's & 22 & 48  \\
\hline
\end{tabular}

\end{table}
\subsection{Component Development and Demonstrations}
Theoretical studies alone will not be sufficient to 
determine the 
practicality of a muon collider. Experimental studies 
are 
essential. Some such studies can be undertaken without 
new 
funding, but the major efforts will require specific 
support. We 
attempt below to estimate what will be required. 
\subsubsection{Proton Driver Experimental R \& D.}
   Beam dynamic experiments at the BNL AGS will be 
needed, but 
should not be expensive. A modification of the AGS to 
avoid 
transition in that machine, and study the resulting 
improvements 
in phase space density would be very desirable, but the 
cost
should probably be justified as an AGS improvement, 
rather than as 
a muon collider experiment, and it has not been included 
in this 
estimate.
\subsubsection{Target, Capture and Decay Channel 
Experimental R \& D.}

An experiment\cite{ref15a,ref15b} has taken data and is 
currently being 
analyzed, to determine pion production at its low energy 
maximum. 
This data, together with assumptions on pion 
reabsorbtion should 
allow more realistic Monte-Carlo calculations of total 
pion yield 
and capture in a solenoid. Nevertheless, there are 
several 
reasons why a demonstration of such capture is 
desirable:

 \begin{itemize}
 \item
Thermal cooling requirements dictate that the target be 
liquid: liquid 
lead and gallium are under consideration. In order to 
avoid shock 
damage to a container, the liquid may need to be in the 
form of a 
jet. Since the magnetic field over the target will 
effect both 
the heat distribution in, and forces on, such a jet, an 
experiment is required. 
 \item
The simulation must make assumptions on the cross 
sections for 
secondary pion production by products of the primary 
interaction. 
This information is needed at low final 
energies and large angles where data is inadequate. A 
conventional 
experiment to determine all such cross sections would be 
expensive.
 \item
We need to know the total radiation directed towards the 
capture 
and focusing solenoids. Shielding will have to be 
provided to 
protect the insulation of the inner resistive solenoid, 
and limit 
heating of the outer superconducting magnets. Only 
direct measurement of such radiation can provide a reliable 
determination. 
 \item
In the current design of phase rotation, the 
first rf 
cavity is placed 3 m from the target. If unshielded, the 
radiation 
level at this point will be very high. We have little data on the performance of a cavity 
under such 
conditions and thus have difficulty calculating the 
shielding 
requirements.
 \end{itemize} 
\subsubsection{Ionization Cooling Experimental R \& D.}
Although the principals of ionization cooling are 
relatively 
simple, there are practical problems in designing 
lattices that 
can transport, and focus the large emittances without 
exciting betatron resonances that blow up the emittance 
and 
attenuate the beam. There will also be problems with 
space charge 
and wake field effects. 

After a design has been defined and simulated, 
demonstrations
will be required. They will 
require significant rf acceleration ($\approx$ 100 MeV) 
and several 
meters of high field solenoids interspersed with bending 
magnets 
and, for a final stage demonstration, current carrying 
lithium rods.
Such an experiment has not been designed yet. It has been suggested that 
this 
experiment might be carried out at FNAL.

An R \& D program would also be required to develop the 
current 
carrying rods. This could be undertaken in a 
collaboration 
between BINP, Novosibirsk, and FNAL.
\subsubsection{Magnet Design and Acceleration 
Experimental R \& D.}
R \& D programs are required both for the high field 
pulsed cosine theta 
magnets and for the lower field pulsed field magnets. The R \& D on the former  is somewhat more urgent since they 
are less 
conventional.

Some R \& D work is also needed to determine the 
performance of 
the required superconducting cavities when excited for 
the 
relatively short pulse durations required. Studies of 
their 
sensitivity to muon decay electrons may also be needed. 
\subsubsection{Collider Ring Experimental R \& D.}
The insertion quadrupoles need urgent R \& D because the 
lattice 
design work depends on the gradients that are achieved. 
${\rm Nb_3Sn},$ or other higher field conductor will be 
prefered. Since 
the magnets operate at a constant field, metallic 
insulation may be acceptable, which would obviate the need for 
impregnation and thus provide better cooling. High $T_c$ 
materials should be considered.

The dipole magnets, if of cosine theta design, may develop excessive mid plane compression in their coils. 
Block 
conductor arrangements may need to be developed. The 
use of 
${\rm Nb_3Sn}$ will again be prefered for its high field 
capability. 	    
\subsubsection{Detector Experimental R \& D.}
Detector R \& D is required to develop the required 
detectors and 
confirm that they can both withstand the expected 
radiation and 
separate the tracks of interest from the background. 
\section{CONCLUSION}
\begin{itemize}
 \item    The initial motive for the study of \mumu colliders was: 
 \begin{itemize}
 \item The lack of beamstrahlung constraints allowing the circulation of 
muons, and the suppression of beamstrahlung which could, in principle, give a 
luminosity an advantage relative to \ee at high energies. 
   \item The realization that, despite the problems in using 
muons whose lifetime is short and production diffuse, it is 
possible to sketch a design for \mumu collider with parameters:
 \begin{itemize}
 \item  Energy = 4 TeV 
 \item  Luminosity = $10^{35}$   ($ cm^{-2}  sec^{-1}$) 
 \item  More moderate power requirements and tolerances than those in an \ee 
collider with the same specification. 
 \end{itemize}
 \end{itemize}
 
 \item  A \mumu collider would have some unique Physics Advantages:

\begin{itemize}
 \item  Because of the lack of beamstrahlung, a \mumu collider  could have 
very narrow energy spread: dE/E = 0.1 - .01 \% 
 \item  Observed reactions would have low Radiative Corrections;
 \item  Cross Section of \mumu to S-channel production of any Higgs Boson (h, 
H, A) would be approximately 40,000 times higher than for \ee. This together 
with the above items, would allow precision measurements of masses and widths 
not possible with \ee.
 \end{itemize}
 
 \item We note that:

\begin{itemize}
 \item  Although a \mumu collider is radically different from existing 
machines, yet it requires no ``exotic" technology; rather, its components would 
be modest extensions of existing technology, though used in an unusual manner. 
 \item  A \mumu collider would be a multipurpose facility: besides \mumu 
collisions, $\mu$-p and $\mu$-Ion collisions could be possible. Its proton 
driver could be a substantial source of spallation neutrons, and intense beams 
of  pions,  kaons, neutrinos and muons would be available.
 \item  A \mumu collider would be an order of magnitude smaller in overall 
size, and about a factor of 6 less in total tunnel length, than current \ee 
collider designs. Because of its small size, it would fit on one of several 
existing lab sites. 
 \item Consistent with its smaller size, it is estimated that a \mumu 
collider would be significantly cheaper to construct. It might thus 
become affordable in a fiscal enviroment which may not allow larger ``mega-
science" projects.
\end{itemize}

 \item   But we recognize disadvantages compared to an \ee machine:
\begin{itemize}
 \item  A \mumu collider would have more background than \ee.
 \item  The muons would have less polarization than electons, although, in 
partial compensation, both \mup and \mum would be equally polarized.
 \item  A gamma-gamma capability would not be possible.
 \item  Although much progress has been made, the concept of  a \mumu collider 
is immature and there could yet be a fatal flaw or some problem could make it 
impossibly expensive. 
    \end{itemize}
\item  The \mumu collider needs much R \& D.
     \begin{itemize}
\item Both theoretical, the highest proiority items being:

     \begin{itemize}
\item  Design and simulate  a complete lattice for cooling. 
\item continue to study instabilities in the collider ring.
\item design collider injection and beam halo scraping. 
     \end{itemize}
\item and experimental: the highest priority items being

     \begin{itemize}
\item  
A demonstrate of muon cooling cooling is essential to show that
hardware can operate and be stable.
  \item Pion capture and  rf phase rotation must be demonstrated.
  \item Many components need modelling, in particular:
lithium 
lenses for cooling, pulsed magnets for acceleration, high field quadrupoles 
for the final focus, large aperture diploles for the collider ring, and muon 
collimators.
     \end{itemize}
\item We estimate that
about five years of R \& D is needed.
 \end{itemize}
\item If this R \& D is successful then we believe a ).5 TeV demonstration collider, with significant physics potential, could be built by 2010; and a 4 TeV collider might be possible a decade later.
     \end{itemize}

\section{Acknowledgment}
We acknowledge important contributions from many colleagues, especially 
those that contributed to the feasibitity study submitted to the Snowmass Workshop 96  Proceedings\cite{book} from which much of the material and some text, for this report has been taken: 
  C. Ankenbrandt (FermiLab)
, A. Baltz (BNL)
, V. Barger (Univ. of Wisconsin)
, O. Benary (Tel-Aviv Univ.)
, M. S. Berger (Indiana Univ.)
, A. Bogacz (UC, Los Angeles)
, W-H Cheng (LBNL)
, D. Cline (UC, Los Angeles)
, E. Courant (BNL)
, D. Ehst (ANL)
, T. Diehl (Univ. of Illinois, Urbana)
, R. C. Fernow (BNL)
, M. Furman (LBNL)
, J. C. Gallardo (BNL)
, A. Garren (LBNL)
, S. Geer (FermiLab)
,  I. Ginzburg (Inst. of Math., Novosibirsk)
, H. Gordon (BNL)
, M. Green (LBNL)
, J. Griffin (FermiLab)
, J. F. Gunion (UC, Davis)
, T. Han (UC, Davis)
, C. Johnstone (FermiLab)
, D. Kahana (BNL)
, S. Kahn (BNL)
, H. G. Kirk (BNL)
, P. Lebrun (FermiLab)
, D. Lissauer (BNL)
~, A. Luccio (BNL)
, H. Ma (BNL)
, A. McInturff (LBNL)
, F. Mills (FermiLab)
, N. Mokhov (FermiLab)
, A. Moretti (FermiLab)
, G. Morgan (BNL)
, M. Murtagh (BNL)
, D. Neuffer (FermiLab)
, K-Y. Ng (FermiLab)
, R. J. Noble (FermiLab)
, J. Norem (ANL)
, B. Norum (Univ. Virginia)
, I. Novitski (FermiLab),
  K. Oide (KEK),
 F. Paige (BNL)
, J. Peterson (LBNL)
, V. Polychronakos (BNL)
, M. Popovic (FermiLab)
, S. Protopopescu (BNL)
, Z. Qian (FermiLab)
, P. Rehak (BNL)
, R. Roser (Univ. of Illinois, Urbana)
, T. Roser (BNL)
, R. Rossmanith (DESY)
, Q-S Shu, (CEBAF) 
, A. Skrinsky (BINP)
, I. Stumer (BNL)
, S. Simrock (CEBAF)
, D. Summers (Univ. of Mississippi)
, H. Takahashi (BNL)
, H. Takai (BNL)
, V. Tchernatine (BNL)
, Y. Torun (SUNY, Stony Brook), D. Trbojevic (BNL)
, W. C. Turner (LBNL), A. Van Ginneken (FermiLab)
,  E. Willen (BNL)
, W. Willis (Columbia Univ.)
, D. Winn (Fairfield Univ.)
,  J. S. Wurtele (UC, Berkeley)
, Y. Zhao (BNL). In particular we acknowledge the contributions of  the Editors of each one of the chapters of the $\mu^+\mu^-$ Collider: A Feasibility Study: V. Barger, 
J. Norem, R. Noble, H. Kirk, R. Fernow, D. Neuffer, J. Wurtele, D. Lissauer, 
M. Murtagh, S. Geer, N. Mokhov and D. Cline.
\medskip
This research was supported by the U.S. Department of Energy under Contract No.
DE-ACO2-76-CH00016 and DE-AC03-76SF00515.

\end{document}